\documentclass[12pt]{iopart}
\usepackage{epsfig,graphicx}
\newcommand{\qav}{{\bra q \ket}}
\newcommand{\intw}{{\int \frac{d\omega}{2\pi}}}
\newcommand{\half}{{\frac{1}{2}}}
\newcommand{\kav}{{\bra k \ket}}
\newcommand{\kvar}{{\bra k^2\ket}}
\newcommand{\qvar}{{\bra q^2\ket}}
\newcommand{\dav}{{\bra d \ket}}
\newcommand{\dprime}{{\prime\prime}}
\newcommand{\tprime}{{\prime\prime\prime}}
\newcommand{\intwwww}{{\int \frac{d\omega d\omega^\prime d\omega^\dprime
d\omega^\tprime}{12\pi^4}}}
\newcommand{\intwww}{{\int \frac{d\omega d\omega^\prime d\omega^\dprime}{8\pi^3}}}
\newcommand{\intww}{{\int \frac{d\omega d\omega^\prime}{4\pi^2}}}
\newcommand{\N}{{\rm I\!N}}

\newcommand{\pprime}{{\prime\prime}}

\newcommand{\be}{\begin{equation}}
\newcommand{\ee}{\end{equation}}
\newcommand{\bea}{\begin{eqnarray}}
\newcommand{\eea}{\end{eqnarray}}

\newcommand{\ed}{\end{displaymath}}

\newcommand{\bra}{\langle}
\newcommand{\ket}{\rangle}

\newcommand{\order}{{\cal O}}

\newcommand{\ba}{\mbox{\boldmath $a$}}
\newcommand{\bc}{\mbox{\boldmath $c$}}

\newcommand{\bxi}{\mbox{\boldmath $\xi$}}

\newcommand{\cij}{{c_{ij}}}

\newcommand{\bcirc}{\circle*{10}}

\newcommand{\bsqr}{\rule{10\unitlength}{10\unitlength}}

\begin{document}

\title[Protein interaction networks and biology: towards the connection]{Protein interaction networks and biology: \\towards the connection}

\author{A Annibale$^\dag$, ACC Coolen$^{\dag\ddag\S}$, N Planell-Morell$^\dag$}
\address{$\dag$ ~ Department of Mathematics, King's College London, The Strand,
London WC2R 2LS, UK}
\address{$\ddag$ Institute for Mathematical and Molecular Biomedicine, King's College London,  Hodgkin Building,
London SE1 1UL, UK}
\address{$\S$ London Institute for Mathematical Sciences, 22 South Audley St, London W1K 2NY, UK}

\begin{abstract}
Protein interaction networks (PIN) are popular means to visualize the proteome. 
However, PIN datasets are known to be noisy, incomplete and biased by the experimental protocols used 
to detect protein interactions.
This paper aims at 
understanding the connection between true protein interactions and the protein 
interaction datasets that have been obtained using 
the most popular experimental techniques, i.e. mass spectronomy (MS) and 
yeast two-hybrid (Y2H).
We show that the most natural adjacency matrix of protein interaction networks has a 
separable form, and this induces precise relations between moments of the 
degree distribution and the number of short loops. These relations 
provide powerful tools to test the reliability of datasets and 
hint at the underlying biological mechanism with which proteins and complexes 
recruit each other.
\end{abstract}


\section{Introduction}

A protein interaction network (PIN) is a graph where nodes $i=1\ldots N$ represent proteins and 
links represent their interactions. 
This graph is encoded in an adjacency matrix ${\bf a}= \{a_{ij}\} $,
whose entries denote whether  there is 
a link between proteins $i$ and $j$ ($a_{ij}\!=\!1$) or not ($a_{ij}\!=\!0$).
However, there is ambiguity in its definition, arising from the 
non-binarity of the underlying biochemistry. For example, three proteins may form 
a complex, but may not interact in pairs. Assigning binary values to intrinsically non-binary interactions requires further prescriptions, which vary across experimental protocols 
and lead in practice to different graphs. 
Moreover, different experiments 
measure protein interactions in different ways, which causes further 
biases 
\cite{HakPinRobLov08,Han2005,Silva2006}. For quantitative studies of the effects of sampling biases on 
networks see 
e.g. \cite{PLOS,Stumpf2005a,Stumpf2005b,Lee2006,Viger,Solokov,AnnCoo11}.
 
In this paper we seek to establish the connection between 
true biological protein interactions and protein interaction datasets 
produced by the most 
popular experimental techniques, mass spectronomy (MS) 
and yeast two-hybrid (Y2H). We argue that 
the most natural network matrix representation of the proteome has a separable form, which induces 
precise relations between the degree distribution and the
density of short loops. These relations provide 
simple tests to assess
the reliability and quality of different data sets, and provide hints on the underlying (evolutionary) mechanisms with which proteins and complexes recruit each other. 
Our study also provides a theoretical framework to discriminate between `party' and `date' hubs in protein interaction networks,  
see e.g. \cite{ChangETAL13} and references therein, and 
addresses several intriguing questions 
concerning the universality of 
protein and complex statistics across species. 
For example, 
given $N$ protein species in a cell, what is the number of complexes they 
typically form, i.e. to what extent is the ratio complexes/proteins conserved 
across different species?
Is the distribution of complex sizes peaked around `typical' values, or does it have long tails? How is this mirrored in the protein 
promiscuities, i.e. the propensities of proteins to participate in multiple
complexes?
Does the power law behaviour of the 
degree distribution of protein interaction networks perhaps result from 
 tails in the distribution of complex sizes and protein promiscuities? 

We tackle the above questions using an approach that is entirely based 
on statistical properties of graph ensembles. In section \ref{sec:def} we first define our models. Sections \ref{sec:q}, 
\ref{sec:d} and \ref{sec:mix} are devoted to the derivation of properties of distinct separable graph 
ensembles which mimic protein interaction networks, each reflecting different possible mechanisms for 
complex genesis. In section \ref{sec:syn} we test these properties in synthetically generated graphs, 
and in section \ref{sec:real} we do the same for protein interaction networks measured by MS and Y2H experiments. 
We end our paper with a summary of our conclusions, and suggest pathways for further research.

\section{Definitions and basic properties}
\label{sec:def}

\subsection{The bipartite graph representation of the proteome}

\unitlength=0.28mm
\begin{figure}[t]
\hspace*{45mm}
\begin{picture}(400,140)
\thinlines
\put(-120,102){\small\em protein species}
\put(-120,84){\small\em $i=1\ldots N$}
\put(-120,15){\small\em complexes}
\put(-120,-2){\small\em $\mu=1\ldots \alpha N$}

\put(10,117){\small $d_1\!=\!4$}
\put(63,117){\small $d_2\!=\!2$}
\put(190,117){\small $\cdots\cdots\cdots$}
\put(361,117){\small $d_N\!=\!3$}

\put(-12,-19){\small $q_1\!=\!2$}
\put(40,-19){\small $q_2\!=\!2$}
\put(210,-19){\small $\cdots\cdots\cdots$}
\put(386,-19){\small $q_{\alpha N}\!=\!3$}

\multiput(30,100)(50,0){8}{\bcirc}
\multiput(0,0)(50,0){9}{\bsqr}
\put(30,100){\line(-1,-4){24}} \put(30,100){\line(1,-4){24}} \put(30,100){\line(3,-1){280}}
   \put(30,100){\line(5,-2){230}} \put(30,100){\line(3,-1){280}}
\put(80,100){\line(-4,-5){80}} \put(80,100){\line(4,-5){80}} 
\put(130,100){\line(-1,-4){24}} \put(130,100){\line(-4,-5){80}} 
 \put(130,100){\line(5,-2){230}} 
\put(180,100){\line(-1,-4){24}} \put(180,100){\line(1,-4){24}}\put(180,100){\line(-4,-5){80}} \put(180,100){\line(4,-5){80}}
 \put(180,100){\line(4,-3){120}}
\put(230,100){\line(4,-5){80}} \put(230,100){\line(-1,-4){24}}
 \put(280,100){\line(1,-4){24}} \put(280,100){\line(4,-5){80}}\put(280,100){\line(4,-3){120}}
\put(330,100){\line(-1,-4){24}} \put(330,100){\line(4,-5){80}}
\put(380,100){\line(-4,-5){80}}\put(380,100){\line(1,-4){24}}\put(380,100){\line(-1,-4){24}}

\end{picture}
\vspace*{8mm}

\caption{Bipartite graph (or `factor graph') representation of protein interactions. The protein species $i=1\ldots N$ are drawn as circles, and their complexes $\mu=1\ldots \alpha N$ as squares. We write the degree of protein $i$ as $d_i$ (the number of complexes it participates in), and the degree of complex $\mu$ as $q_\mu$ (the number of protein species it contains).  The bipartite graph gives more detailed information than the conventional PIN with protein nodes and pairwise links only. For instance, one distinguishes easily between different types of `hub' proteins: `date hub' proteins connect to many degree-2 complexes, whereas `party hub' proteins connect to a high degree complex.   }
\label{fig:bipartite}
\end{figure}
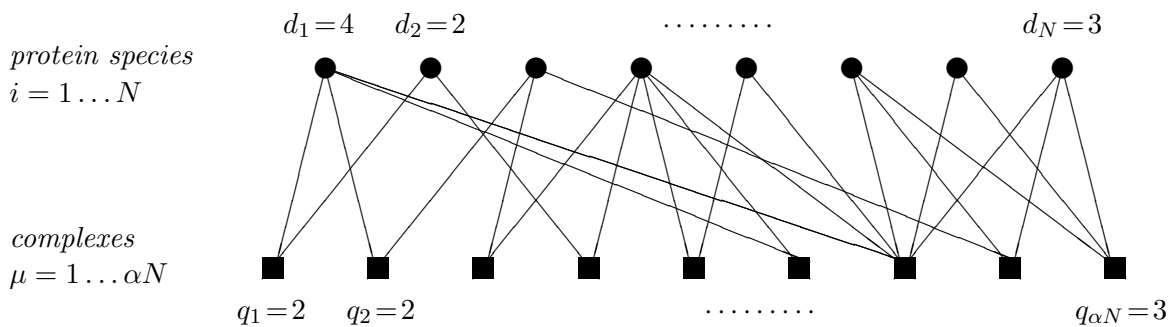

Proteins are large and complicated heteropolymers, which can bind in specific combinations to form stable molecular complexes.
We consider a set of $N$ protein species, labelled by $i=1\ldots N$. We assume 
that the number of stable complexes $p$ scales as $p=\alpha N$ where 
$\alpha >0$, and we label the complexes by $\mu=1\ldots \alpha N$.
We can represent this system as a bi-partite 
graph \cite{NewStrWat02}, see Figure \ref{fig:bipartite},  with 
 two sets of nodes. The set $\nu_p$ represents
proteins (drawn as circles), the set $\nu_c$ represents complexes 
(drawn as squares), and 
a link between protein $i\in\nu_p$ and complex $\mu\in \nu_c$ is drawn
 if protein $i$  participates  in complex $\mu$. 
This graph is defined by the $N\!\times\! \alpha N$ connectivity matrix $\bxi=\{\xi_i^\mu\}$, 
where 
$\xi_i^\mu=1$ if there is a link between $i$ and $\mu$, and 
$\xi_i^\mu=0$ otherwise. For simplicity we do not allow for 
complexes with more than one occurrence of any given protein species. 

In the bipartite graph one has two types of node degrees: the degree $d_i(\bxi)=\sum_\mu \xi_i^\mu$ 
(or `promiscuity') of each protein $i$ gives the number of different complexes in which it is involved, and the degree  $q_\mu(\bxi)=\sum_i \xi_i^\mu$ (or `size') of each complex 
$\mu$ gives the number of protein species of which it is formed.
We define the distribution of promiscuities in graph $\bxi$
as $p(d|\bxi)=N^{-1}\sum_i \delta_{d,d_i(\bxi)}$,  with the 
average promiscuity  $\bra d(\bxi) \ket=\sum_d d p(d|\bxi)$, 
and the distribution of complex sizes as $p(q|\bxi)=(\alpha N)^{-1}\sum_{\mu=1}^{\alpha N}\delta_{q,q_\mu(\bxi)}$, with the average complex
size $\bra q(\bxi) \ket =\sum_q q p(q|\bxi)$. Since the number of links is conserved, we always have 
$\bra d(\bxi) \ket =\alpha \bra q(\bxi) \ket$ for any bipartite graph $\bxi$.

\subsection{Link distribution in the bipartite graph}

Since we generally do not know the microscopic bipartite graph $\bxi$,
we will regard it as a quenched random object. Several natural choices can be proposed for 
its distribution $p(\bxi)$.
If we assume that complexes recruit proteins,  independently and with the same likelihood, we are led to
\begin{eqnarray}
p_A(\bxi)&=& \prod_{i\mu}\left[\frac{q_\mu}{N}\delta_{\xi_i^\mu,1}+\left(1- \frac{q_\mu}{N}\right)\delta_{\xi_i^\mu,0}\right]
\label{eq:quenched_q}
\end{eqnarray}
with $\delta_{xy}=1$ for $x=y$ and $0$ otherwise, and where the $\{q_\mu\}$ are distributed according to 
$P(q)=(\alpha N)^{-1}\sum_\mu \delta_{q, q_\mu}$. For graphs $\bxi$ drawn from the ensemble (\ref{eq:quenched_q}) and $N\to\infty$, each complex size $q_\mu(\bxi)$ is a Poissonian random variable 
with average $q_\mu$, and all protein promiscuities $d_i(\bxi)$ are Poissonian 
variables with  
average $\bra d \ket=\alpha \qav$, since
\bea
p(d)&=&\lim_{N\to\infty}\bra \delta_{d,\sum_\mu \xi_i^\mu}\ket =\lim_{N\to\infty}
\int_{-\pi}^\pi \frac{\rmd\omega}{2\pi} \rme^{\rmi\omega d} \bra \rme^{-\rmi\omega \sum_\mu \xi_i^\mu}\ket
\nonumber\\
&=&\int_{-\pi}^\pi  \frac{\rmd\omega}{2\pi} \rme^{\rmi\omega d+\alpha \bra q\ket (\rme^{-\rmi\omega}-1)}=\rme^{-\alpha \bra q \ket}
(\alpha \bra q \ket)^d/d! 
\label{eq:pq_dp}
\eea
In the scenario (\ref{eq:quenched_q}) complexes have sizes that are determined e.g. by their functions, and this controls the promiscuities of the recruited proteins. 
Alternatively one could assume that the likelihood of a protein participating in a complex is driven by its promiscuitiy, leading 
to the `dual' ensemble 
\begin{eqnarray}
p_B(\bxi)&=& \prod_{i\mu}\left[\frac{d_i}{\alpha N}\delta_{\xi_i^\mu,1}+\left(1- \frac{d_i}{\alpha N}\right)\delta_{\xi_i^\mu,0}\right]
\label{eq:quenched_d}
\end{eqnarray}
where the $\{d_i\}$ are  distributed according to 
$P(d)=N^{-1}\sum_i \delta_{d, d_i}$.
Here as $N\to\infty$ the protein promiscuities $d_i(\bxi)$ are Poissonian variables  with averages $d_i$, whereas 
all complex sizes $q_\mu(\bxi)$ are Poisson variables with identical average  $\qav=\dav/\alpha$, since
\bea
p(q)&=&\lim_{N\to\infty} \bra \delta_{q,\sum_i \xi_i^\mu}\ket
= \lim_{N\to\infty}\int_{-\pi}^\pi  \frac{\rmd\omega}{2\pi} \rme^{\rmi\omega q} \bra \rme^{-\rmi\omega \sum_i \xi_i^\mu}\ket=
\nonumber\\
&=&\int_{-\pi}^\pi  \frac{\rmd\omega}{2\pi} \rme^{\rmi\omega q+\frac{\bra d \ket}{\alpha}(\rme^{-\rmi\omega}-1)}=\rme^{-\dav/\alpha}
(\dav/\alpha)^q/q! 
\label{eq:pd_qp}
\eea
In this second ensemble proteins have intrinsic promiscuities, determined e.g. by the number of their binding sites, their polarization and so on, 
and these drive their recruitment to complexes. 
A third obvious choice is the `mixed' ensemble 
\begin{eqnarray}
p_C(\bxi)&=& \prod_{i\mu}\left[\frac{d_iq_\mu}{\alpha N \qav}\delta_{\xi_i^\mu,1}+
\left(1- \frac{d_iq_\mu}{\alpha N \qav}\right)\delta_{\xi_i^\mu,0}\right]
\label{eq:quenched_qd}
\end{eqnarray}
where all protein promiscuities and complex sizes are constrained on average, 
i.e.
$\bra d_i(\bxi)\ket=d_i$ and $\bra q_\mu(\bxi)\ket=q_\mu$, with $\{d_i\}$ and $\{q_\mu\}$ 
distributed according to $P(d)$ and $P(q)$. 
Here protein binding  
statistics are driven both by complex functionality and protein promiscuity factors.
The mixed ensemble (\ref{eq:quenched_qd}) reduces to (\ref{eq:quenched_q}) for the choice $P(d)=\delta_{d,\alpha\qav}$, 
and to (\ref{eq:quenched_d}) when $P(q)=\delta_{q,\qav}$. 
By determining which of the above ensemble reflects better biological reality, we will thus learn about 
the mechanisms with which complexes and proteins recruit each 
other. 

The above  three ensembles become equivalent when $q_\mu=\bra q\ket~\forall~\mu$ and $d_i=\alpha\bra q\ket~\forall~i$.  In that case complex sizes and protein 
promiscuities are homogeneous, and the recruitment process between 
proteins and complexes is fully random. 
Bipartite graphs drawn from (\ref{eq:quenched_q}) were found to have 
modular topologies, and to accomplish parallel information processing 
for suitable values of the parameter $\alpha$
\cite{Peter,Roma}. Their ensemble entropy has been 
calculated in \cite{Kate}.
One can show easily that 
if one replaces the soft constraints on the local degrees in our soft-constrained graph ensembles (\ref{eq:quenched_q},\ref{eq:quenched_d})  by hard constraints, then one finds asymptotically the
same distributions (\ref{eq:pq_dp},\ref{eq:pd_qp}).  Finally, we note that all three ensembles (\ref{eq:quenched_q},\ref{eq:quenched_d},\ref{eq:quenched_qd}) are of the form 
$p(\bxi)=\prod_{i\mu}p_{i\mu}(\xi_i^\mu)$, 
so there are no correlations between the entries of $\bxi$. This strong 
assumption of our models 
will need to be checked a posteriori. 

\subsection{Accounting for binding sites}

In all PINs each protein is reduced to a simple network node, 
in spite of the fact that proteins are in reality complex chains of aminoacids with several binding domains.
Here we show that the ensembles introduced in the previous section 
can accommodate the presence of multiple 
binding sites when these are equally reactive.
Let us first assume that each protein has $d$ functional reactive amino-acid endgroups. 
When two such proteins bind, the resulting dimer has $2d-2$ unused reactive 
endgroups, a trimer has $3d-4$ endgroups, and a $k$-mer has 
$kd-2(k-1)=(d-2)k+2$ endgroups. If all endgroups are equally reactive, 
the  a priori probability that a protein $i$ is part of a complex $\mu$ is given 
by 
\bea
p(\xi_i^\mu=1)=\frac{d[(d-2)q_\mu+2]}{Z}\simeq 
\frac{q_\mu d}{\alpha N \bra q \ket}
\eea 
where the last approximate equality holds for $d\gg 1$ and 
$Z=\sum_\mu q_\mu d=\alpha N \bra q \ket d$. This corresponds 
to ensemble (\ref{eq:quenched_q}), with the choice $d=\alpha 
\bra q \ket$.
If proteins have different endgroups $d_i$, 
\bea
p(\xi_i^\mu=1)\simeq \frac{d_i[(d-2)q_\mu+2]}{\alpha N \bra q \ket d}\simeq 
\frac{d_i q_\mu}{\alpha N \bra q \ket}
\eea 
where $d=N^{-1}\sum_i d_i$, leading to ensemble (\ref{eq:quenched_qd}).
If the variability of $q_\mu$ is small, $q_\mu\simeq \bra q \ket$,
\bea
p(\xi_i^\mu=1)
=\frac{d_i}{\alpha N}
\eea 
and we retrieve (\ref{eq:quenched_d}). 
The assumption of unbiased interactions between proteins with varying 
individual binding affinities has been supported in \cite{IvaWalRei08b}.

\subsection{Protein interactions as detected by experiments}

Protein detection experiments seek to measure for each pair $(i,j)$ of protein species whether they interact in any complex, and 
assign an undirected link between nodes $i$ and $j$ if they do. Hence the PIN adjacency matrix $\ba=\{a_{ij}\}$ resulting from such experiments can be expressed in terms of the entries of the bipartite graph $\bxi$ in Figure \ref{fig:bipartite} via
\be
a_{ij}=\theta(\sum_{\mu=1}^{\alpha N} \xi_i^\mu \xi_j^\mu)\quad \forall~ i\neq j
\label{eq:separable}
\ee 
and $a_{ii}=0~\forall~i$, with the convention $\theta(0)=0$ for the step function, defined by $\theta(x>0)=1$ and $\theta(x<0)=0$.
The aim of this paper hence translates into 
studying the properties of the following ensemble of 
nondirected random graphs, in which the $\{\xi_i^\mu\}$ are drawn from either of the ensembles (\ref{eq:quenched_q},\ref{eq:quenched_d},\ref{eq:quenched_qd}):
\begin{eqnarray}
p({\bf a})&=& \Big\bra \left[\prod_{i<j}\delta_{a_{ij},\theta(\sum_{\mu\leq \alpha N}
\xi_i^\mu\xi_j^\mu)}\right]\left[\prod_i \delta_{a_{ii},0}\right]\Big\ket_{\bxi}
\label{eq:ensemble_a}
\end{eqnarray}
Some properties of (\ref{eq:quenched_q},\ref{eq:quenched_d}) will turn out not to depend on the 
choices made for the distributions of complex sizes and protein promiscuities, and this leads to powerful benchmarks against which to test available PIN datasets.  
A key feature we exploit  in our analysis is that averages over (\ref{eq:ensemble_a}) can often be replaced by averages over 
the following related ensemble of {\em weighted} graphs
\begin{eqnarray}
p(\bc)&=& \Big\bra \left[\prod_{i<j}\delta_{c_{ij},\sum_{\mu\leq \alpha N}
\xi_i^\mu\xi_j^\mu}\right]\left[\prod_i \delta_{c_{ii},0}\right]\Big\ket_{\bxi}
\label{eq:ensemble_c}
\end{eqnarray}
Here an entry $c_{ij}=\sum_{\mu\leq \alpha N}\xi_i^\mu\xi_j^\mu\in\N$ represents the {\em number} of complexes in which proteins $i$ and $j$ participate simultaneously.
For finite $q_\mu, d_i$ and $\alpha$, one finds that in large networks generated via (\ref{eq:quenched_q},\ref{eq:quenched_d},\ref{eq:quenched_qd}) the probability of seeing $c_{ij}>1$ is of order $\order{(N^{-2})}$, and the values of many macroscopic 
observables in the ${\bf a}$ and $\bc$ ensembles will, to leading order in $N$, be identical.

\section{Network properties generated by the $q$-ensemble}
\label{sec:q}

In this section we study the statistical properties of the ensembles 
(\ref{eq:ensemble_c}) and (\ref{eq:ensemble_a}) upon generating the bipartite protein interaction graph $\bxi$  from ensemble (\ref{eq:quenched_q}), where complexes recruit proteins. 

\subsection{Link probabilities}

For the graphs $\bc$ of (\ref{eq:ensemble_c})  we find the following expectation values of individual bonds 
\begin{eqnarray}
\bra c_{ij}\ket&=& \sum_{\mu=1}^{\alpha N}\bra \xi^\mu_i\xi^\mu_j\ket_{\bxi} = \sum_{\mu=1}^{\alpha N}\left(\frac{q_\mu}{N}\right)^2=\frac{\alpha}{N}\bra q^2\ket
\label{eq:pij_q}
\end{eqnarray}
where the brackets on the right-hand side denote averaging over the complex size distribution $P(q)$.
The likelihood of an individual bond is (see \ref{app:pij})
\begin{eqnarray}
p(c_{ij})&=& \big\bra \delta_{c_{ij},\sum_{\mu\leq \alpha N}\xi_i^\mu\xi_j^\mu}\big\ket_{\bxi}
\nonumber\\
&=&
\delta_{\cij,0}+\frac{\alpha \bra q^2\ket}{N}(\delta_{\cij,1}-\delta_{\cij,0})
+\Big(\frac{\alpha^2 \bra q^2\ket^2}{2N^2}-\frac{1}{2}\frac{\alpha \bra q^4\ket}{N^3}\Big)(\delta_{\cij,2}-2\delta_{\cij,1}+\delta_{\cij,0})
\nonumber\\
&&+\frac{\alpha^3 \bra q^2\ket^3}{6N^3}(\delta_{\cij,3}-3\delta_{\cij,2}+3\delta_{\cij,1}-\delta_{\cij,0})
+\order{(N^{-4})}
\end{eqnarray}
so we find for the first few probabilities:
\begin{eqnarray}
p(0)&=&1-\frac{\alpha \bra q^2\ket}{N}+\frac{\alpha^2 \bra q^2\ket^2}{2N^2}
-\frac{\alpha \bra q^4\ket}{2N^3}-\frac{\alpha^3 \bra q^2\ket^3}{6N^3}
+\order{(N^{-4})}
\\
p(1)&=& \frac{\alpha \bra q^2\ket}{N}
-\frac{\alpha^2 \bra q^2\ket^2}{N^2}
+\frac{\alpha \bra q^4\ket}{N^3}+\frac{\alpha^3 \bra q^2\ket^3}{2N^3}
+\order{(N^{-4})}
\end{eqnarray}
and hence
\begin{eqnarray}
&&\hspace*{-2mm}
\sum_{\ell>1}p(\ell)=1\!-\!p(0)\!-\!p(1)=\order(N^{-2}),~~~~~~
\sum_{\ell>1}\ell p(\ell)=\bra c_{ij}\ket-p(1)=\order(N^{-2})
\end{eqnarray}
The probability to have $c_{ij}\neq 0$
is of order $\order{(N^{-1})}$, so the graphs generated by (\ref{eq:ensemble_c})   are finitely connected. 
Moreover, although the graphs $\bc$ are in principle weighted, for large $N$ the number of links per node that are not in $\{0,1\}$ will be vanishingly small.

\subsection{Densities of short loops}

We now turn to the calculation of expectation values for different observables in ensemble (\ref{eq:ensemble_c}).
First, we calculate the average number of ordered and oriented loops of length $3$ per node, which are (see \ref{app:pij}):
\begin{eqnarray}
m_3&=& \Big\bra\frac{1}{N}\sum_{ijk}c_{ij}c_{jk}c_{ki}\Big\ket_{\bxi}
=\frac{1}{N}\sum_{\mu\nu\rho=1}^{\alpha N}\sum_{i\neq j\neq k} \Big\bra
\xi^\mu_i\xi^\mu_j\xi^\nu_j\xi^\nu_k\xi^\rho_k\xi^\rho_i
\Big\ket_{\bxi}
\label{eq:m3_def}
\\[-1mm]
&=&\alpha \bra q^3\ket  +\order(N^{-1})
\label{eq:m3q}
\end{eqnarray}
Calculating the density of loops $m_L$ for lengths  $L>3$ 
can be simplified by returning to the bipartite graph $\bxi$. We define a star $S_n$ to be a simple $(n\!+\!1)$-node 
tree in  $\bxi$, of which the central 
node belongs to $\nu_c$ (the complexes), and the $n$ leaves belong to $\nu_p$ (the proteins). Thus  
$S_2$ stars 
represent protein dimers, $S_3$ stars represent protein trimers, and so 
on. Each link in $\bc$ corresponds to at least one
$S_2$ star in the bipartite graph
(which, in turn, can be a subset of any $S_n$ star with $n>2$). 
Therefore, the total number of $S_2$ stars in the bipartite graph, 
\be
\sum_\mu \sum_{i\neq j}\bra \xi_i^\mu \xi_j^\mu\ket=
\sum_\mu \sum_{i\neq j}\bra \xi_i^\mu\ket \bra \xi_j^\mu\ket=
\sum_{i\neq j}\sum_\mu \frac{q_\mu^2}{N^2}=\alpha(N-1)\bra q^2\ket
\ee
has to equate in leading order the total number of links $N\bra k \ket$ in 
graph $\bc$, yielding 
\be
\bra q^2\ket=\frac{\bra k\ket}{\alpha}+\order{(N^{-1})}
\label{eq:kav_q}
\ee
which is indeed in agreement with the result of the direct calculation
$\bra k\ket= N^{-1}\sum_{ij} \bra c_{ij}\ket$, using (\ref{eq:pij_q}). 
Similarly we can obtain the number of loops of length $3$, 
calculated earlier, by realising that these loops arise 
when we have in the bipartite graph
either a star $S_3$ (which can be a subset of any $S_n$ with $n>3$)
or a combination of three $S_2$ stars, where every leaf is shared by two 
stars. The contribution of the number of $S_3$ stars per node to the number 
of loops of length $3$ is 
\bea
\frac{1}{N}\sum_\mu \sum_{i\neq j\neq k(\neq i)}\bra \xi_i^\mu \xi_j^\mu 
\xi_k^\mu\ket&=&\frac{1}{N}\sum_\mu \sum_{i\neq j\neq k(\neq i)}\bra \xi_i^\mu \ket \bra\xi_j^\mu \ket \bra \xi_k^\mu\ket\nonumber\\
&=&\frac{1}{N}\sum_\mu \sum_{i\neq j\neq k(\neq i)}\frac{q_\mu^3}{N^3}
=\alpha \bra q^3\ket+\order{(N^{-1})}
\eea
The contribution of the combination of three $S_2$ stars, where each 
leaf is shared by two stars, is 
\bea
\frac{1}{N}\sum_{[\mu, \nu, \rho]} 
\sum_{[i,j,k]}\bra \xi_i^\mu \xi_j^\mu \xi_j^\nu \xi_k^\nu 
\xi_k^\rho \xi_i^\rho\ket=\frac{1}{N}\sum_{[\mu,\nu, \rho]} 
\sum_{[i,j,k]}\frac{q_\mu^2 q_\nu^2 q_\rho^2}{N^6}=\frac{1}{N}\alpha^3
\bra q^2\ket^3+\order{(N^{-1})}
\eea 
with the square brackets $[i,j,k]$  denoting that the three
indices are distinct.
The expected density of length-$3$ loops is the sum of an $\order{(1)}$ contribution 
 from $S_3$ stars, plus an $\order{(N^{-1})}$ contribution 
 from combinations of three $S_2$ stars that share leaves.  
For large $N$ the second contribution vanishes, and we recover $m_3=\alpha \bra q^3\ket$.
Likewise, the $\order{(1)}$ contribution to the 
density of length-$4$ loops comes from $S_4$ stars in the 
bi-partite graph, which consist of five sites (four leaves and one central 
node) and four links, each with probability $\order{(N^{-1})}$.
Combinations of two $S_3$ stars with  two shared leaves, 
or of $S_2$ stars, always 
involve a number of links at least equal to the number of nodes and therefore 
yield sub-leading contributions. Hence, the density of loops of 
length $4$ is
\be
m_4=\frac{1}{N}\sum_\mu \sum_{[i,j,k,\ell]}
\bra \xi_i^\mu \xi_j^\mu 
\xi_k^\mu \xi_\ell^\mu \ket=\alpha \bra q^4\ket+\order{(N^{-1})}
\ee
More generally, the average density of loops of arbitrary 
length $L$ is given by 
\be
m_L=\alpha \bra q^L\ket+\order{(N^{-1})}
\label{eq:mn_q}
\ee
For large $N$ the ratio $\alpha$ and the distribution $P(q)$ 
of complex sizes apparently determine in full 
the statistics of loops of arbitrary length in $\bc$, if the protein interactions are described by 
 (\ref{eq:quenched_q}). 

Finally, we note that if $m_L$ gives  
the number of ordered and oriented loops of length $L$ per node,  the number of 
unordered and unoriented closed paths of length $L$ equals $\bar m_L=m_L/6$, since  there are $L$ possible nodes to start a closed path from, and two possible orientations. 


\subsection{The degree distribution}

It follows from (\ref{eq:kav_q}, \ref{eq:mn_q}) that by measuring the average degree 
$\kav$ and the densities  $m_L$ of 
loops of length $L$
we can compute all the moments of the distribution of complex 
sizes $P(q)$:
\begin{eqnarray}
\bra q^2\ket= \kav/\alpha,~~~~~~~~\forall L>2:~~\bra q^L\ket=m_L/\alpha 
\label{eq:qmoments}
\end{eqnarray}
This would allow us to calculate 
$P(q)$ in full via its generating function, provided $\alpha$ and $\bra q \ket$ are known. However, counting the number of loops of arbitrary length in a graph is computationally challenging, 
and $\alpha$ and $\bra q \ket$ are generally unknown. 
However, it is possible to express $P(q)$  for large $N$ 
in terms of the degree distribution $p(k)$ of $\bc$. Specifically,  
in \ref{app:pk} we show that 
\begin{eqnarray}
\lim_{N\to\infty}p(k)
&=& \int_0^\infty\!\rmd y~P(y)~\rme^{-y}y^k/k!
\label{eq:pkc}
\end{eqnarray}
where 
\bea
P(y)&=& \rme^{-\alpha\bra q\ket}\sum_{\ell\geq 0}\frac{(\alpha\bra q\ket)^\ell}{\ell !}
\sum_{q_1\ldots q_\ell\geq 0} W(q_1)\ldots W(q_\ell)~\delta[y-\sum_{r\leq \ell }q_r ]
\eea
and $W(q)=qP(q)/\bra q\ket$ is the likelihood to draw a link attached to a complex-node of degree $q$ 
in the bipartite graph $\bxi$. Formula (\ref{eq:pkc}) is 
easily interpreted.  The degree of node $i$ in  $\bc$ is given by 
the second neighbours of $i$ in $\bxi$; the number $\ell$ of first neighbours  of 
node $i$ will thus be a Poissonian variable with average $\alpha\qav$, and each of its $\ell$ first neighbours 
will have a degree $q_r$ drawn from $W(q_r)$. 
Clearly, any tail in the distribution $W(q)$ will induce a tail in the distribution $p(k)$, with (as we will show below)  the same exponent, but 
an amplitude that is reduced by a factor $\alpha \qav$.

One can complement  (\ref{eq:pkc}) with a reciprocal relation that gives $P(q)$ in terms of $p(k)$. To achieve this 
we  define the generating functions $Q_1(z)=\sum_{k}p(k)\rme^{-kz}$, $Q_2(z)=\int_0^\infty\!\rmd y~P(y)\rme^{-yz}$
and $Q_3(z)=\sum_q W(q)\rme^{-zq}$. 
We then see from expression (\ref{eq:pkc}) for $p(k)$ that
\begin{eqnarray}
Q_1(z)
&=& \int_0^\infty\!\rmd y~P(y)~\rme^{-y}\sum_{k\geq 0}\frac{(y\rme^{-z})^k}{k!}
=\int_0^\infty\!\rmd y~P(y)~\rme^{y[\rme^{-z}-1]} =Q_2(1-\rme^{-z})
\\
Q_2(z)&=&\rme^{-\alpha\bra q\ket}\sum_{\ell\geq 0}\frac{(\alpha\bra q\ket)^\ell}{\ell !}
\sum_{q_1\ldots q_\ell\geq 0} W(q_1)\ldots W(q_\ell)~\rme^{-z\sum_{r\leq \ell }q_r }
\nonumber
\\&=& \rme^{-\alpha\bra q\ket}\sum_{\ell\geq 0}\frac{(\alpha\bra q\ket Q_3(z))^\ell}{\ell !} =
 \rme^{\alpha\bra q\ket [Q_3(z)-1]} 
\label{eq:Q2}
\end{eqnarray}
The first identity can be rewritten as $Q_1(-\log (1-y))=Q_2(y)$. Inserting this 
into (\ref{eq:Q2}), allows us to express the desired $Q_3(z)$ as 
\begin{eqnarray}
Q_3(z)&=&
1+\frac{\log Q_2(z)}{\alpha\bra q\ket }=1+\frac{\log Q_1(-\log (1-z))}{\alpha\bra q\ket }
\label{eq:generating_Q}
\end{eqnarray}
which translates into 
\begin{eqnarray}
\sum_{q>0}P(q)q\rme^{-zq}
&=&
\bra q\ket+\frac{1}{\alpha}\log \sum_{k}p(k)(1-z)^k
\label{eq:link_kq}
\end{eqnarray}
We can now extract the asymptotic form of $P(q)$ from that of 
$p(k)$. The generating functions $Q_1(z)$ of degree distributions that exhibit prominent tails, i.e. $p(k)\simeq Ck^{-\mu}$ for large $k$ with $2\!<\! \mu\!<\!3$ (as observed  in protein interaction networks \cite{BarAlb99,AlbBar02,JunSch08,DorogovtsevMendesBook}), are for small $z$ 
of the form 
\bea
Q_1(z)=1-\kav z+C\Gamma(1\!-\!\mu)z^{\mu-1}+\ldots
\label{eq:euler}
\eea
where $\Gamma$ is Euler's gamma function \cite{AbramStegun}.
For small $z$ we may use $1-z\simeq \rme^{-z}$ to rewrite (\ref{eq:generating_Q}) as 
\bea
\log Q_1(z)\simeq \alpha \bra q \ket [Q_3(z)-1]
\eea
Combining this with (\ref{eq:euler}) then gives, for small $z$,  
\bea
-\bra k\ket z+C \Gamma(1-\mu) z^{\mu-1}\simeq \alpha \bra q \ket [Q_3(z)-1]
\eea
Hence, for small $z$, $Q_3(z)$ has the same form as $Q_1(z)$, 
\be
Q_3(z)=1-\frac{\kav}{\alpha \qav}z+\frac{C}{\alpha \bra q \ket}\Gamma(1\!-\!\mu) z^{\mu-1}
\ee
Therefore $W(q)$ behaves asymptotically in the same way as $p(k)$, i.e.
$W(q)\simeq  (C/\alpha \bra q \ket) q^{-\mu}$. 
This, in turn,  gives 
\bea 
P(q)\simeq (C/\alpha)~q^{-\mu-1}
\eea
The complex size distribution $P(q)$ in (\ref{eq:quenched_q}) decays faster than the degree distribution of the associated $\bc$, so  fat tails in the degree distribution of protein interaction networks can emerge from
less heterogeneous complex size distributions.  In particular, complex size distributions with a finite 
second moment (but diverging higher moments) give scale-free degree distributions in $\bc$.
This is consistent with the intuition that, while large hubs are 
often observed in protein interaction networks, super-complexes 
of the same number of proteins are unlikely to be stable. Indeed, many interactions in hubs are `date'  type, as opposed to `party' type \cite{ChangETAL13}.  
Our framework allows us to discriminate between different type of hub proteins, and  suggests that heterogeneities in PINs may emerge from homogeneous 
protein `dating' and moderately heterogenous protein `partying'.

\subsection{Relations that are independent of $P(q)$ and $\alpha$}

\begin{figure}[t]
\setlength{\unitlength}{0.69mm}
\hspace*{30mm}

\begin{picture}(200,152)
\put(-3.5,0){\includegraphics[width=97\unitlength,height=70\unitlength]{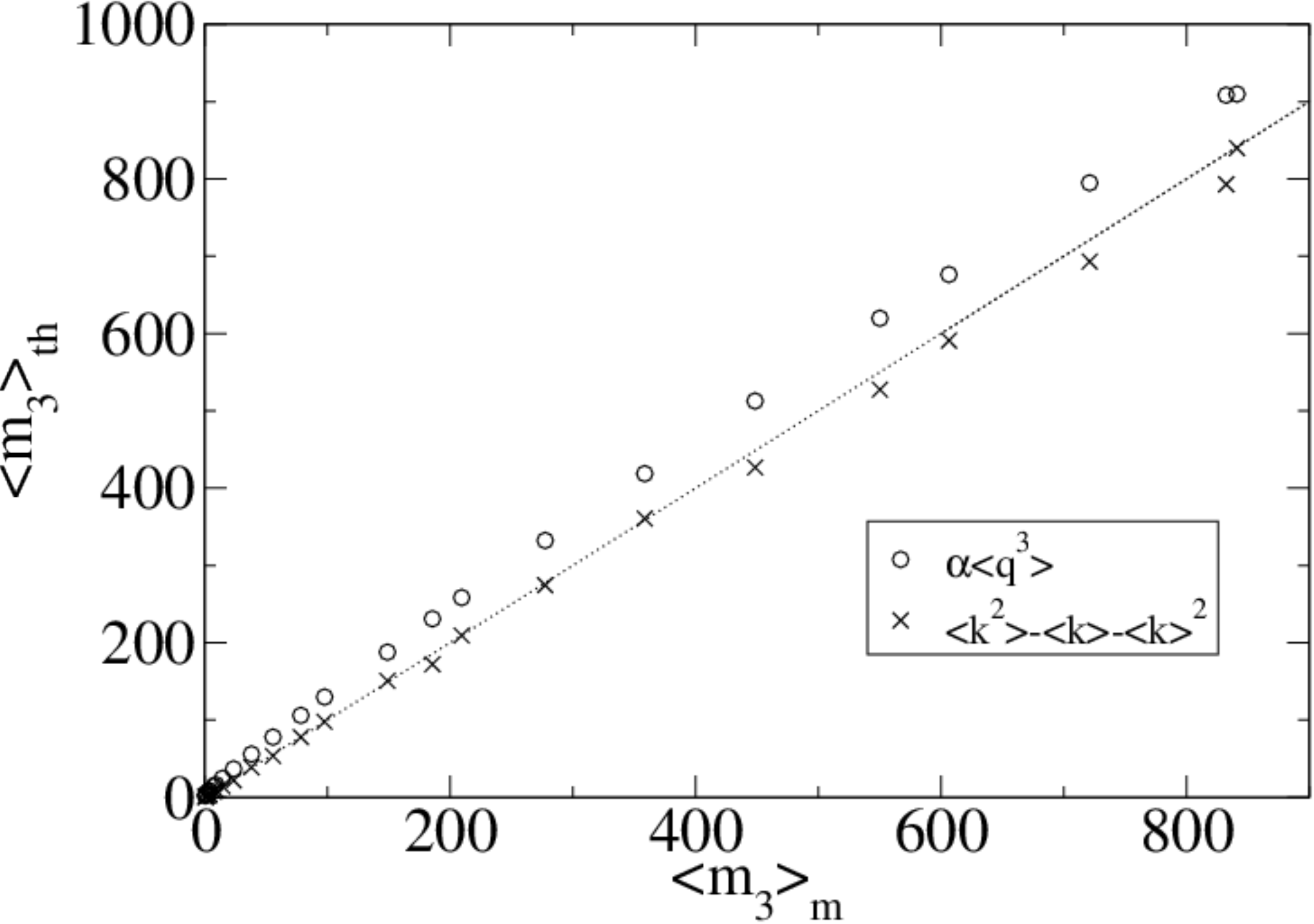}}
\put(100,0){\includegraphics[width=105\unitlength,height=70\unitlength]{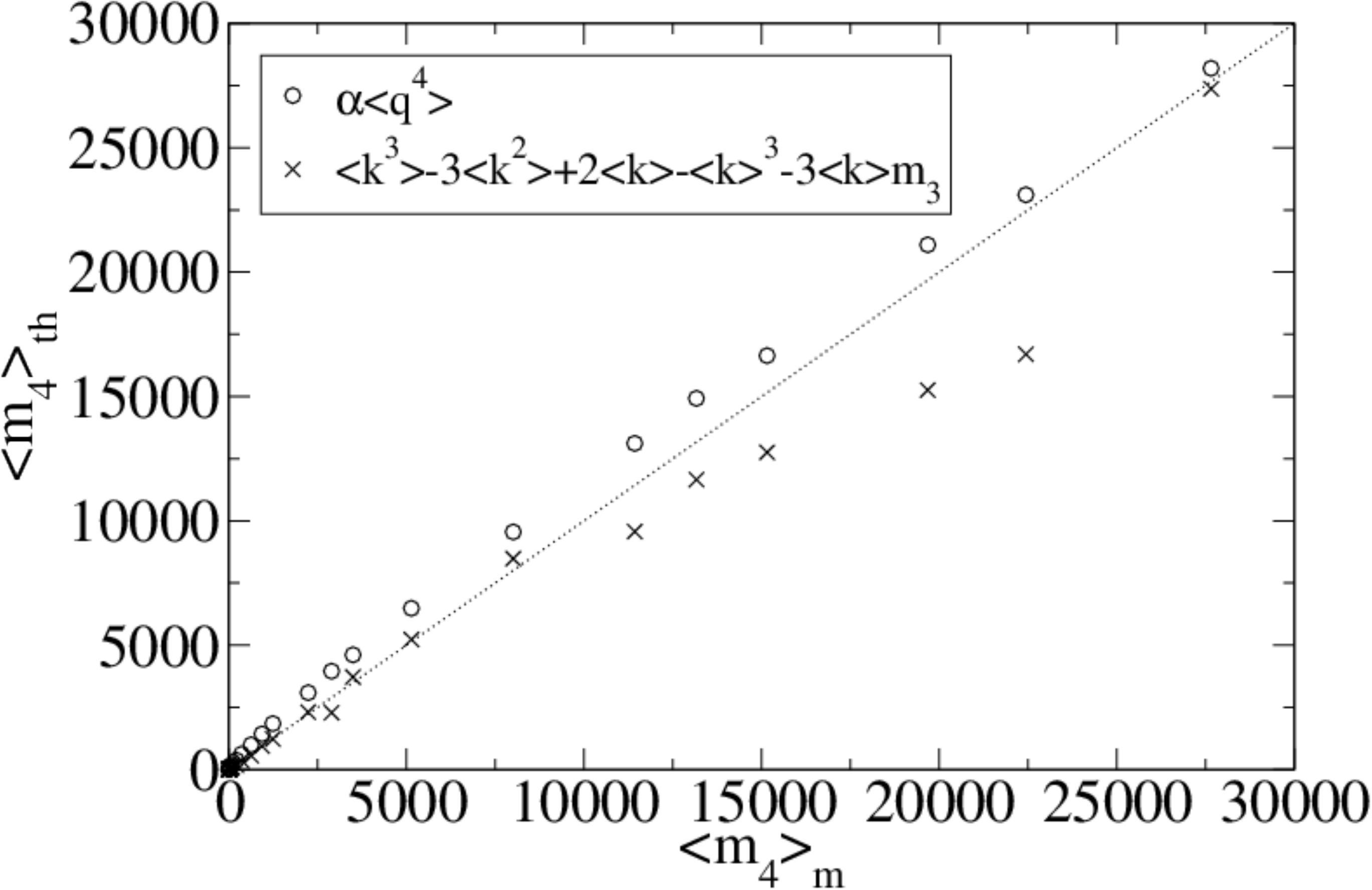}}

\put(0,78){\includegraphics[width=96\unitlength,height=70\unitlength]{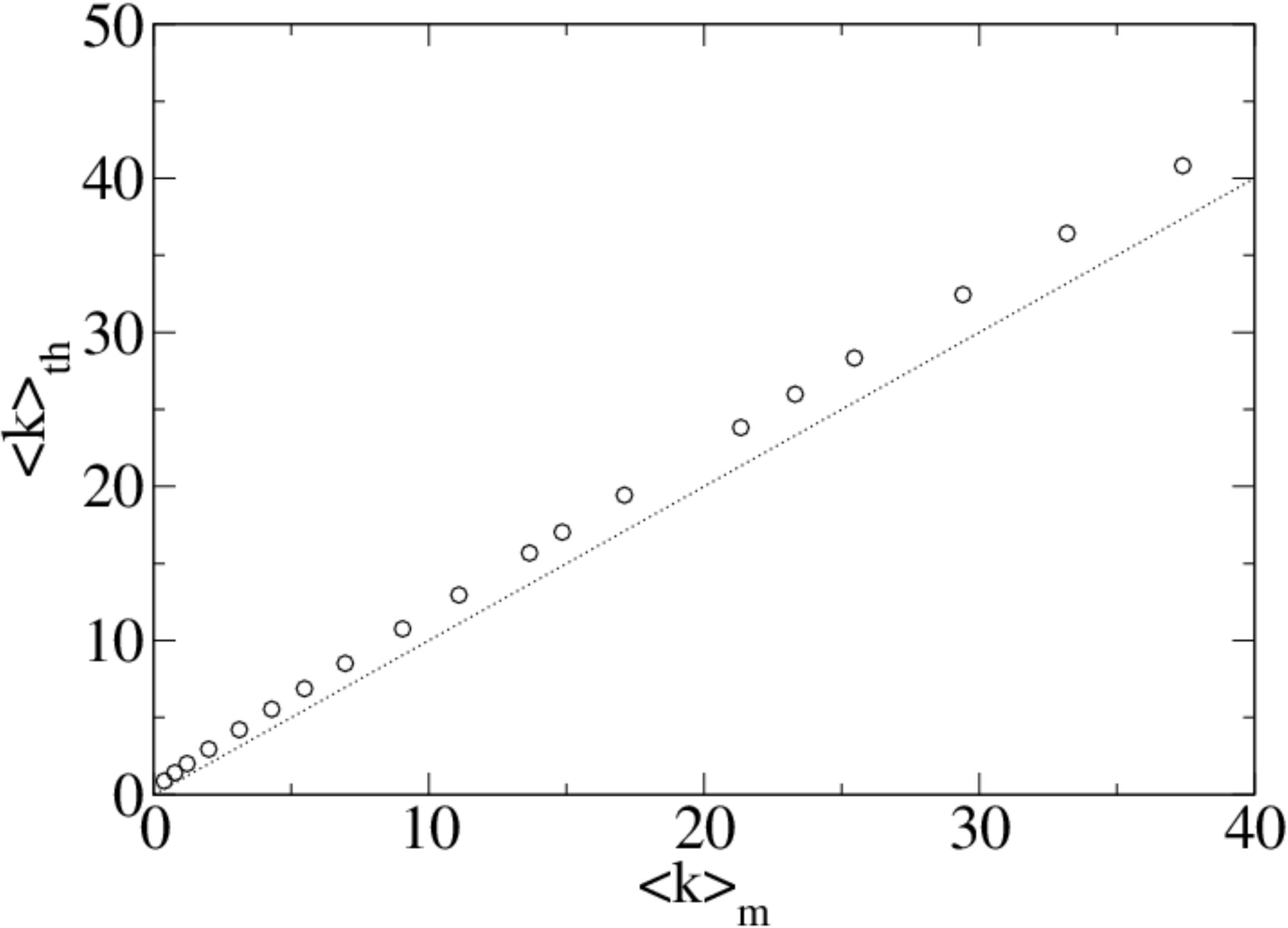}}
\put(100,76){\includegraphics[width=104\unitlength,height=72\unitlength]{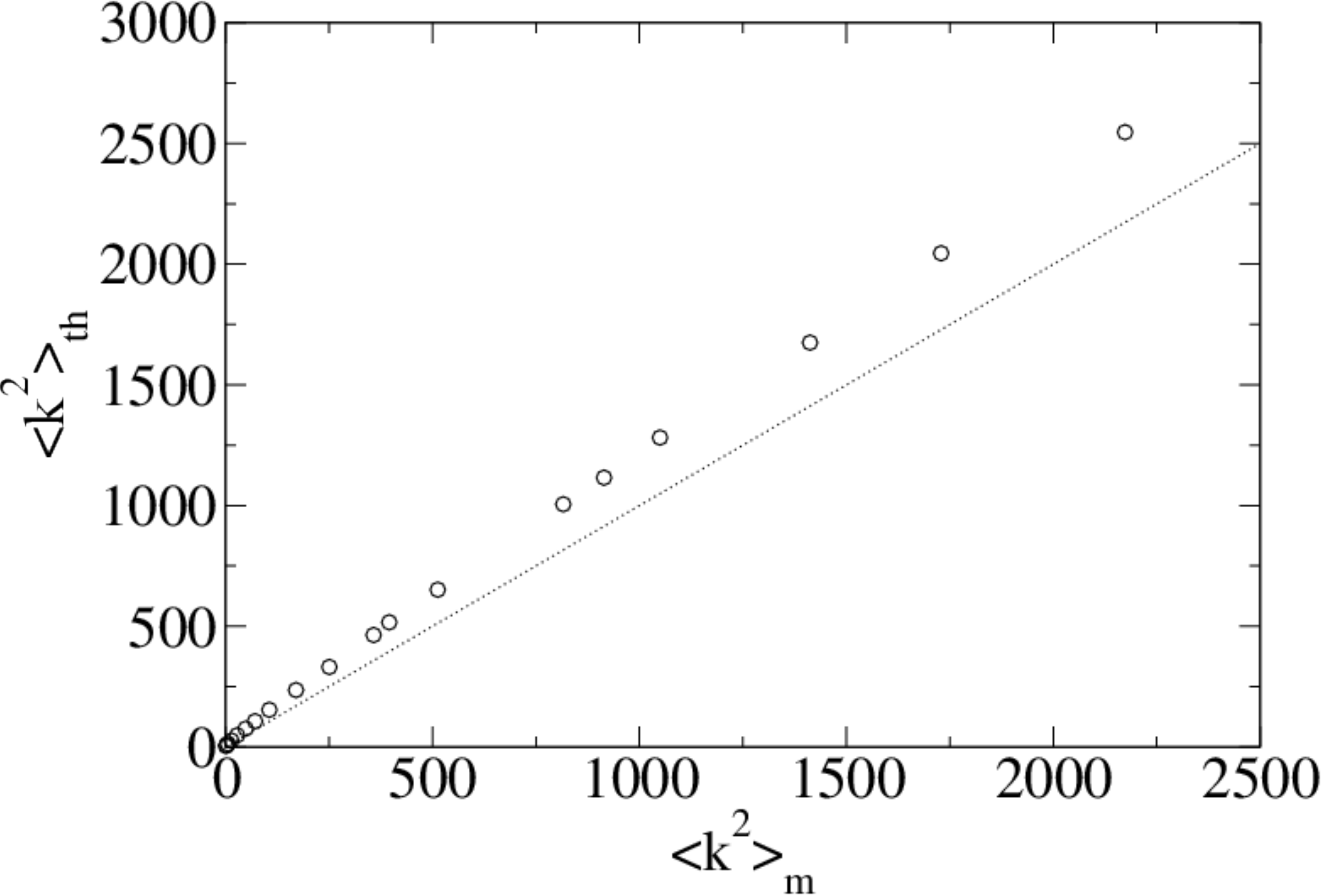}}
\end{picture}

\caption{Symbols: theoretical $\bra \ldots\ket_{\rm th}$ versus   measured $\bra \ldots\ket_{\rm m}$  values 
of observables  $\kav$, $\kvar$, $m_3$ and $m_4$ in synthetically random graphs 
$\bc$ with $N=3000$, defined via (\ref{eq:quenched_q},\ref{eq:ensemble_c}) for a power-law distributed complex size distribution
$P(q)$. Theoretical values are given by formulae 
(\ref{eq:kav_q2}) for $\kav$, (\ref{eq:kvar_q}) for $\kvar$,  
(\ref{eq:mn_q}) and (\ref{eq:m3_q}) for $m_3$ and (\ref{eq:mn_q}) and (\ref{eq:m4_q}) for $m_4$. 
 Dotted lines: the diagonals (as a guides to the eye).
} 
\label{fig:qth}
\end{figure}

The first two moments of $p(k)$ are given, to leading order 
in $N$, by (see \ref{app:pk})
\begin{eqnarray}
\bra k\ket&=&  \alpha \bra q^2\ket +\order{(N^{-1})}
\label{eq:kav_q2}
\end{eqnarray}
which is in agreement with (\ref{eq:kav_q}), and 
\begin{eqnarray}
\bra k^2\ket&=& 
\alpha \bra q^2\ket+\alpha
\bra q^3\ket+
\alpha^2\bra q^2\ket^2
\label{eq:kvar_q}
\end{eqnarray}
The latter is easily interpreted in terms 
of the underlying bipartite graph: $\bra k^2\ket$ is the average density of paths of 
length two, so it has a contribution from $\bra k\ket=\alpha 
\bra q^2 \ket$ due to backtracking, plus a contribution from pairs of 
$S_2$ stars that share a node, whose density is 
\bea
&&\frac{1}{N}\sum_{[ijk]}\sum_{\mu\neq \nu} \bra \xi_i^\mu \xi_j^\mu
\xi_j^\nu \xi_k^\nu\ket=
\frac{1}{N}\sum_{[ijk]}\sum_{\mu\neq \nu}\frac{q_\mu^2}{N^2}
\frac{q_\nu^2}{N^2}=\alpha^2\bra q^2\ket^2,
\eea
plus a contribution from $S_3$ stars, whose density is $\alpha \bra q^3\ket$ (as shown earlier).
Combining (\ref{eq:kvar_q}) with (\ref{eq:qmoments}) gives us a relation between average and width of the degree distribution of $\bc$
and its density of length-3 loops. Remarkably, this relation is  completey
independent of $\alpha$ and $P(q)$:
\begin{eqnarray}
m_3&=& \bra k^2\ket-\bra k\ket^2-\kav 
\label{eq:m3_q}
\end{eqnarray}
This identity and others, which all depend only on the separable underlying nature of 
the PIN and the assumption of complex-driven recruitment of proteins to complexes, can be derived more systematically from (\ref{eq:link_kq}) by expanding both sides as power series in $z$ and comparing the expansion coefficients.  This gives a hierarchy of relations between moments of $p(k)$ and $P(q)$, and hence (via (\ref{eq:mn_q})) between  moments of $p(k)$ and densities of  loops of increasing length, that 
are all completely independent of $\alpha$ and $P(q)$. 
At order $z^2$ one recovers (\ref{eq:m3_q}). 
The next order $z^3$  leads to 
\bea
m_4&=&\bra k^3\ket -3\bra k^2\ket +2\bra k\ket +\bra k \ket
(\bra k^2\ket -\bra k\ket-2\kav^2)
\nonumber\\
&=&\bra k^3\ket -3\bra k^2\ket +2\bra k\ket-\kav^3 -3\bra k \ket m_3
\label{eq:m4_q}
\eea 
To test these asymptotic identities in finite systems, we generate random graphs 
$\bc$ of size $N=3000$ according to (\ref{eq:quenched_q},\ref{eq:ensemble_c}), and we compared the measured values of $m_3$ and $m_4$ in these random graphs with the predictions of  
formulae (\ref{eq:m3_q}) and (\ref{eq:m4_q}), respectively. We show the results in 
Figure \ref{fig:qth}.


\subsection{Link between ${\bf a}$ and $\bc$ graph definitions}

\begin{figure}[t]

\setlength{\unitlength}{0.68mm}
\hspace*{30mm}
\begin{picture}(200,280)

\put(1.2,210){\includegraphics[width=95\unitlength,height=67\unitlength]{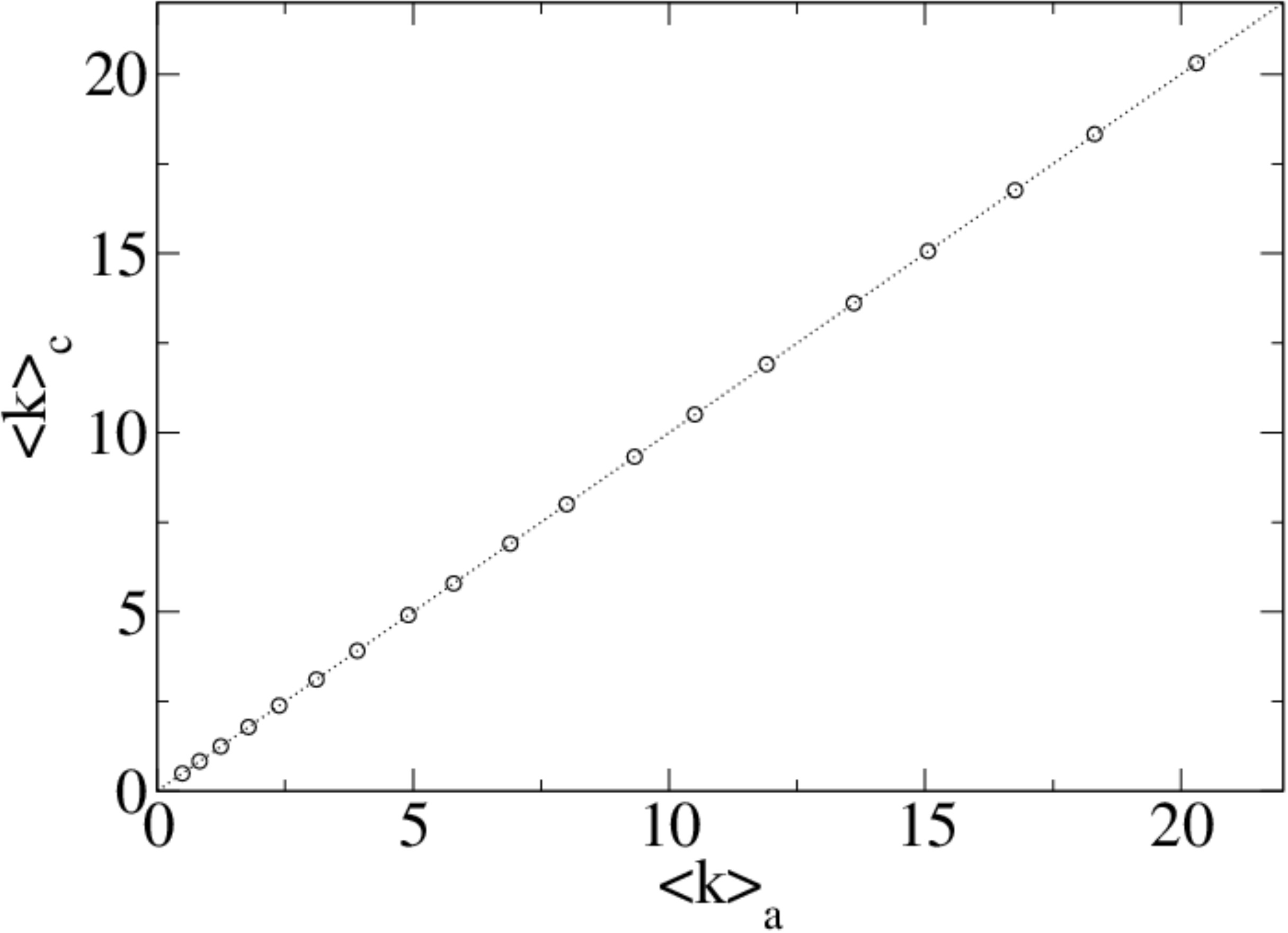}}
\put(103,210.2){\includegraphics[width=97\unitlength,height=68\unitlength]{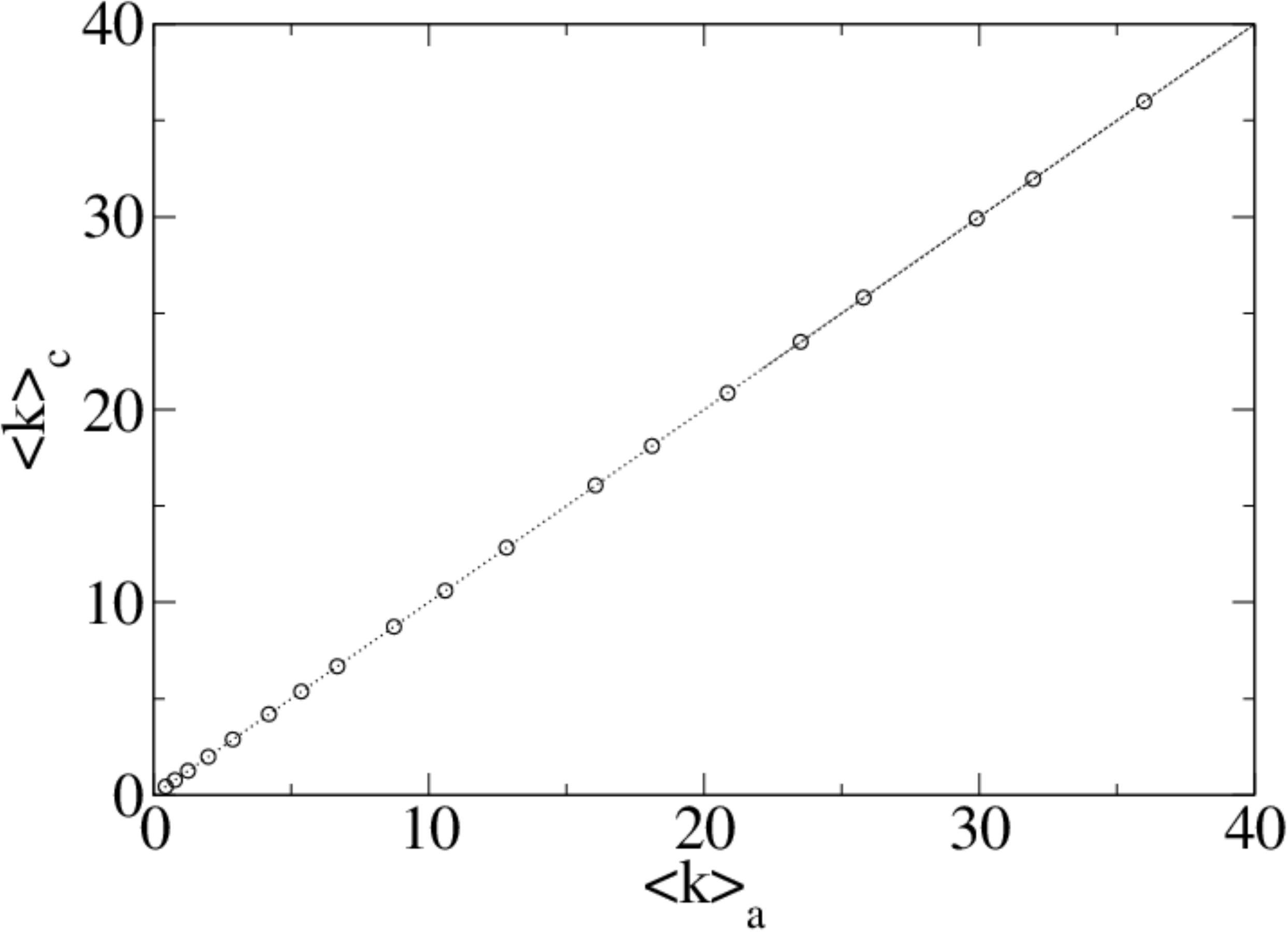}}

\put(-1,140){\includegraphics[width=100\unitlength,height=69\unitlength]{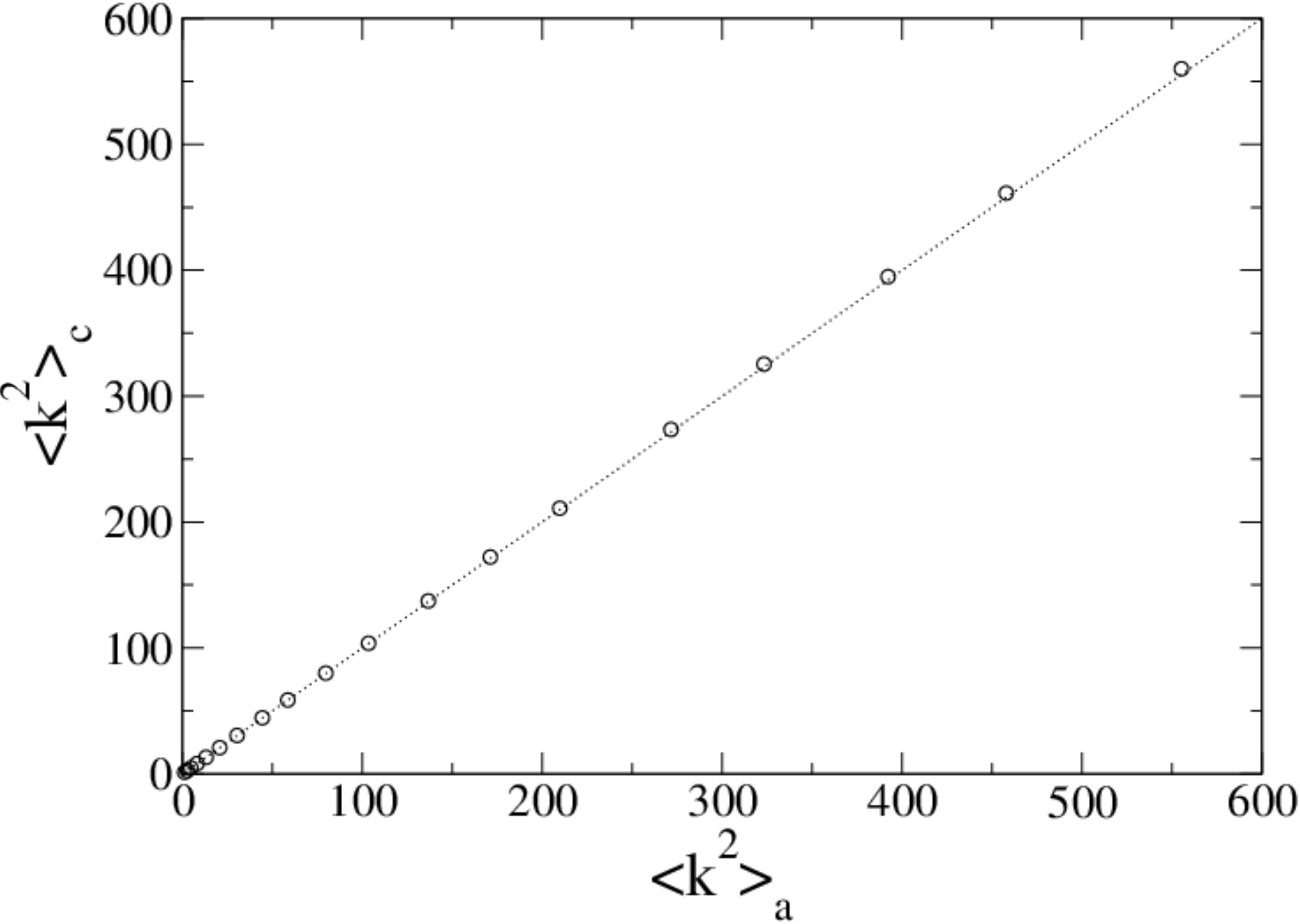}}
\put(99,140){\includegraphics[width=102\unitlength,height=69\unitlength]{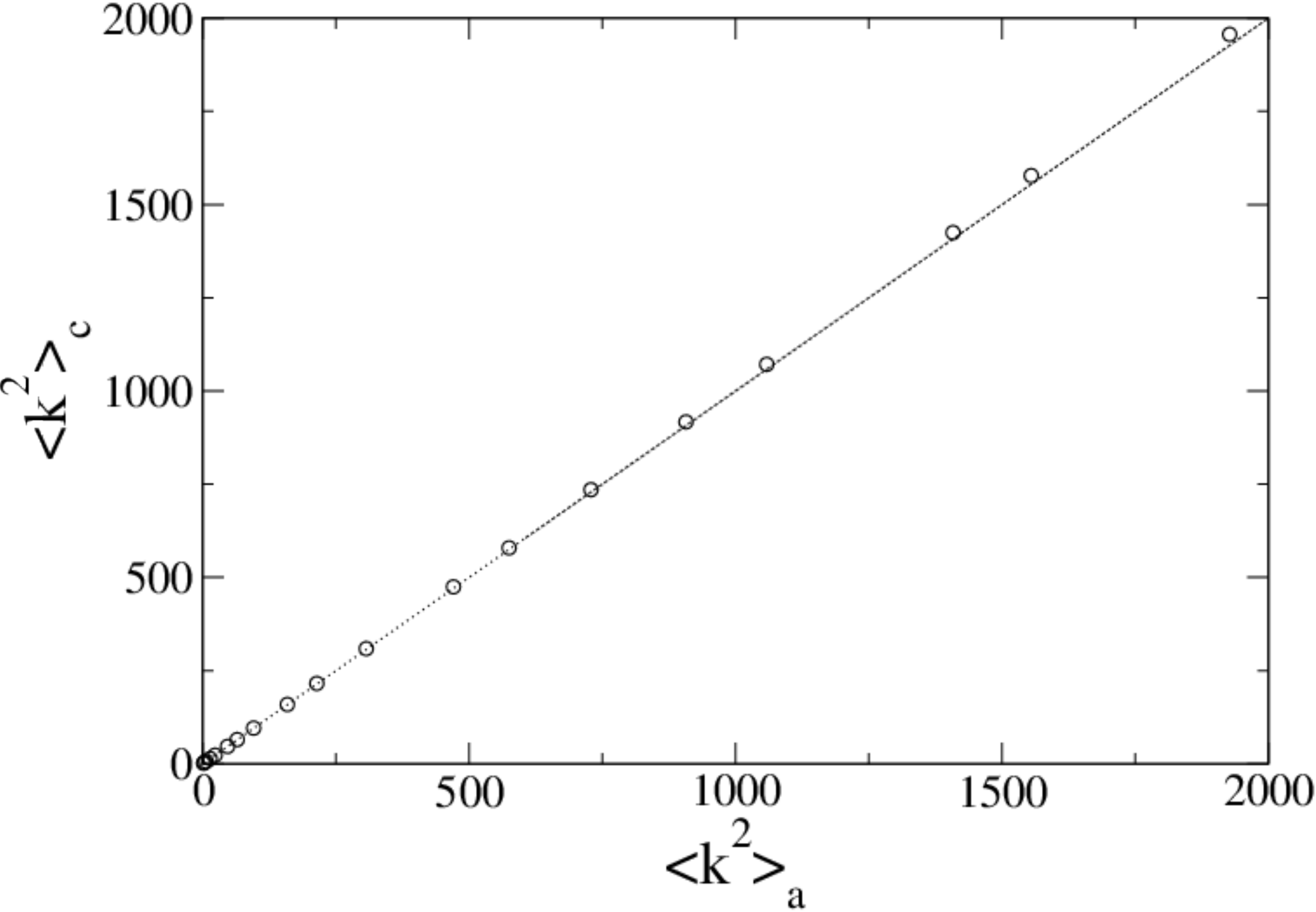}}

\put(1.5,70){\includegraphics[width=98\unitlength,height=69\unitlength]{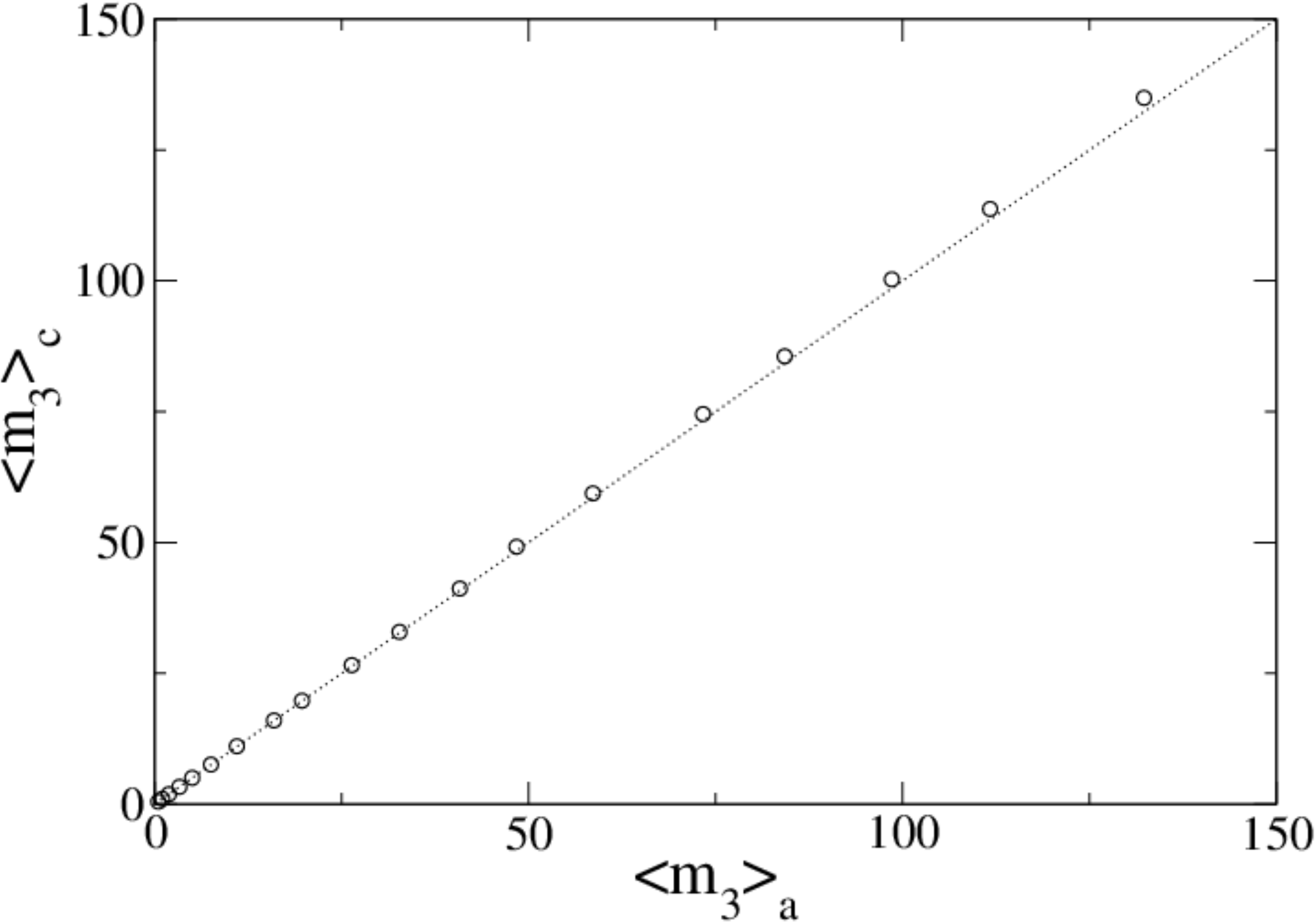}}
\put(103,70){\includegraphics[width=97.3\unitlength,height=69\unitlength]{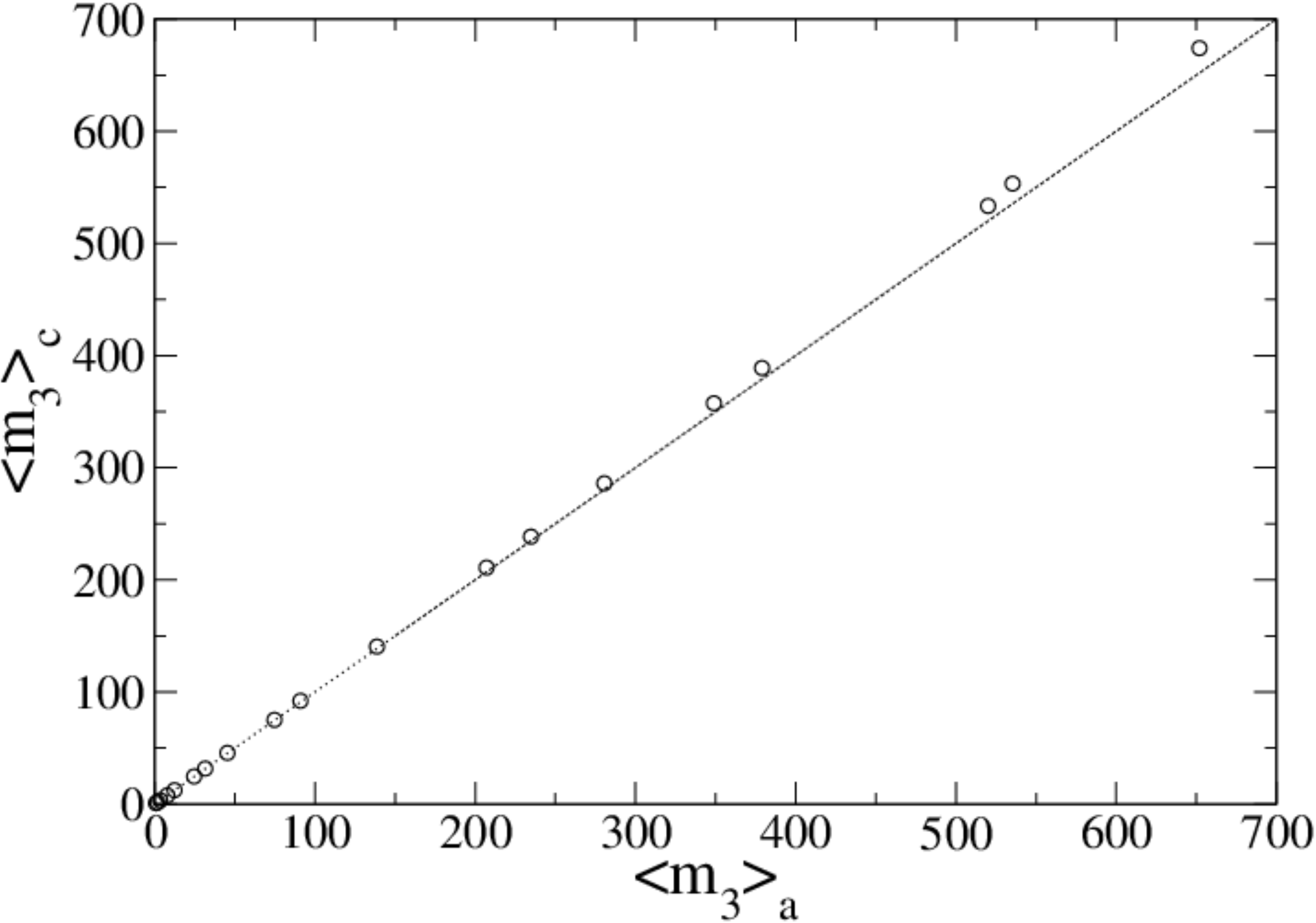}}

\put(0,0){\includegraphics[width=100\unitlength,height=67\unitlength]{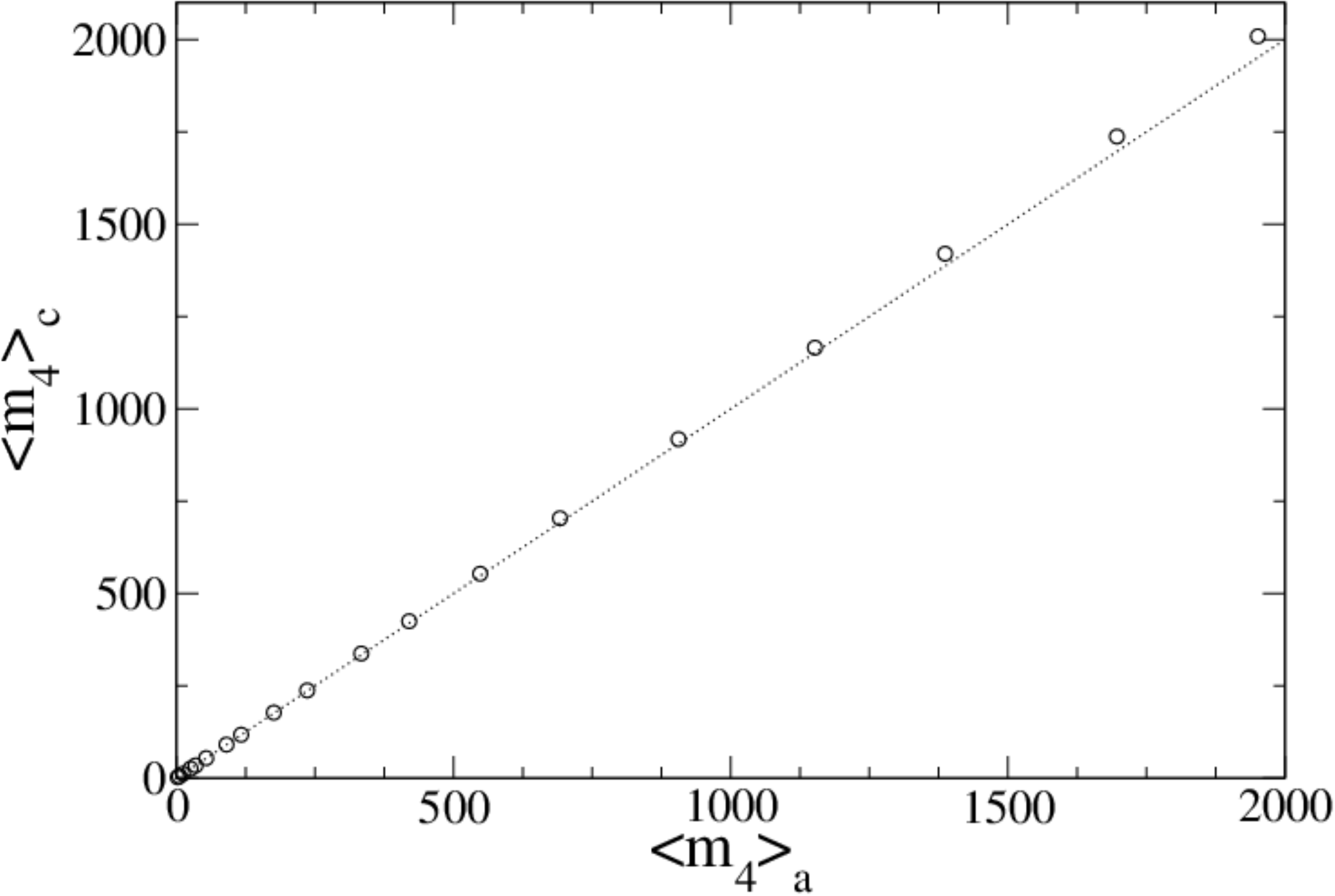}}
\put(100,0.2){\includegraphics[width=102\unitlength,height=67.7\unitlength]{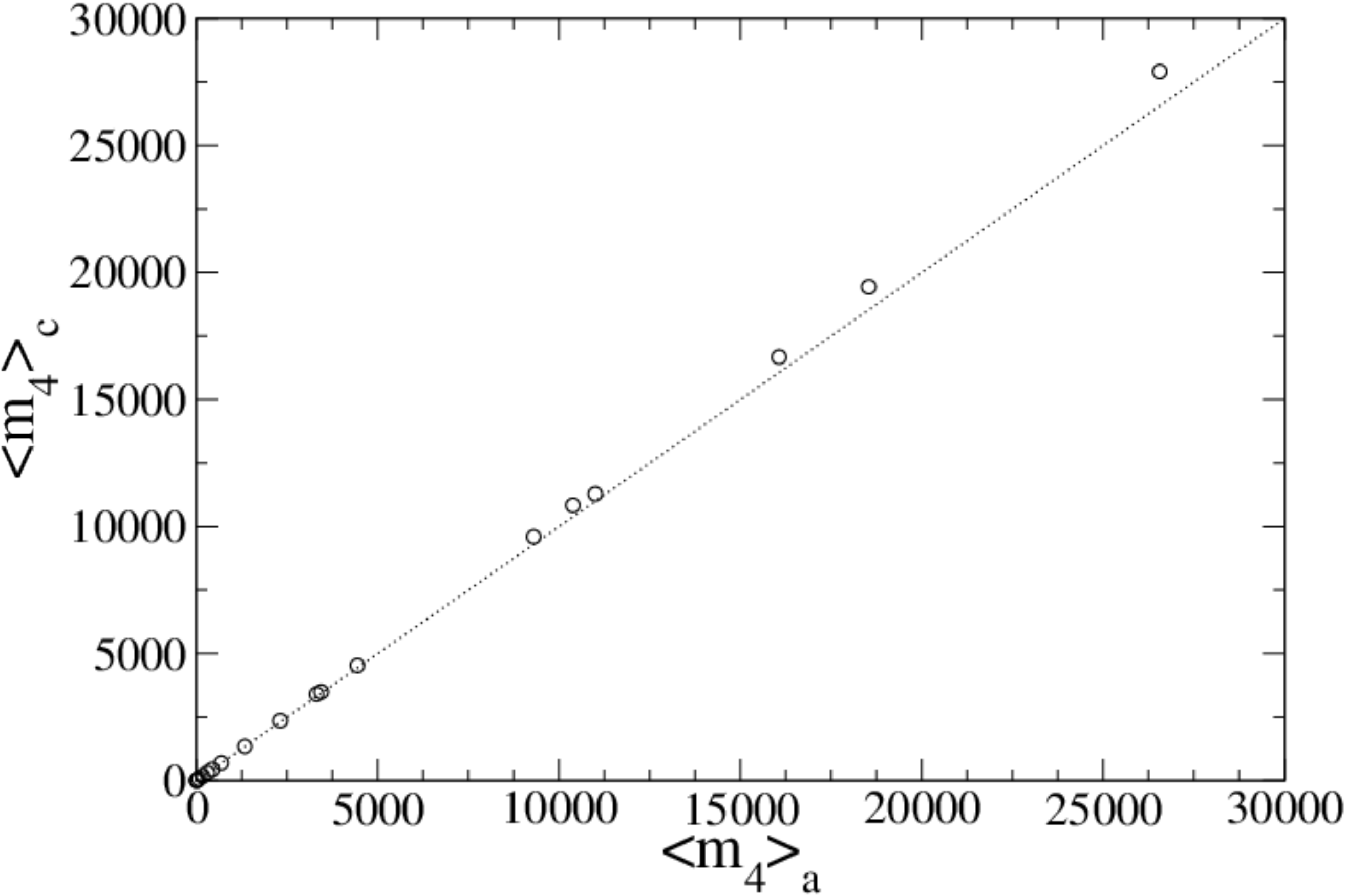}}

\end{picture}
\vspace*{-0mm}

\caption{Symbols:  $\kav$, $\bra k^2\ket$, $m_3$ and $m_4$ as measured 
in synthetic graphs $\bc$ drawn from (\ref{eq:ensemble_c}) with $N=3000$, shown versus corresponding values found in the binary graphs $\ba$ drawn from (\ref{eq:ensemble_a}). Bipartite interaction graphs $\bxi$ are drawn 
from (\ref{eq:quenched_q}), with complex size distributions $P(q)$   that are Poissonian (left panels) or 
power law (right panels). Dotted lines: the diagonals (shown as guides to the eye).  As expected, the values measured in the weighted graphs 
$\bc$ are consistently higher than in the binary ones, but one finds that these deviations 
get smaller for increasing network sizes $N$.} 
\label{fig:avacq}
\end{figure}

In conventional experimental PIN data bases one records only whether or not protein pairs 
interact, not the {\em number} of complexes in which they 
interact. Hence, protein interactions are normally represented in terms 
of the adjacency matrix $\ba=\{a_{ij}\}$, which is related to the weighted 
matrix $\bc=\{c_{ij}\}$ via 
$a_{ij}= \theta(c_{ij})~\forall ~(i\neq j)$, with the convention for the step function $\theta(0)=0$.
We therefore have 
$p(a_{ij})=
\bra \delta_{c_{ij},0}\ket \delta_{a_{ij,0}}+(1-\bra \delta_{c_{ij},0}\ket)\delta_{a_{ij},1}$. 
However, the links $\{a_{ij}\}$ are correlated. In \ref{app:a_c} we derive 
the relation between the expected values of different graph observables
for the two graph ensembles $p(\ba)$ and $p(\bc)$. 
Denoting averages in the $\ba$ ensemble as $\bra \ldots\ket_a$, and using the usual notation $\bra \ldots \ket$ for averages 
in the $\bc$ ensemble,  one finds that for large $N$  the first two moments of the degree distributions and the first two loop densities in the two ensembles are identical:
\begin{eqnarray}
\bra k\ket_a&=& \frac{1}{N}\sum_{ij}\bra a_{ij}\ket_a= 
\frac{1}{N}\sum_{ij}[1-\bra \delta_{c_{ij},0}\ket]= \alpha\bra q^2\ket+\order(N^{-1})
\nonumber\\
&=& \bra k\ket+\order(N^{-1})
\\
\bra k^2\ket_a&=& \frac{1}{N}\sum_{i\neq j\neq k}\bra a_{ij}a_{jk}\ket
=\alpha \bra q^2\ket+\alpha \bra q^3\ket +\alpha^2 \bra q^2\ket^2+\order(N^{-1})
\nonumber\\
&=& \bra k^2\ket+\order(N^{-1})
\\
m_3^a&=&\frac{1}{N}\sum_{i\neq j\neq k(\neq i)}\bra a_{ij}a_{jk}a_{ki}\ket
=\alpha \bra q^3\ket +\order{(N^{-1})}
\nonumber\\
&=& m_3+\order{(N^{-1})}
\\
m_4^a&=&\frac{1}{N}\sum_{[i,j,k,\ell]}\bra a_{ij}a_{jk}a_{k\ell}a_{\ell i}\ket
=\alpha\bra q^4\ket+\order{(N^{-1})}
\nonumber\\
&=& m_4 +\order{(N^{-1})}
\end{eqnarray}
Square brackets underneath summations again indicates distinct indices, which excludes  
backtracking in the 
counting of length-$4$ loops. 
Apparently, the ensembles $p({\bf a})$ and 
$p({\bf c})$ are asymptotically equivalent with regard to the  statistics of these four quantities. 
We will see in the next section that this equivalence holds 
also for the `dual' ensemble (\ref{eq:quenched_d}). To test the above claims we compute and 
show in Figure \ref{fig:avacq}
the above observables 
in synthetic graphs $\bc$ and $\ba$ generated  randomly from (\ref{eq:ensemble_a},\ref{eq:ensemble_c}), where the random bipartite interaction graph $\bxi$ is 
drawn from (\ref{eq:quenched_q}).

\section{Network properties generated by the $d$-ensemble}
\label{sec:d}

In this section we will derive properties for the network ensembles (\ref{eq:ensemble_a},\ref{eq:ensemble_c}) upon 
assuming that the statistics 
of the underlying bipartite protein interaction network are given by  (\ref{eq:quenched_d}), i.e. are protein-driven as opposed to complex-driven. 
In spite of the superficial similarity between definitions (\ref{eq:pq_dp}) and 
(\ref{eq:pd_qp}), the  
expectations of graph observables in
the two ensembles are found to be remarkably different.

\subsection{Link probabilities}

We start by calculating the link expectation values in the weighted graphs
$c_{ij}=\sum_\mu \xi_i^\mu \xi_j^\mu$:
\bea
\bra c_{ij}\ket=\sum_\mu \bra \xi_i^\mu \xi_j^\mu\ket=\frac{d_i d_j}{\alpha N}
\eea
Hence the random graphs $\bc$ are again finitely connected, now with 
\bea
\bra k\ket=\frac{1}{N}\sum_{ij} \bra c_{ij}\ket=\frac{\bra d \ket^2}{\alpha}
\label{eq:av_k_d}
\eea
Averages over $d$ refer to the distribution $P(d)$ of protein promiscuities in the bipartite graph $\bxi$. 
The result (\ref{eq:av_k_d}) can also be written as $\kav=\alpha \qav^2$, and is 
thus notably different from the earlier expression $\bra k\ket =\alpha \bra q^2\ket$ found in the $q$-ensemble. The link likelihood is calculated in \ref{app:pij}, and 
shows again that $p(c_{ij}>1)=\order{(N^{-2})}$.

\subsection{Densities of short loops}

We can calculate the density of length-$3$ loops similar to how this was done 
 for the $q$-ensemble in the previous section. Again these 
are given, to order $\order{(1)}$, by the $S_3$ stars in the bi-partite graph, 
since the contribution from combinations of $S_2$ stars is as before $\order{(N^{-1})}$. 
Here we obtain
\bea
m_3=\frac{1}{N}\sum_{[ijk]}\sum_\mu \bra \xi_i^\mu \xi_j^\mu \xi_k^\mu\ket=\frac{1}{N}\sum_{[ijk]}\sum_\mu 
\frac{d_id_jd_k}{\alpha^3N^3}= \frac{\dav^3}{\alpha^2}
\label{eq:m3_d1}
\eea
For loops of arbitrary 
length $L$
this generalises to 
 \be
m_L=\bra d\ket^L/\alpha^{L-1}
\label{eq:mL_d}
\ee
Interestingly, the densities $m_L$ of short loops  and the average connectivity $\bra k\ket$ depend on $P(d)$ only through its first moment. 
Promiscuity heterogeneity apparently cannot affect the densities of short loops. In the present ensemble these densities must therefore be identical to what would be found in 
a randomly wired bipartite graph. This prediction will be confirmed in simulations.

\subsection{The degree distribution}

In \ref{app:pk} 
we calculate the asymptotic degree distribution of $\bc$ for the protein-driven complex recruitment model (\ref{eq:quenched_d}), giving 
\bea
p(k)&=&\lim_{N\to\infty}\frac{1}{N}\sum_i \delta_{k,\sum_j c_{ij}}
=\sum_{d\geq 0} P(d)\sum_\ell \left(\rme^{-d}d^\ell/\ell!\right)\left(\rme^{-\ell\frac{\dav}{\alpha}}\left(\frac{\ell \dav}{\alpha}\right)^k/k!\right)
\label{eq:pk_d}
\eea
This result is again understood easily: the number of neighbours of a node $i$ is a  
Poissonian variable $\ell$, with average $d$, where $d$ is now drawn from $P(d)$. Each of the $\ell$ first neighbours will have
a degree which is a Poissonian variable with average $\dav/\alpha$, so the number $k$ of second neighbours of $i$ in the bipartite 
graph is a Poisson variable with average $\ell \dav/\alpha$.
Equation (\ref{eq:pk_d}) shows that a tail in the promiscuity distribution 
$P(d)$ will induce a tail in the degree distribution $p(k)$ of $\bc$. The link between the two 
distributions is again most easily expressed via generating functions.
Upon defining $Q_1(z)=\sum_k p(k)\rme^{-zk}$ and $Q_4(z)=\sum_d P(d)\rme^{-zd}$, 
we obtain from (\ref{eq:pk_d}):
\begin{eqnarray}
Q_1(z)&=&\sum_{d\geq 0} P(d) \rme^{-d} \sum_\ell \Big(d \rme^{
 \dav(\rme^{-z}-1)/\alpha}\Big)^\ell\!/\ell!
 ~=~Q_4(1- \rme^{
 \dav(\rme^{-z}-1)/\alpha})
\end{eqnarray}
For $z\simeq 0$ this gives 
\bea
Q_1(z)\simeq Q_4(z\dav/\alpha)
\eea
Hence, if $p(k)$ decays for large $k$ as  
$p(k)\simeq Ck^{-\mu}$ with $2<\mu<3$, then via (\ref{eq:euler}) we infer that
\bea
Q_4(z\dav/\alpha)\simeq 1-\kav z +C\Gamma(1\!-\!\mu)z^{\mu-1}
\eea
Equivalently, 
\bea
Q_4(x)\simeq 1-\alpha \kav x/\dav+C\Gamma(1\!-\!\mu)(\alpha/\dav)^{\mu-1}x^{\mu-1}
\eea
This implies that for large $d$ the promiscuity distribution will be of the form 
$P(d)\simeq C' d^{-\mu}$, where 
\bea
C' =C(\alpha/\dav)^{\mu-1}=C\qav^{1-\mu}
\eea
Any tail in the promiscuity distribution 
will produce the same tail in the degree distribution of $\bc$, 
but with a rescaled amplitude.
Fat tails in the degree distribution of 
protein interaction networks can thus arise from equally heterogeneous `dating' 
interactions between proteins, combined with a homogeneous distribution of `party' interactions.
Short loops are boosted by broad distributions of complex sizes, since large 
complexes in the bipartite graph induce large cliques in the network $\bc$. The $d$-ensemble (\ref{eq:quenched_d}), 
which attributes any heterogeneity in $p(k)$ to heterogeneity of protein binding promiscuities, 
  generates separable PIN graphs $\bc$ with the least number of loops.  
Conversely, the $q$-ensemble (\ref{eq:quenched_q}), which attributes all heterogeneity in $p(k)$ to heterogeneity in complex sizes, 
generates separable PIN graphs $\bc$ with the largest number of loops.

\subsection{Relations that are independent of $P(d)$ and $\alpha$}

\begin{figure}[t]
\setlength{\unitlength}{0.69mm}
\hspace*{30mm}
\begin{picture}(200,152)
\put(-2.3,78){\includegraphics[width=96\unitlength,height=70\unitlength]{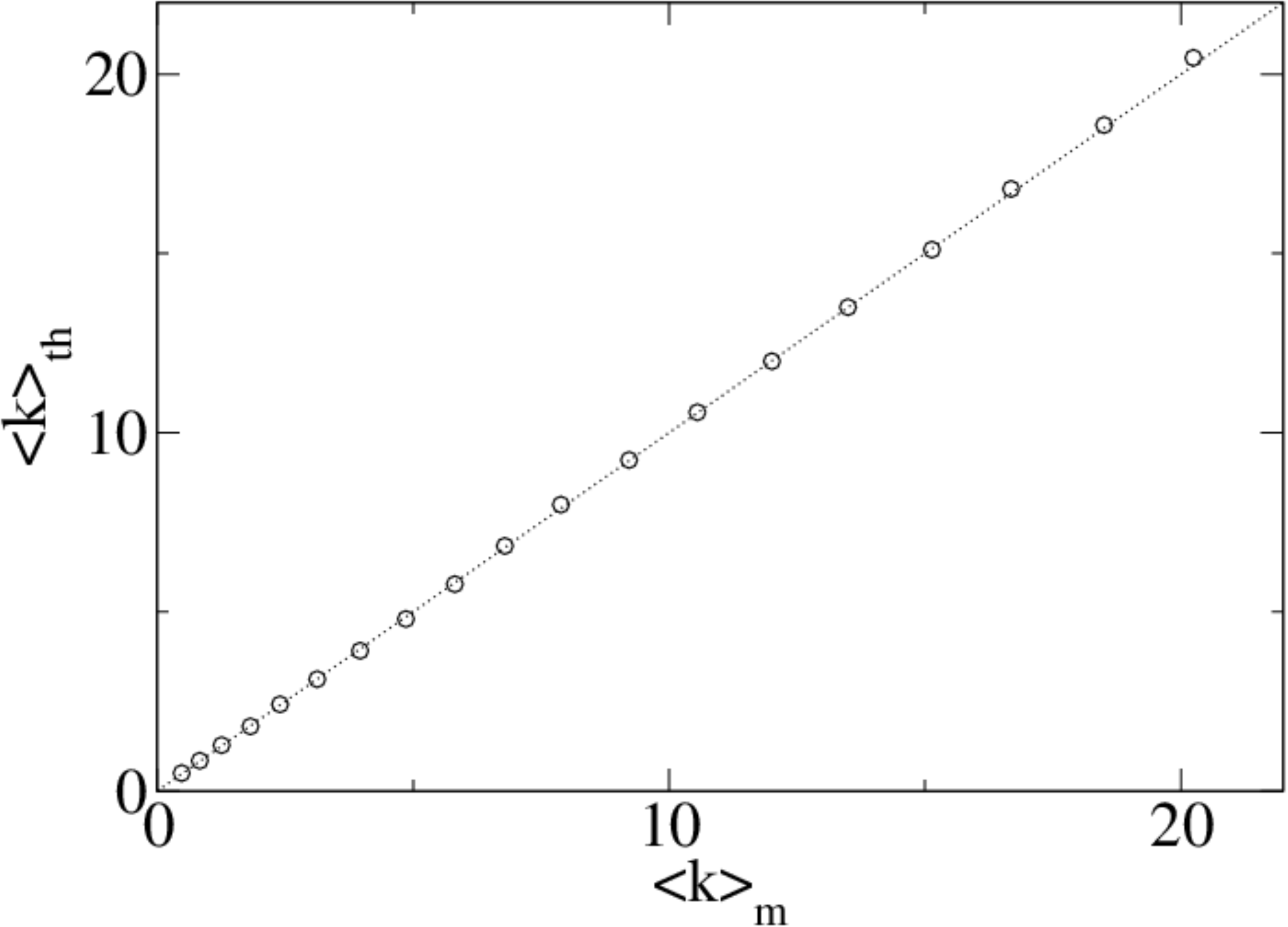}}
\put(98,76.5){\includegraphics[width=107\unitlength,height=73\unitlength]{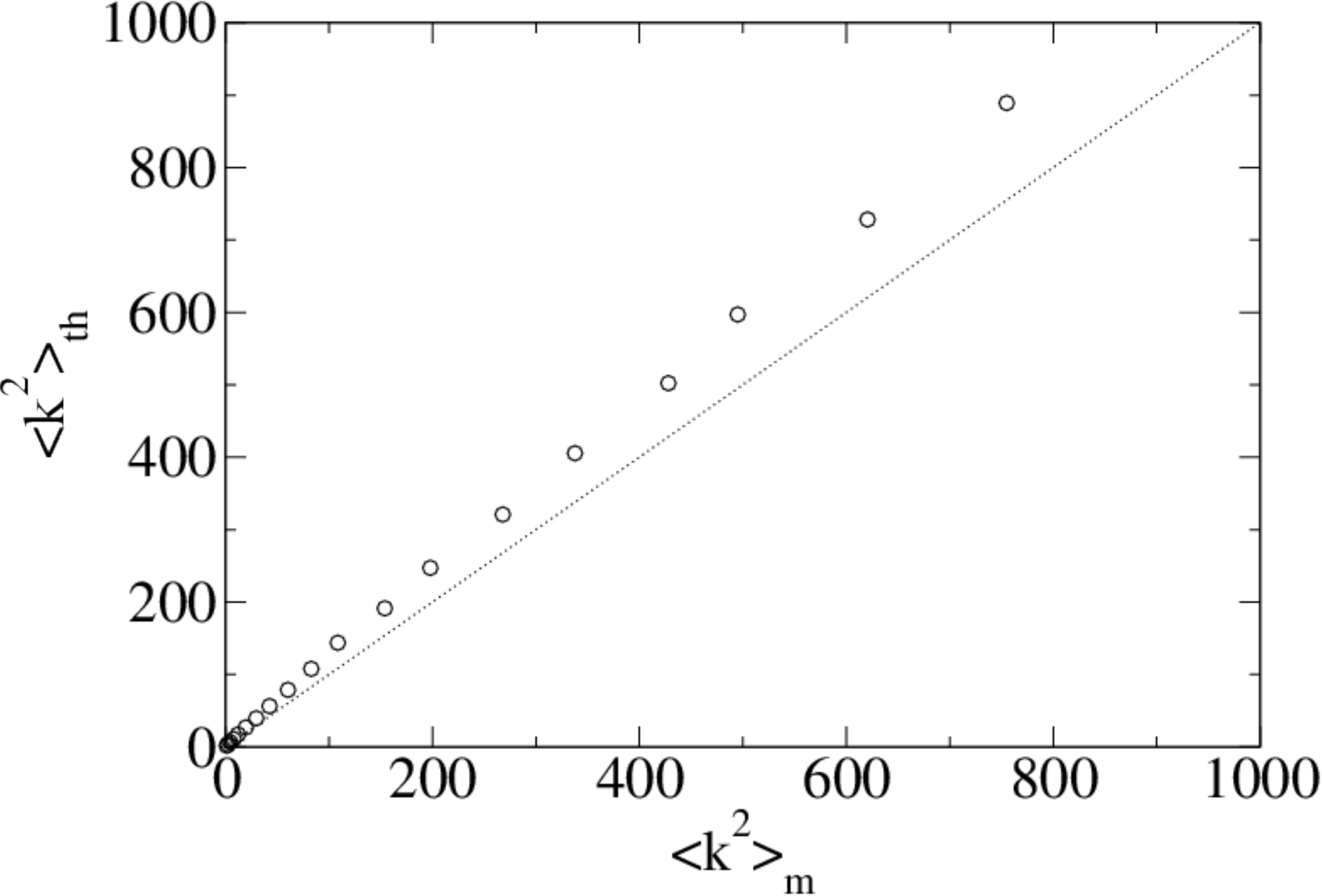}}

\put(-3.5,0){\includegraphics[width=97\unitlength,height=70\unitlength]{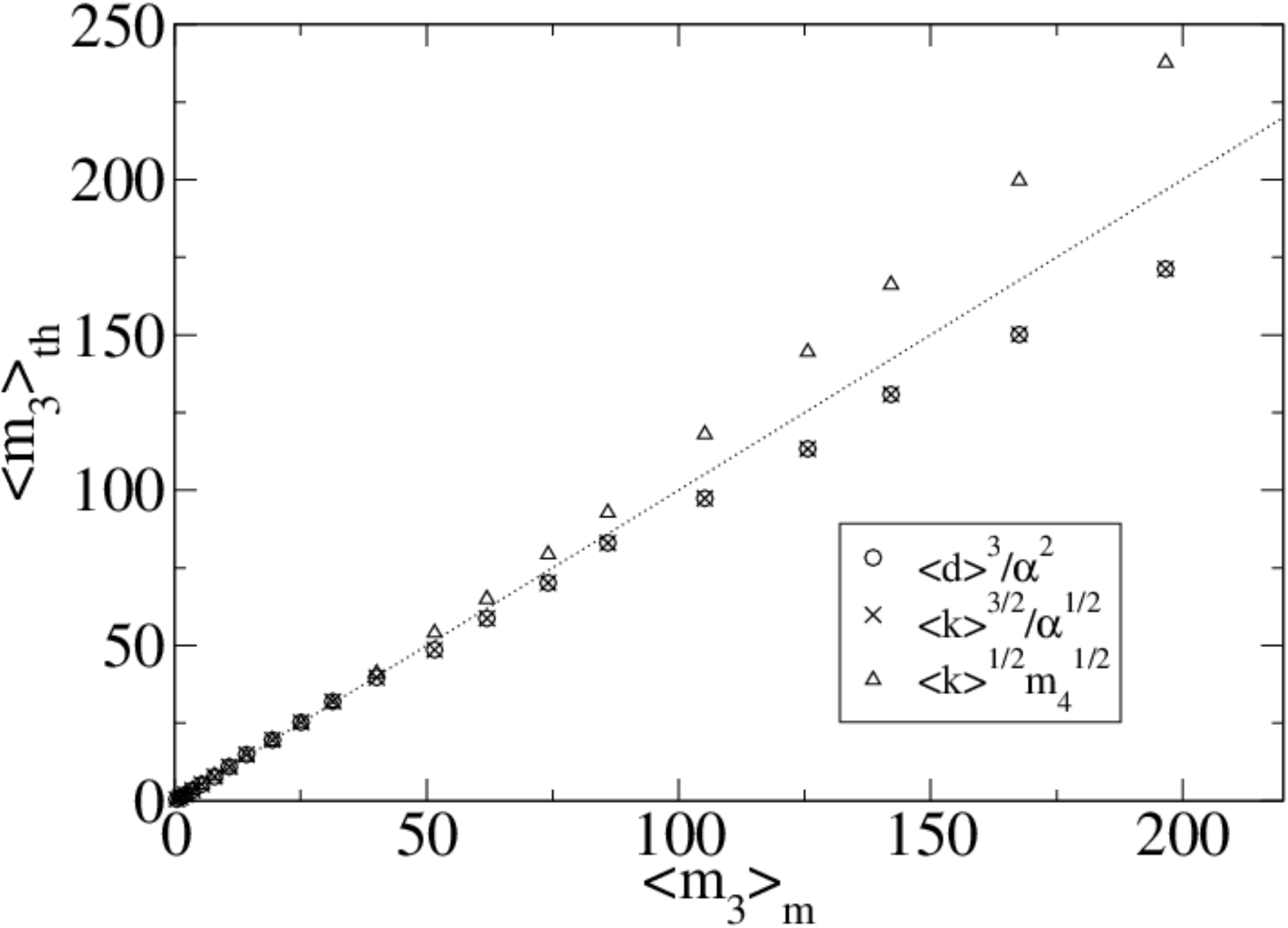}}
\put(100,0.3){\includegraphics[width=105\unitlength,height=70\unitlength]{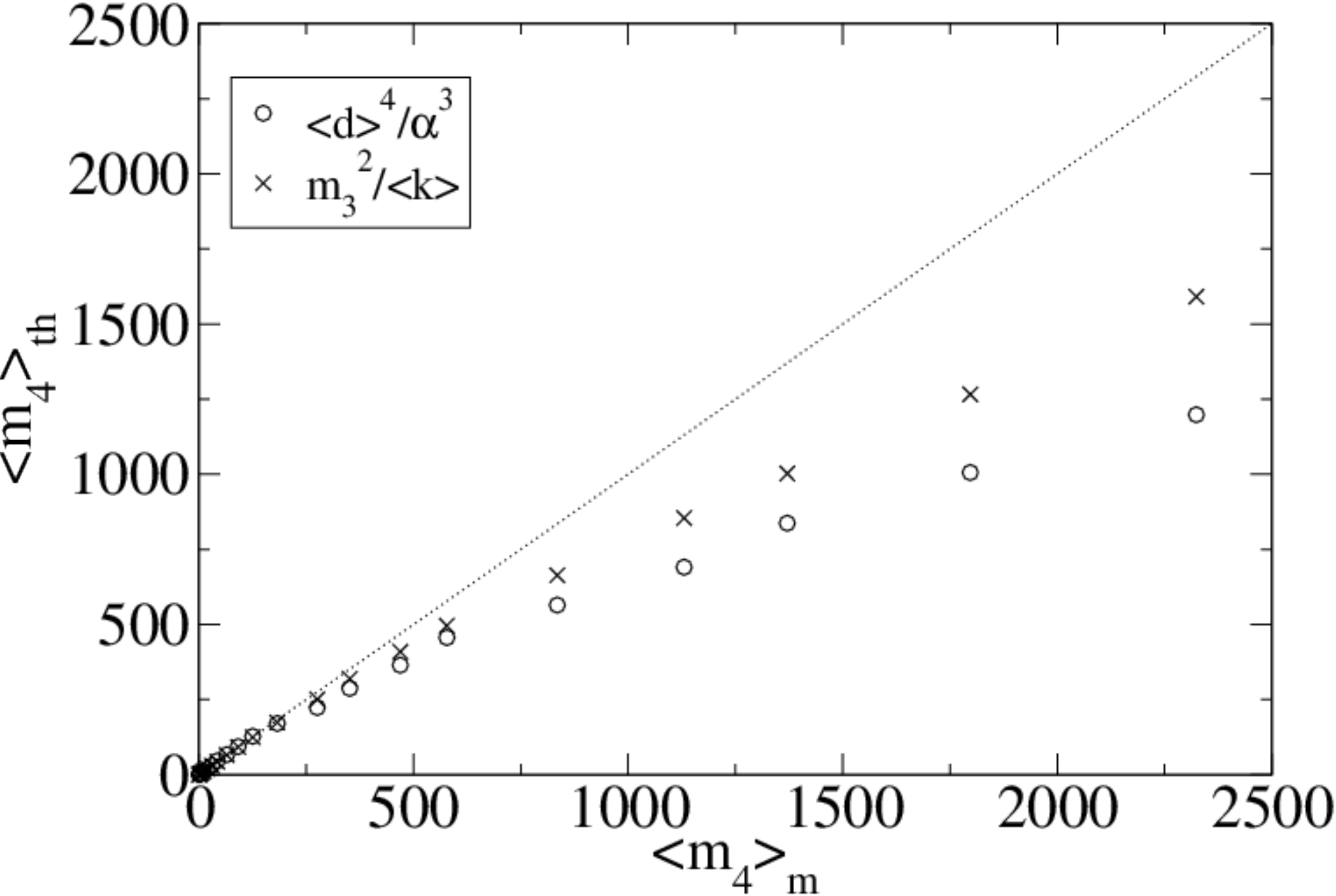}}
\end{picture}

\caption{
Symbols: theoretical $\bra \ldots\ket_{\rm th}$ versus   measured $\bra \ldots\ket_{\rm m}$  values 
of observables  $\kav$, $\kvar$, $m_3$ and $m_4$ in synthetic random graphs 
$\bc$ with $N=3000$, defined via (\ref{eq:quenched_q},\ref{eq:ensemble_c}) for a power-law distributed promiscuity distribution
$P(d)$. Theoretical values are given by formulae 
(\ref{eq:kav_d}) for $\kav$, (\ref{eq:kvar_d}) for $\kvar$,  
(\ref{eq:m3_d1}), (\ref{eq:m3_d3}) and (\ref{eq:m4_d}) for $m_3$ and (\ref{eq:mL_d}) and (\ref{eq:m4_d}) for $m_4$.
Dotted lines: the diagonals (shown as guides to the eye).
}

\label{fig:dth}
\end{figure}

The first two moments of the degree distribution $p(k)$ of the separable PIN networks $\bc$ are
\bea
\bra k\ket &=&\sum_k k p(k)=\sum_d P(d)\sum_\ell \rme^{-d}\frac{d^\ell}{\ell!} \frac{\ell \dav}{\alpha} 
=\dav^2/\alpha
\label{eq:kav_d}
\\
\bra k^2\ket&=&
\sum_k k^2 p(k)=\sum_d P(d)\sum_\ell e^{-d}\frac{d^\ell}{\ell!}
\Big[
\Big(\frac{\ell \dav}{\alpha}\Big)^2+\frac{\ell \dav}{\alpha}
\Big]
\nonumber\\
&=&
\dav^2/\alpha+\dav^3/\alpha^2+\dav^2\bra d^2\ket/\alpha^2
\label{eq:kvar_d}
\eea
Combination of (\ref{eq:kav_d}), (\ref{eq:kvar_d}) and (\ref{eq:m3_d1}) now yields the relation
\be
\bra d^2\ket/\alpha=(\bra k^2\ket -\kav -m_3)/\kav
\label{eq:m3_d2}
\ee
which still involves $\bra d^2\ket$ and $\alpha$.
We can also find an alternative expression for the density of loops of length $3$ 
by combining (\ref{eq:kav_d}) and (\ref{eq:m3_d1}) 
\be
m_3=\bra k\ket^{3/2}/\sqrt{\alpha}
\label{eq:m3_d3}
\ee
Unfortunately, neither of our two expressions for $m_3$, (\ref{eq:m3_d2}) nor (\ref{eq:m3_d3}), are useful, because the protein 
promiscuities distribution $P(d)$ and the ratio $\alpha$ 
are generally unknown. Access to 
information on these quantities via future detection experiments may therefore be 
extremely welcome in support of theoretical modelling of protein interaction datasets. 
To make progress, we need to derive relations for graph observables that are independent of $\alpha$ and 
$P(d)$. We note that (\ref{eq:mL_d}) yields
\be
\forall L\geq 3:~~~~
m_{L+1}/m_L=\bra d \ket/\alpha
\ee
This can be rewritten using (\ref{eq:kav_d}), as
\bea
\forall L\geq 3:~~~~
m_{L+1}/m_L=\sqrt{\kav/\alpha}
\eea
On the other hand, we know from (\ref{eq:m3_d3}) 
that 
$m_3/\kav=\sqrt{\kav/\alpha}$. 
Combining the above formulae allows us to establish the following relation, that now is  completely 
independent of $P(d)$ and $\alpha$:
\be
m_4=m_3^2/\kav
\label{eq:m4_d}
\ee
Again we have tested the various formulae in synthetically generated graphs, see Figure \ref{fig:dth}.

\subsection{Link between ${\bf a}$ and $\bc$ graph definitions}

\begin{figure}[t]
\setlength{\unitlength}{0.69mm}
\hspace*{30mm}
\begin{picture}(200,152)

\put(-3.6,78){\includegraphics[width=94.5\unitlength,height=70\unitlength]{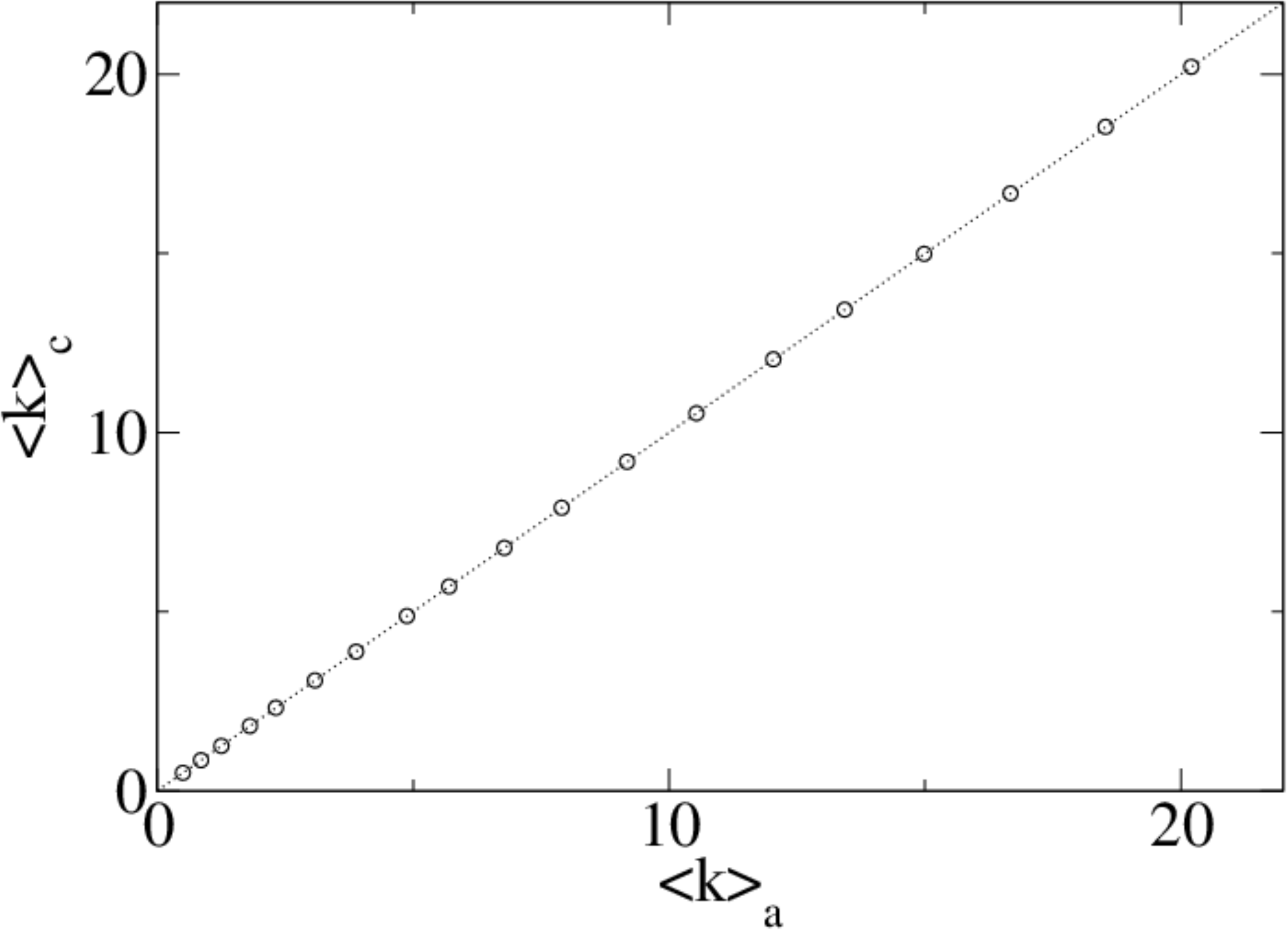}}
\put(97.5,76.5){\includegraphics[width=102.3\unitlength,height=73\unitlength]{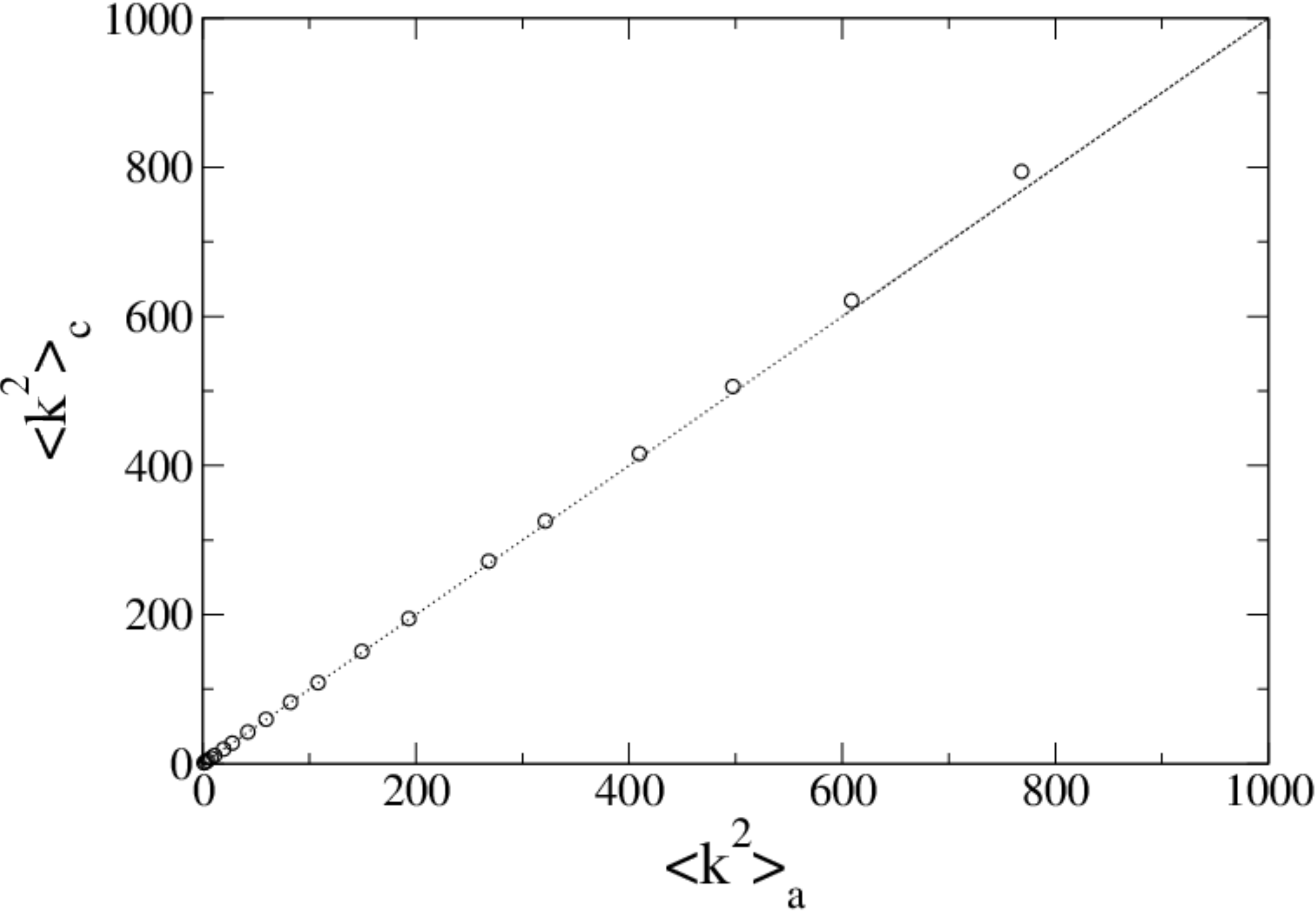}}

\put(-3.5,0){\includegraphics[width=97\unitlength,height=70\unitlength]{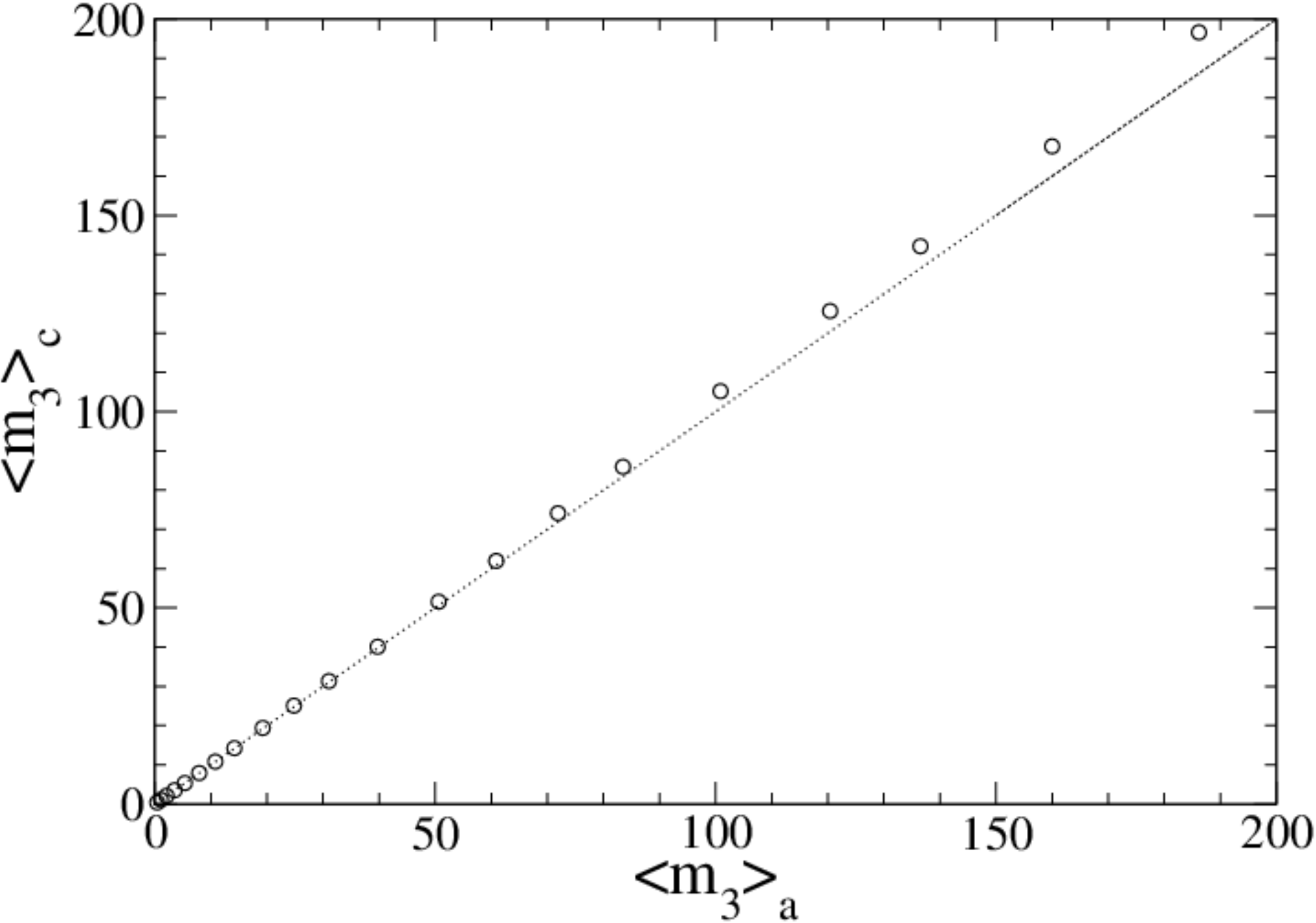}}
\put(100,0.3){\includegraphics[width=100\unitlength,height=70\unitlength]{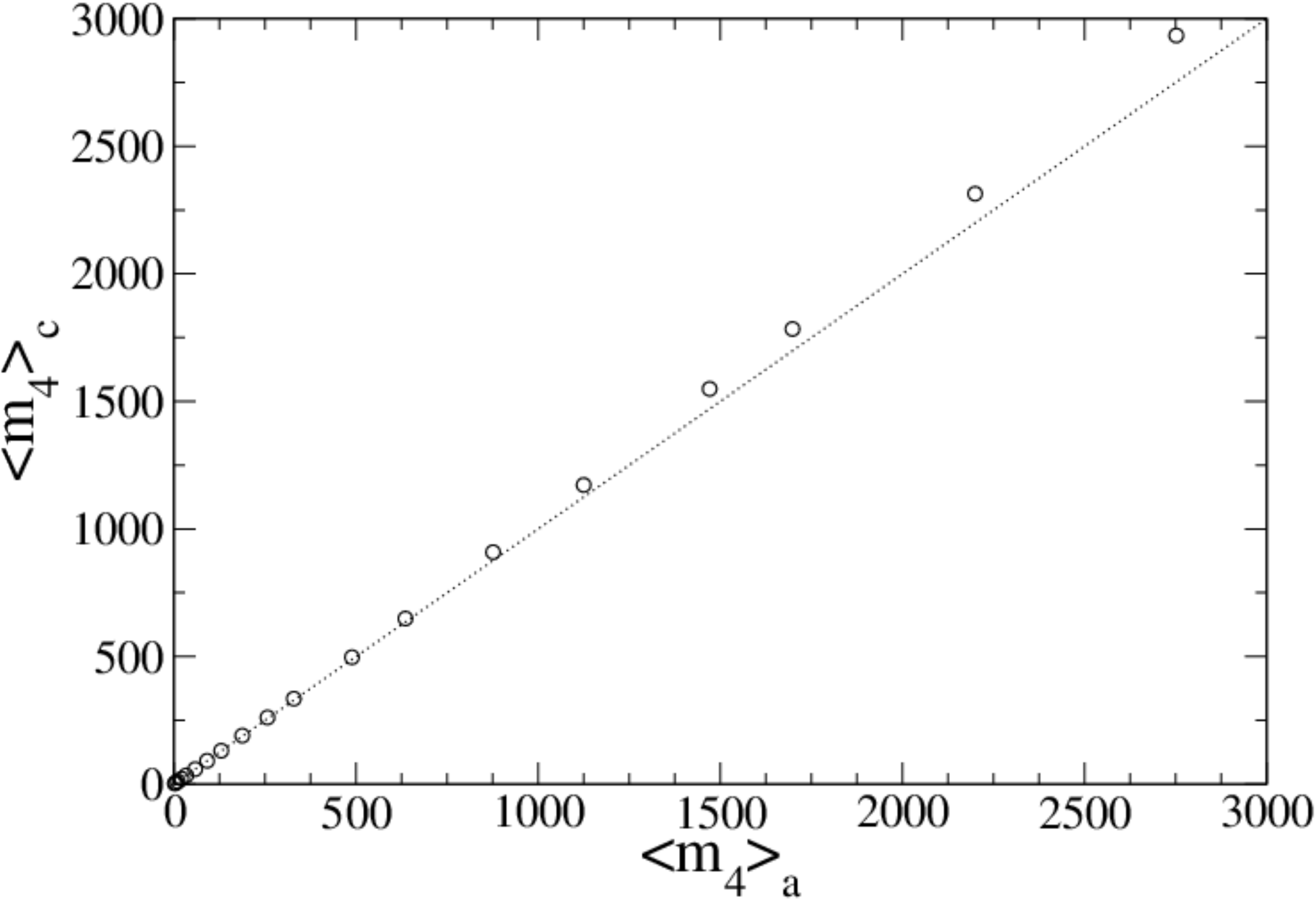}}
\end{picture}

\begin{caption}{
Symbols:  $\kav$, $\bra k^2\ket$, $m_3$ and $m_4$ as measured 
in synthetic graphs $\bc$ drawn from (\ref{eq:ensemble_c}) with $N=3000$, shown versus corresponding values found in the binary graphs $\ba$ drawn from (\ref{eq:ensemble_a}). 
Bipartite interaction graphs $\bxi$ are drawn 
from (\ref{eq:quenched_d}), with protein promiscuity distributions $P(d)$   that have 
a power law form. Dotted line: the diagonals (shown as guides to the eye).  As expected, the values measured in the weighted graphs 
$\bc$ are consistently higher than in the binary ones, but these deviations 
get smaller for increasing network sizes $N$.} 
\end{caption}

\label{fig:acd}
\end{figure}

As a final step, we check whether the observables $m_3$ and $m_4$ are indeed the same for the two PIN definitions (\ref{eq:ensemble_a}, \ref{eq:ensemble_c}), with the bipartite graph of our protein-driven ensemble (\ref{eq:quenched_d}), since 
protein detection experiments provide the binary matrix 
$\ba$ as opposed to the weighted graph $\bc$ for which (\ref{eq:m4_d}) was
derived. Again we denote averages relating to $\ba$ as $\bra \ldots\ket_a$, and those relating to $\bc$ 
as $\bra \ldots \ket$. 
For the moments of the degree distributions we find the differences to be negligible:
\begin{eqnarray}
\bra k\ket_a &=& \frac{1}{N}\sum_{ij}\bra a_{ij}\ket_a=\frac{\bra d\ket^2}{\alpha}+\order(N^{-1})
= \bra k\ket+\order(N^{-1})
\label{eq:kav_d}
\\
\bra k^2\ket_a&=& \frac{1}{N}\sum_{i\neq j\neq k}\bra a_{ij}a_{jk}\ket
=\frac{\bra d\ket^2}{\alpha}
+\frac{\bra d\ket^3}{\alpha^2}
+\frac{\bra d^2\ket \bra d \ket^2}{\alpha^2}+\order(N^{-1})
= \bra k^2\ket+\order(N^{-1})
\label{eq:kvar_da}
\eea
The same is true for the densities of 
 loops of length $3$ and $4$:
\bea
m_3^a&=&\frac{1}{N}\sum_{i\neq j\neq k(\neq i)}\bra a_{ij}a_{jk}a_{ki}\ket
=\frac{\bra d\ket^3}{\alpha^2} +\order{(N^{-1})}
= m_3 +\order{(N^{-1})}
\label{eq:m3_d}
\\
 m_4^a&=&\frac{1}{N}\sum_{[i,j,k,\ell]}\bra a_{ij}a_{jk}a_{k\ell}a_{\ell i}\ket
=\frac{\bra d\ket^4}{\alpha^3}+\order{(N^{-1})}
= m_4+\order{(N^{-1})}
\label{eq:m4_d}
\end{eqnarray}
This equivalence between the ensembles $p(\ba)$ and $p(\bc)$  when calculating the main average values of 
graph observables for large $N$ implies that large protein interaction adjacency matrices can in practice be regarded as having a separable structure. 
Again, we check our relations (\ref{eq:kav_d}, \ref{eq:kvar_d}, \ref{eq:m3_d}, \ref{eq:m4_d}), 
against synthetically generated graphs and show results in figure \ref{fig:acd}.

\section{Macroscopic observables in the mixed ensemble}

\label{sec:mix}
The two bipartite graph ensembles (\ref{eq:quenched_q}, \ref{eq:quenched_d}) considered so far led to Poissonian distributions  either  for the protein promiscuities $d_i$ (in the $q$-ensemble), or 
for the complex sizes $q_\mu$ (in the $d$ ensemble). 
It is possible to model heterogeneity in both $d_i$ and $q_\mu$ using the mixed ensemble 
(\ref{eq:quenched_qd}).
Due to the similarities with previous calculations we can and will be more brief in this section. 
For ensemble (\ref{eq:quenched_qd}) the expectation values of individual links in the weighted graph $\bc$ are
\bea
\bra c_{ij}\ket=\sum_\mu \bra \xi_i^\mu \xi_j^\mu\ket
=\sum_\mu \frac{d_i d_j q_\mu^2}{\alpha^2 \bra q \ket^2 N^2}
=\frac{d_i d_j \bra q^2\ket}{\alpha\bra q \ket^2 N}+\order(N^{-3/2})
\eea
and the average connectivity follows as
\bea
\bra k\ket=\frac{1}{N}\sum_{ij} \bra c_{ij}\ket
=\frac{\dav^2 \bra q^2\ket}{\alpha \bra q \ket^2 }+\order{(N^{-1/2})}
=\alpha \bra q^2\ket +\order{(N^{-1/2})}
\label{eq:kav_mix}
\eea
Full details are found in \ref{app:pij}.
As in previous ensembles, the leading contribution to the density of 
length-$3$ loops comes from the $S_3$ stars 
in the bipartite graphs, now giving
\bea
m_3=\frac{1}{N}\sum_{[ijk]}\sum_\mu \bra \xi_i^\mu \xi_j^\mu \xi_k^\mu\ket=\frac{1}{N}\sum_{[ijk]}\sum_\mu 
\frac{d_id_jd_k q_\mu^3}{\alpha^3\qav^3 N^3}\simeq
\frac{\dav^3 \bra q^3\ket}{\alpha^2 \qav^3}=\alpha \bra q^3\ket
\label{eq:m3_qd}
\eea
As before, the heterogeneity in the $d$ affects neither the average connectivity $\kav$ nor the density of triangles $m_3$, both are 
as they were in the $q$-ensemble. This is confirmed numerically, see Figure \ref{fig:m3}.
The degree distribution for large $N$ in the ensemble $p(\bc)$  
is calculated in \ref{app:pk}, giving
\begin{eqnarray}
p(k)
&=& \int_0^\infty\!\rmd y~P(y)~\rme^{-y} y^k/k!
\label{eq:pk_mix}
\eea
where
\bea
P(y)&=& 
 \sum_d P(d)\rme^{-d}
\sum_{\ell\geq 0}\frac{d^\ell}{\ell !}
\sum_{q_1\ldots q_\ell\geq 0} W(q_1)\ldots W(q_\ell)
~\delta[y-\sum_{r\leq \ell }q_r ]
\label{eq:py_mix}
\end{eqnarray}
Again it is possible to relate the asymptotic behaviour of $p(k)$ to 
that of $P(d)$ and $W(q)$, by inspecting the relation between the 
relevant generating functions. Using our previous definitions 
for $Q_1(z),~ Q_2(z),~ Q_3(z)$, and $Q_4(z)$, we obtain via (\ref{eq:pk_mix}) and (\ref{eq:py_mix}):
\bea
Q_1(z)&=&\int \rmd y\, P(y)\sum_k \rme^{-y}(ye^{-z})^k/k!
=\int \rmd y\, P(y)\rme^{-y(1-\rme^{-z})}
\nonumber\\
&=&Q_2(1-e^{-z})
\label{eq:Q12}
\\
Q_2(z)&=&\sum_d P(d)\rme^{-d}\sum_\ell \frac{d^\ell}{\ell!}
\prod_{r=1}^\ell \Big(\sum_{q_r} W(q_r) \rme^{-z q_r}\Big)
=\sum_d P(d)\rme^{-d}\sum_\ell \frac{d^\ell}{\ell!}Q_3^\ell(z)
\nonumber\\
&=&\sum_d P(d)\rme^{-d[1-Q_3(z)]}=Q_4(1-Q_3(z))
\label{eq:Q24}
\eea
Expanding (\ref{eq:Q12}) for small $z$ tells us that 
$Q_1(z)\simeq Q_2(z)$. Substitution into (\ref{eq:Q24}) subsequently gives
\be
Q_1(z)\simeq Q_4(1-Q_3(z))
\label{eq:Q14}
\ee
Assuming $W(q)$ to have a power-law tail, but with a finite first moment (as in all cases previously considered), i.e. $W(q)\simeq K q^{-\gamma}$ with $\gamma>2$, 
its generating function $Q_3(z)$ can be written as 
\be
Q_3(z)=1-\bra q^2\ket z/\qav+\order{(z^\delta)}
\ee
where $\delta={\rm min}\{2,\gamma-1\}$.
Insertion into (\ref{eq:Q14}) then leads to
\be
Q_1(z)\simeq Q_4(z\bra q^2\ket/\qav-\order{(z^\delta)})
\label{eq:Q14_mix}
\ee
If $p(k)=Ck^{-\mu}$, with $2<\mu<3$, we may use our earlier result (\ref{eq:euler}) and get 
\be
Q_4(x-\order{(x\qav/\qvar)^\delta})\simeq 1-\kav \qav x/\bra q^2\ket+C\Gamma(1\!-\!\mu)(\qav/\bra q^2\ket)^{\mu-1}x^{\mu-1}
\label{eq:Q4_mix}
\ee
If $\gamma>\mu$ we have $\delta>\mu-1$, so we can 
neglect the second term in the argument of $Q_4$ and conclude that   
the promiscuity distribution has the asymptotic form $P(d)=C'd^{-\mu}$
where $C'=C(\qvar/\qav)^{1-\mu}$. This means that if $W(q)$ decays faster than $p(k)$ (as in Section \ref{sec:d}), 
then the tail in $p(k)$ must arise from the tail in $P(d)$. Note, however, that heterogeneities in $P(q)$ will affect the amplitude 
of the power law tail in $P(d)$, which will be smaller  by a factor 
$(\qvar/\qav^2)^{1-\mu}$ compared to the case where 
$P(q)=\delta_{q,\qav}$, where we had $C'=C\qav^{1-\mu} $.
Conversely, if $\gamma=\mu$ we have $\delta=\mu-1$, and writing the $\order{(z^\delta)}$ 
term explicitely in (\ref{eq:Q14_mix}) gives
\bea
Q_4(z\qvar/\qav-K \Gamma(1\!-\!\mu)z^{\mu-1})=1-\kav z +C\Gamma(1\!-\!\mu)z^{\mu-1}
\eea
Expanding both sides in powers of $z$ and equating prefactors tells us that either $C'=0$ and $C=K\dav$ 
(i.e. $K=C/\alpha\qav$, which retrieves the case in Section \ref{sec:q}), or 
$\delta=\mu$ with $K\dav+C'(\qvar/\qav)^{\mu-1}=C$. Hence, if $P(d)$ is as broad as $W(q)$, then 
both contribute to the tail in $p(k)$, whose amplitude will be the sum of the amplitudes of the tails in $P(q)$ and $P(d)$.
We see in (\ref{eq:Q4_mix}) that $\gamma<\mu$ is not possible, i.e. $W(q)$ needs to decay at least as fast as $p(k)$.

In \ref{app:pk}  we calculate the first two moments of the degree distribution $p(k)$ of the ensemble $p(\bc)$. This recovers (\ref{eq:kav_mix}) for the first moment, and for the second moment gives
\begin{eqnarray}
\bra k^2\ket&=& \alpha \bra q^2\ket+ \alpha \bra q^3\ket+\bra d^2\ket \kav^2/\dav^2
\label{eq:kvar_mix}
\end{eqnarray}
Substituting (\ref{eq:kav_mix}) and (\ref{eq:m3_qd}) into (\ref{eq:kvar_mix})
then leads to
\bea
m_3&=&\bra k^2\ket- 
\kav-\kav^2 \bra d^2\ket/\dav^2
\label{eq:m3_mix}
\end{eqnarray}
The density of length-$3$ loops depends again on the first two moments of the degree distribution $p(k)$, but is also seen to depend 
on the first two moments 
of the promiscuity distribution $P(d)$, which is unknown. Hence, this relation cannot serve as a test of PIN data quality. 
It is nevertheless useful for comparing the 
mixed ensemble to the $d$- and the $q$-ensembles in synthetically generated data. 

\section{Numerical comparison of the three bipartite generative ensembles}
\label{sec:syn}

\begin{figure}[t]
\setlength{\unitlength}{0.345mm}
\hspace*{10mm}
\begin{picture}(300,368)

\put(-40,250){\includegraphics[width=160\unitlength,height=110\unitlength]{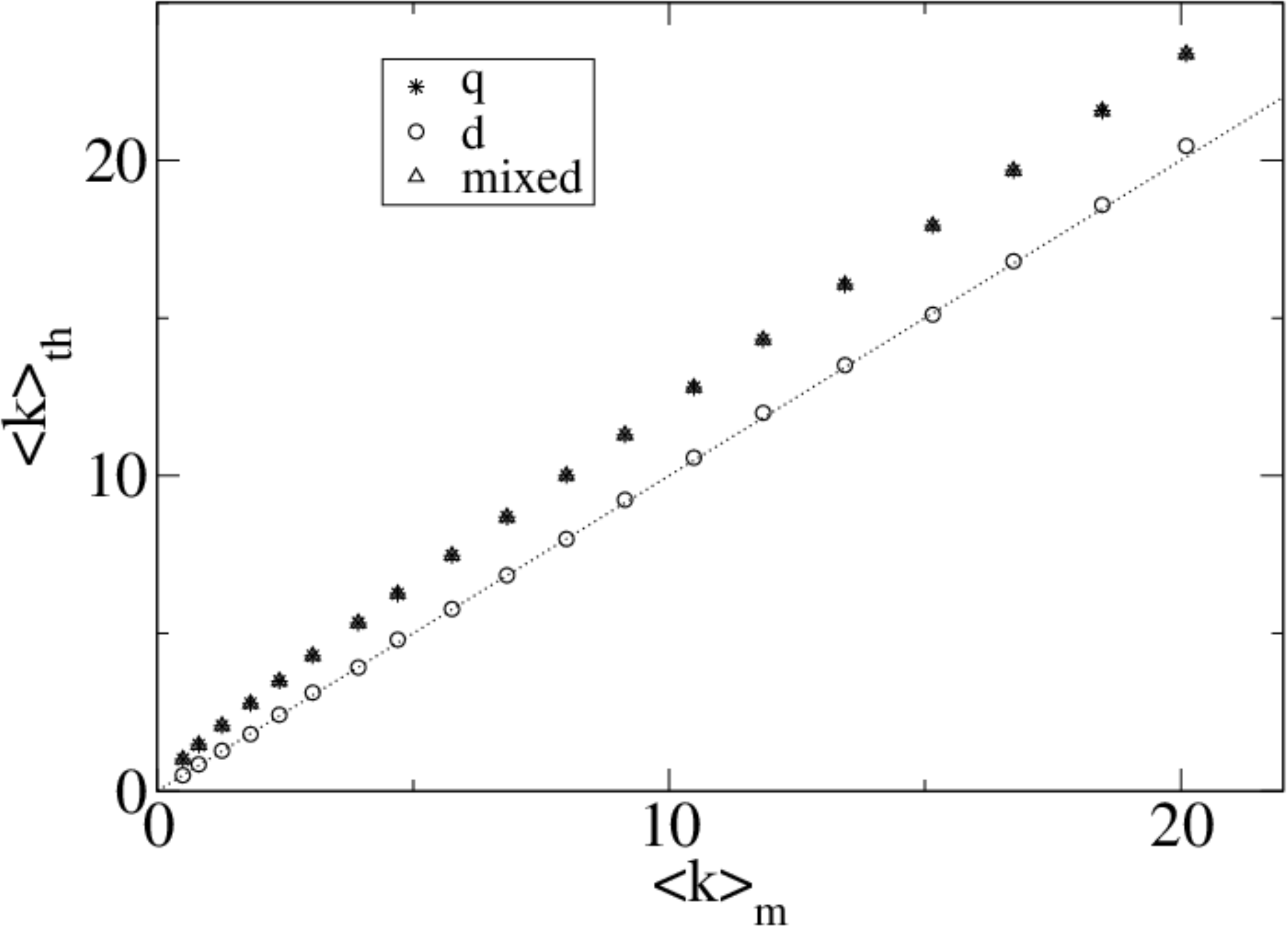}}
\put(125,247){\includegraphics[width=175\unitlength,height=115\unitlength]{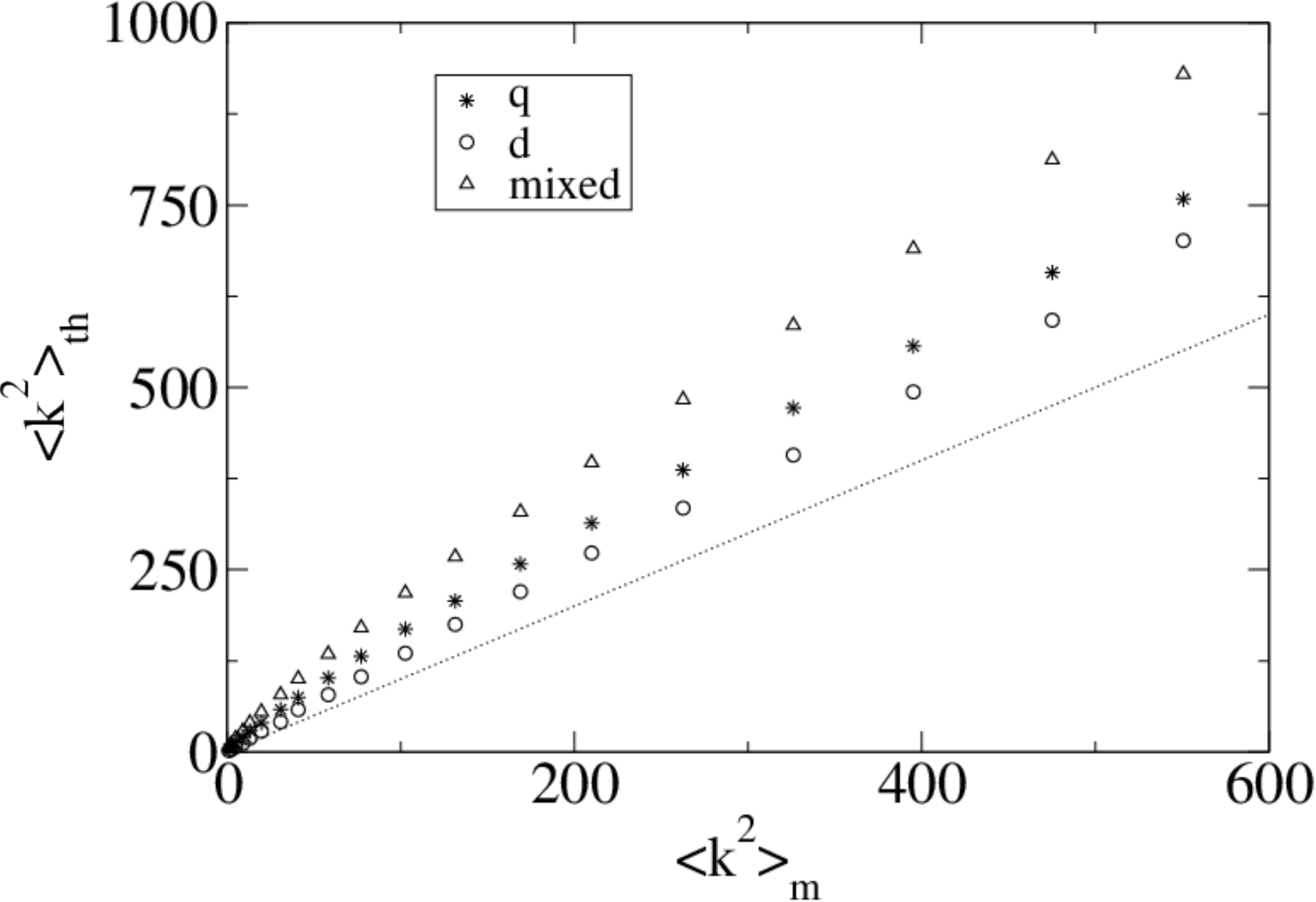}}
\put(310,251.3){\includegraphics[width=168\unitlength,height=111\unitlength]{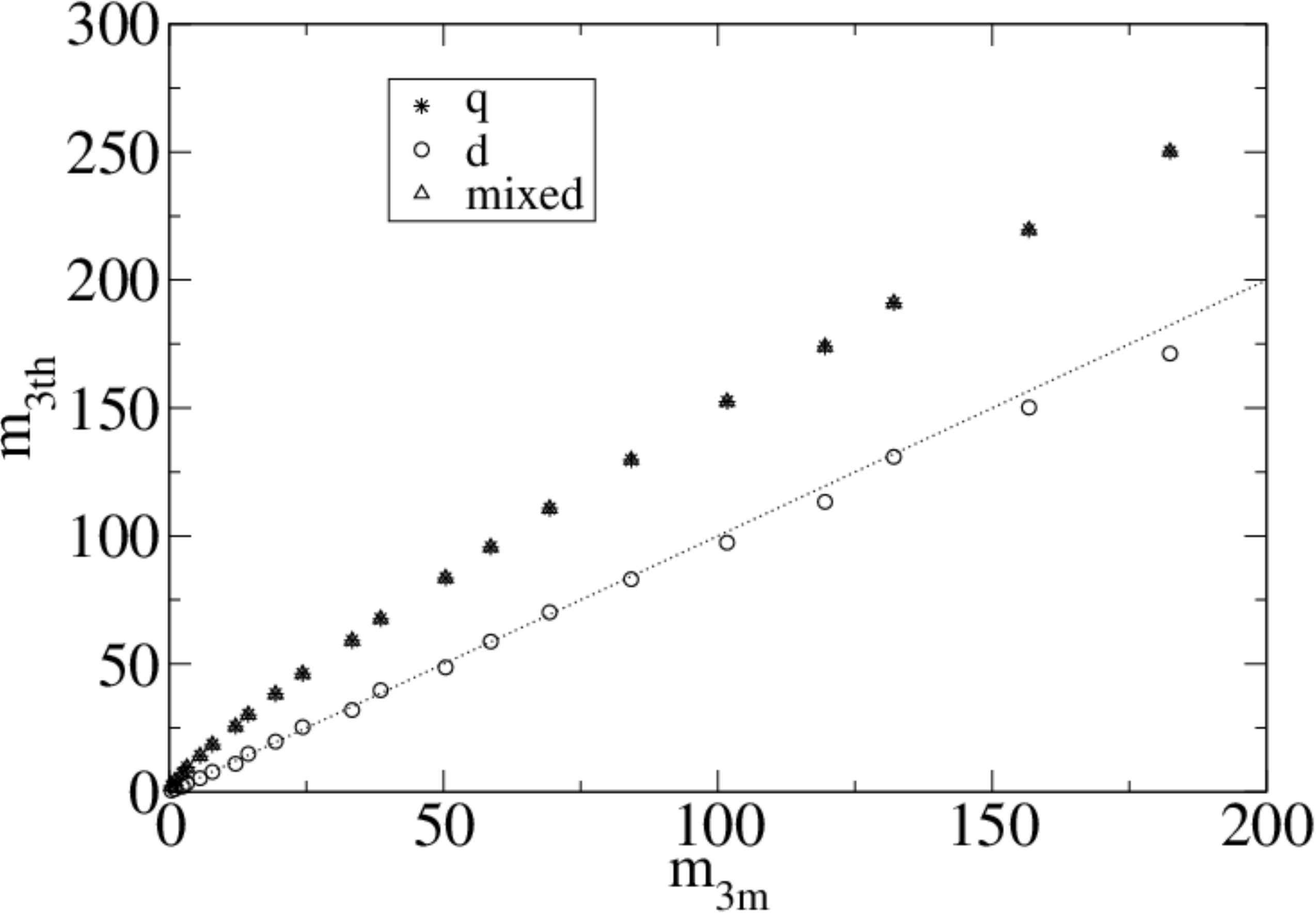}}

\put(-40,125){\includegraphics[width=162.5\unitlength,height=113\unitlength]{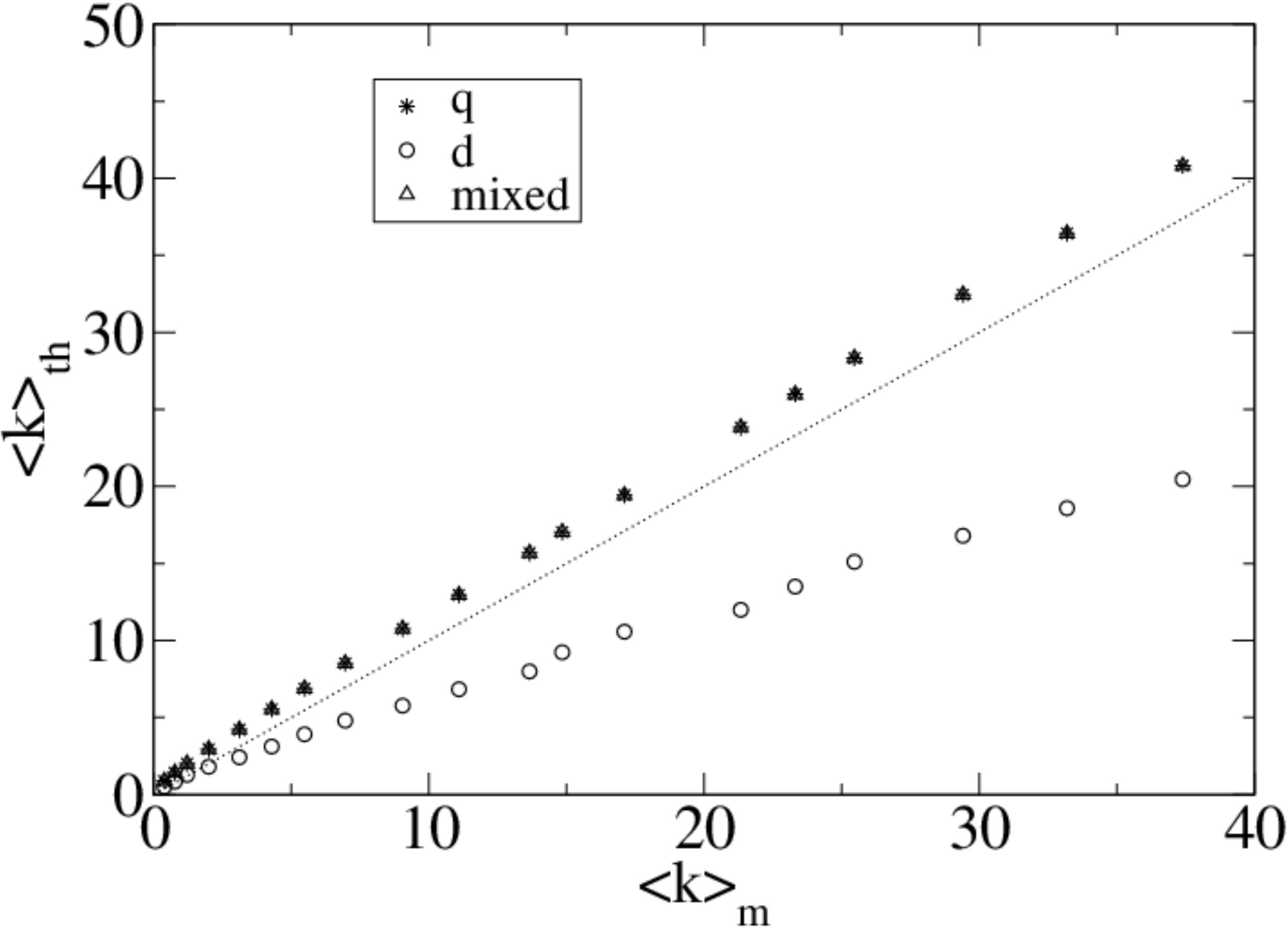}}
\put(125,122){\includegraphics[width=170\unitlength,height=113\unitlength]{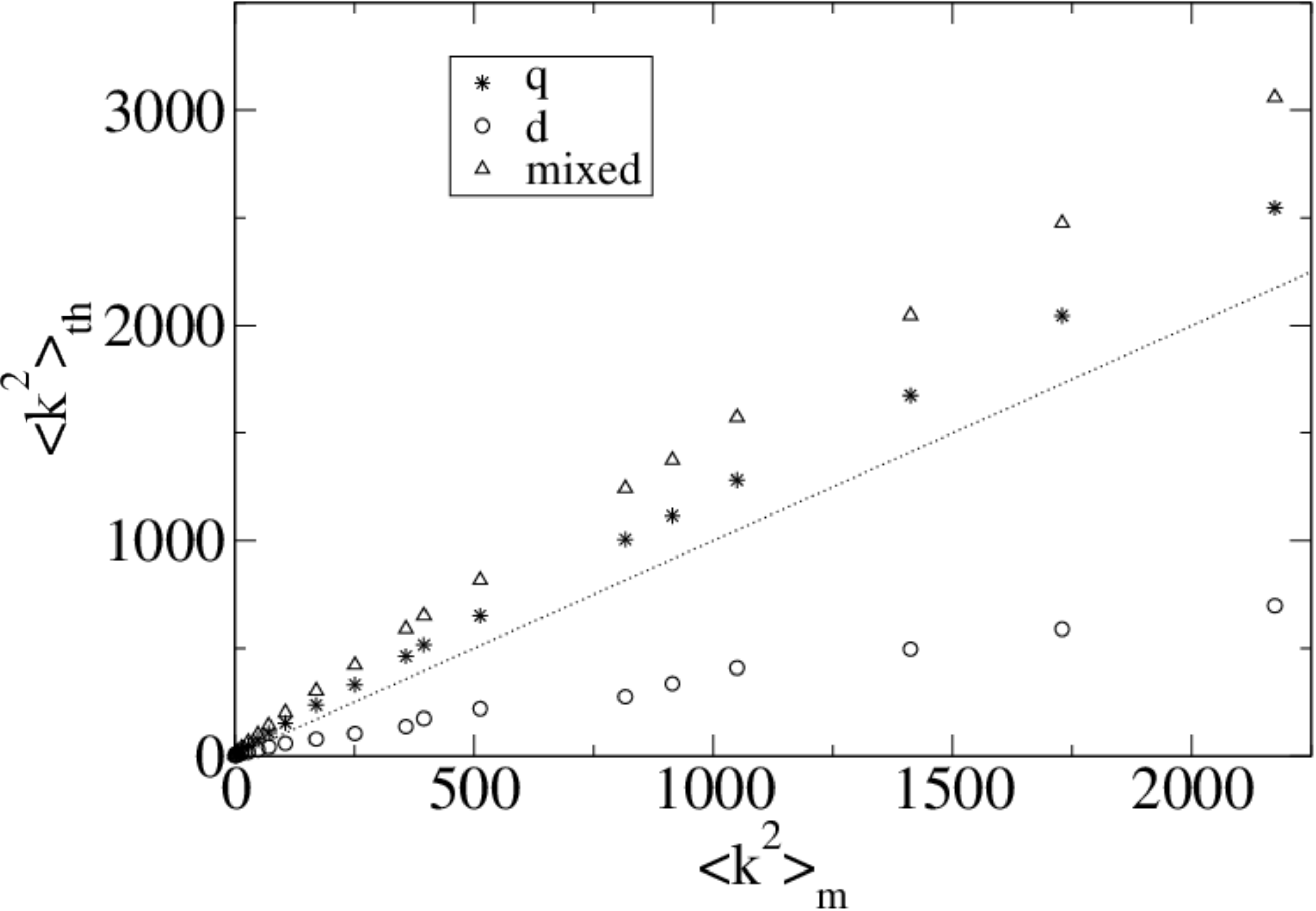}}
\put(306,126){\includegraphics[width=174\unitlength,height=112\unitlength]{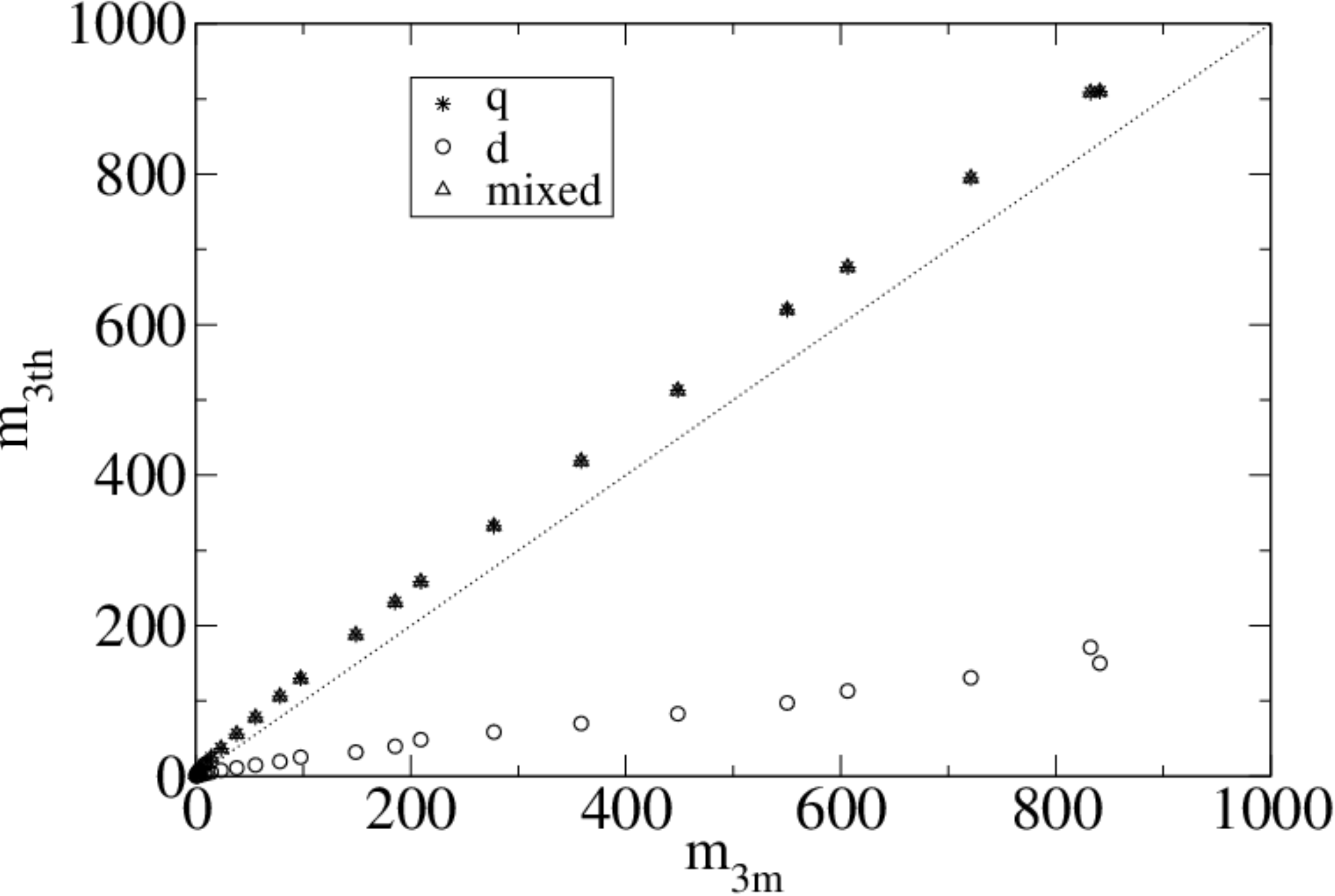}}

\put(-40,0){\includegraphics[width=160\unitlength,height=110\unitlength]{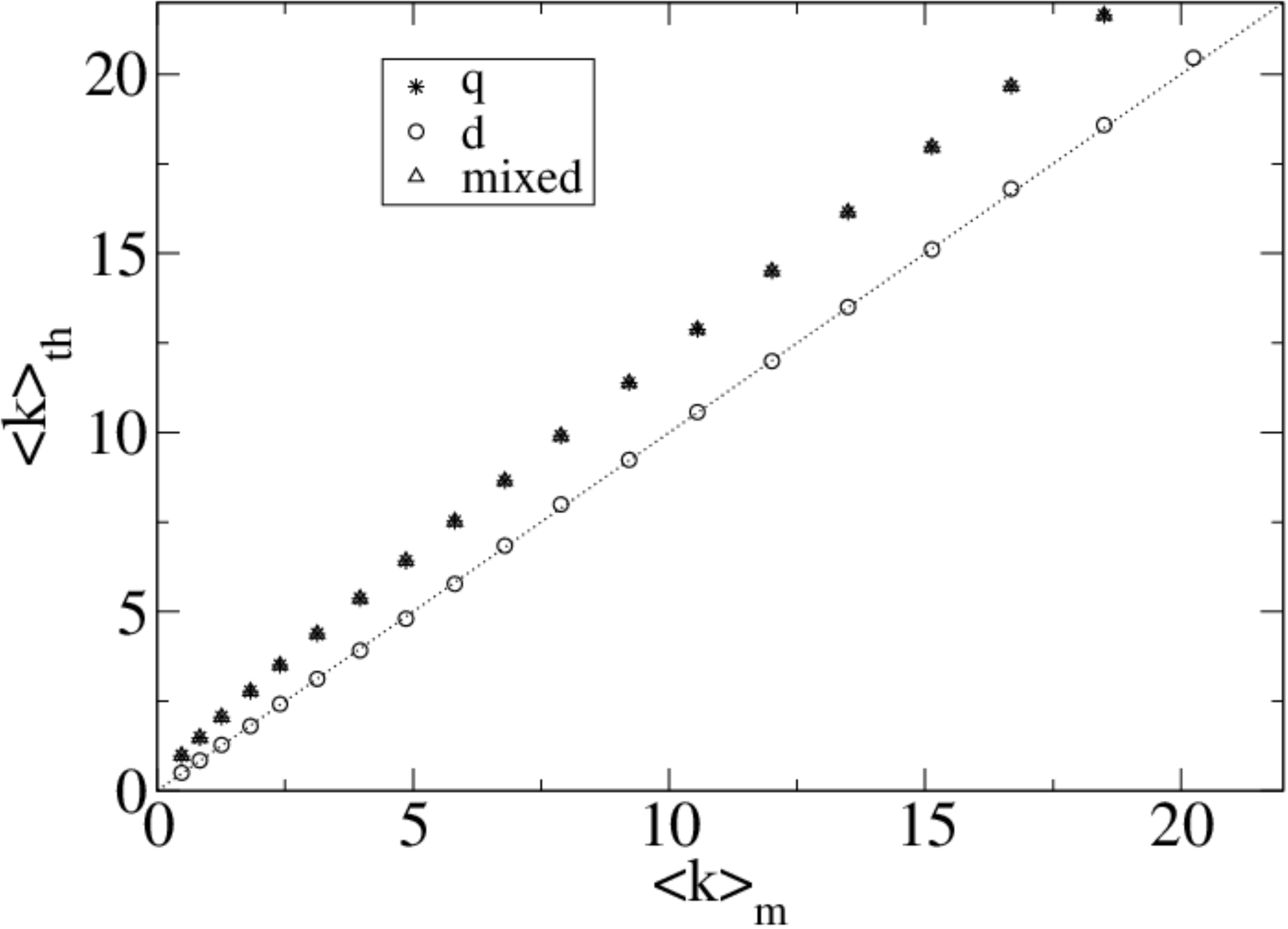}}
\put(125,-2.5){\includegraphics[width=177\unitlength,height=115.3\unitlength]{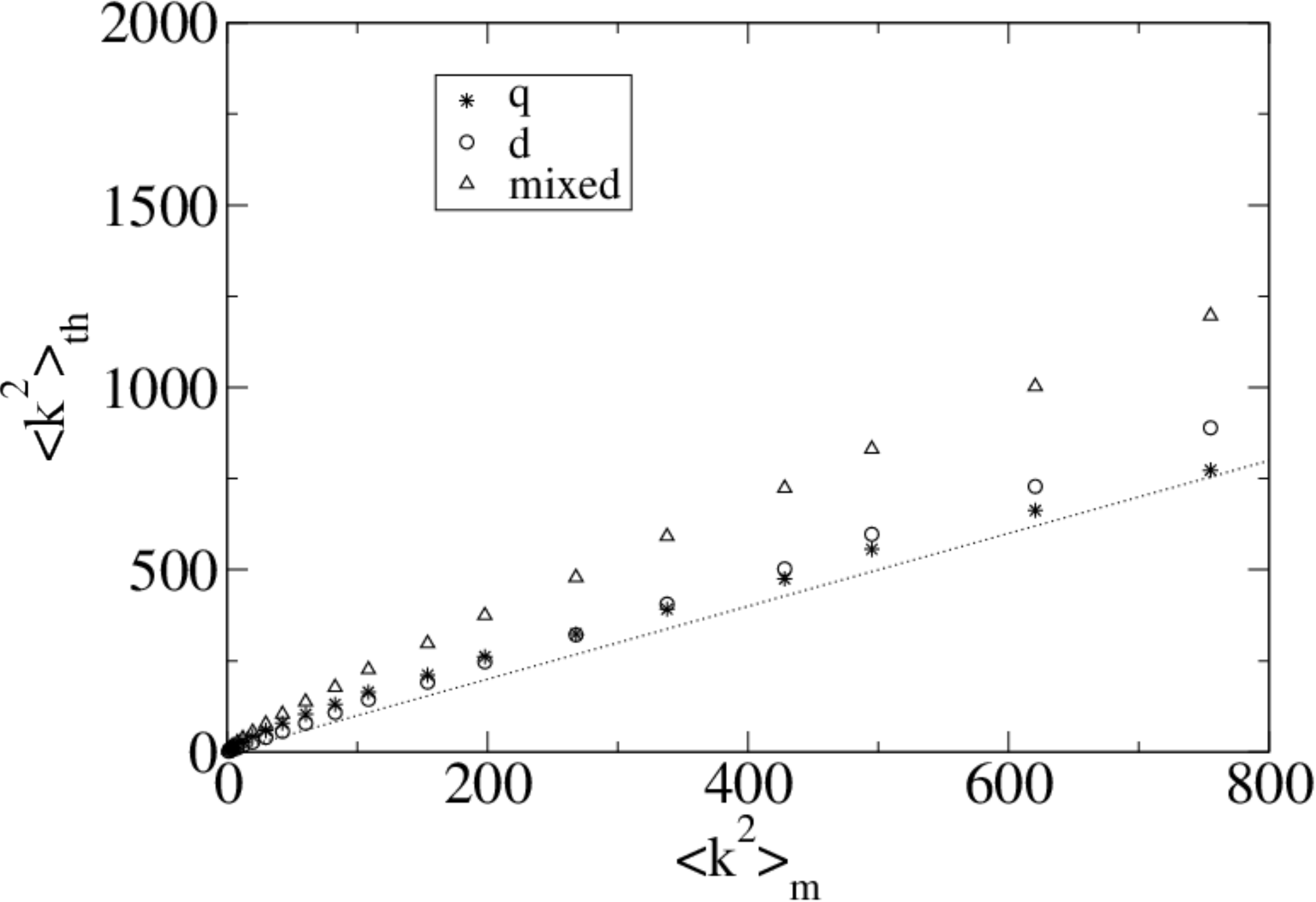}}
\put(308,1.5){\includegraphics[width=170\unitlength,height=111.5\unitlength]{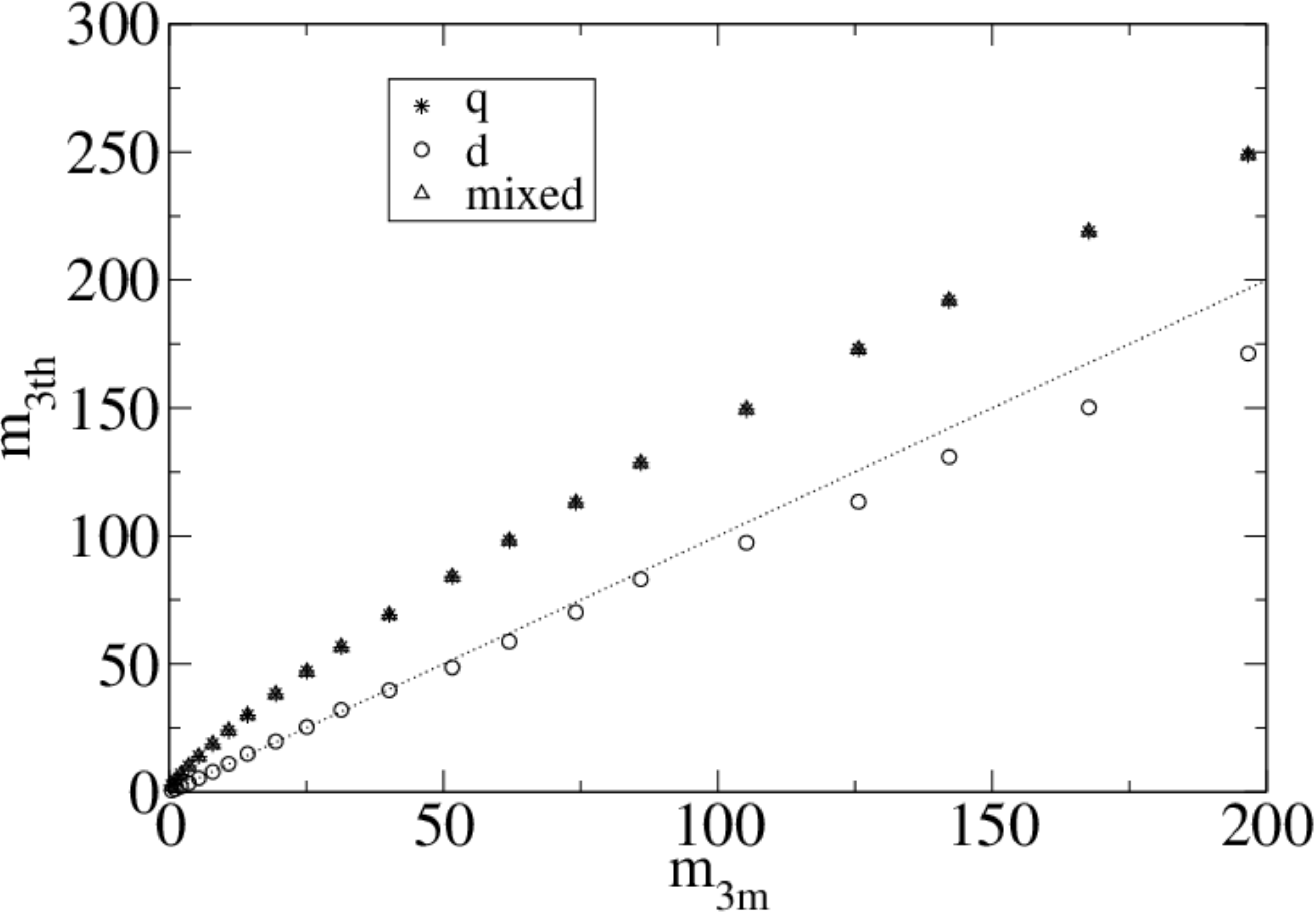}}
\end{picture}
\vspace*{-1mm}

\caption{
Symbols: theoretical $\bra \ldots\ket_{\rm th}$ versus   measured $\bra \ldots\ket_{\rm m}$  values 
of observables  $\kav$, $\kvar$, and $m_3$  in synthetic random graphs 
$\ba$ with $N=3000$ and and $\alpha=0.5$,  generated either via 
random wiring (top panels), $q$-preferential attachment (middle panels) or $d$-preferential attachment (bottom panels).
Dotted lines: the diagonals (shown as guides to the eye).
}

\label{fig:m3}
\end{figure}

Here we compare the ability of our bipartite ensembles (\ref{eq:quenched_q}, \ref{eq:quenched_d}, \ref{eq:quenched_qd}) to predict properties of the associated 
binary PIN graphs, for synthetic networks that are generated from any of these
ensembles. We focus on comparing homologous fomulae for the observables 
$\bra k\ket$, $\bra k^2\ket$, $m_3$ and $m_4$. The  synthetic matrices $\ba=\{a_{ij}\}$ with $a_{ij}\in\{0,1\}$ are defined as before via $a_{ij}=\theta(\sum_{\mu}\xi_i^\mu\xi_j^\mu)$, with 
$\theta(0)=0$,   
and the links of the bipartite graph $\bxi$ are generated from the following three protocols. In the first protocol, links between nodes $(i,\mu)$ 
are drawn randomly and independently, until their total number 
reaches a prescribed limit. In the second protocol, we assign the links 
prefentially 
to complexes with large sizes.  In a third protocol we assign links 
preferentially to proteins with large promiscuities. 

In Figure \ref{fig:m3} we show along the vertical axes the values of 
$\kav$ (left) predicted by the three ensembles, via 
formulae (\ref{eq:kav_q2}), (\ref{eq:av_k_d}) and (\ref{eq:kav_mix}), 
the predicted values of 
$\bra k^2\ket$ (middle), via  
 (\ref{eq:kvar_q}), (\ref{eq:kvar_d}), and (\ref{eq:kvar_mix}), and 
the predicted triangle density $m_3$ (right), 
via (\ref{eq:m3_q}), (\ref{eq:m3_d2}) and (\ref{eq:m3_mix}). All are shown together  with the corresponding values that were measured in $\ba$, along the horizontal axis. 
As expected, the $d$-ensemble outperforms the other 
ensembles when links are drawn according to $d$-preferential attachment, whereas the
$q$-ensemble performs better for graphs generated via $q$-preferential 
attachment. The mixed ensemble performs 
very similar to the $q$-ensemble in terms of 
counting triangles, as expected from the reasoning in Section \ref{sec:mix}. 
Deviations between the $q$ and the mixed ensembles are most evident in 
the second moment of the degree distribution, where the mixed ensemble always leads to values well above those
of the $q$- and the $d$-ensembles. 
We found in Section \ref{sec:d} that the $d$-ensemble is indistinguishable from 
a fully random ensemble when calculating $\kav$ and $m_3$, which explains why the $d$-ensemble predicts the values of these two observables perfectly. 
The other two ensembles are more sensitive to finite size effects, as any heterogeneity in the $q$ will boost the number of loops.

\begin{figure}[t]
\setlength{\unitlength}{0.345mm}
\hspace*{40mm}
\begin{picture}(250,363)

\put(-40,249){\includegraphics[width=165\unitlength,height=111\unitlength]{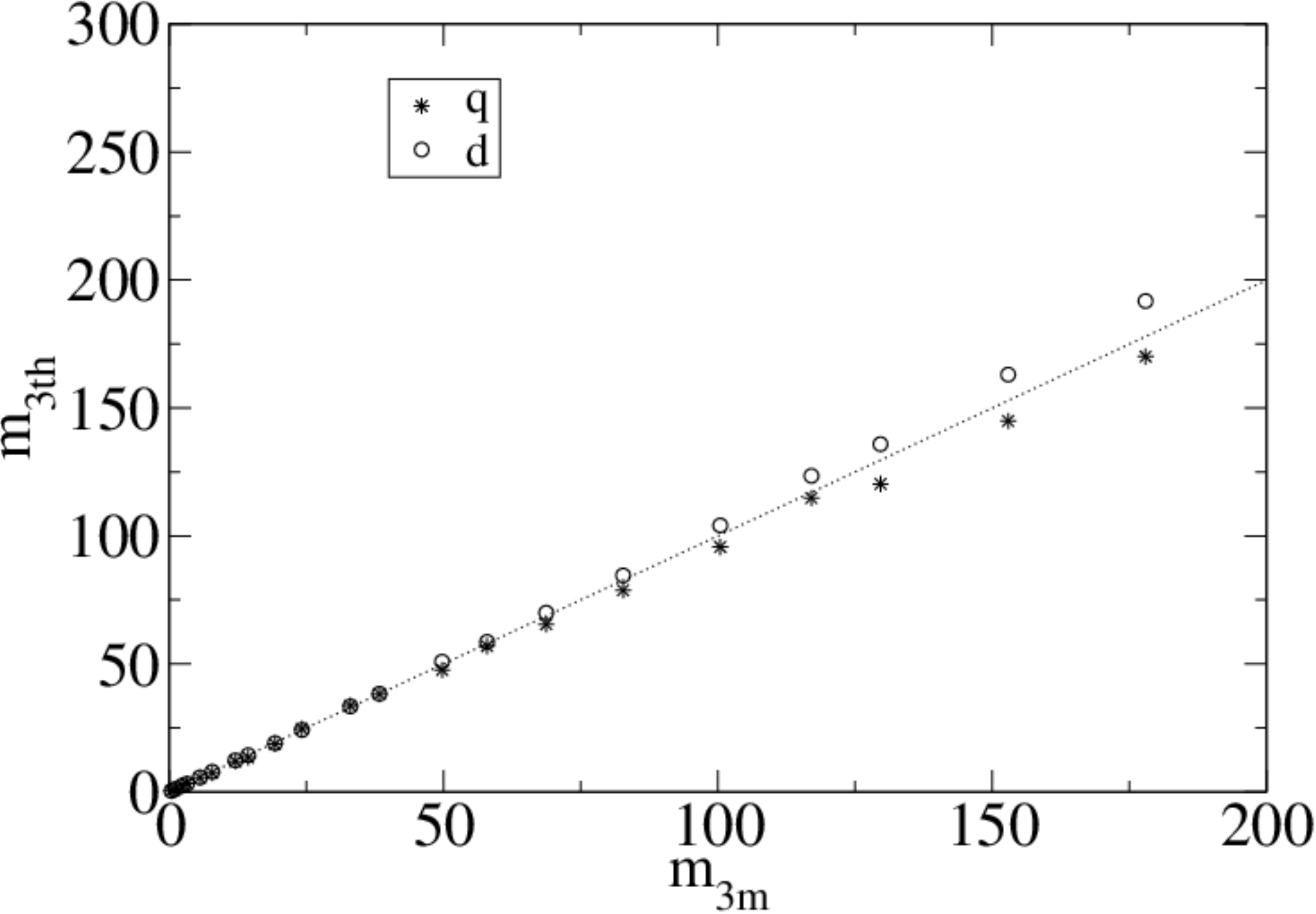}}
\put(126,247){\includegraphics[width=164\unitlength,height=110\unitlength]{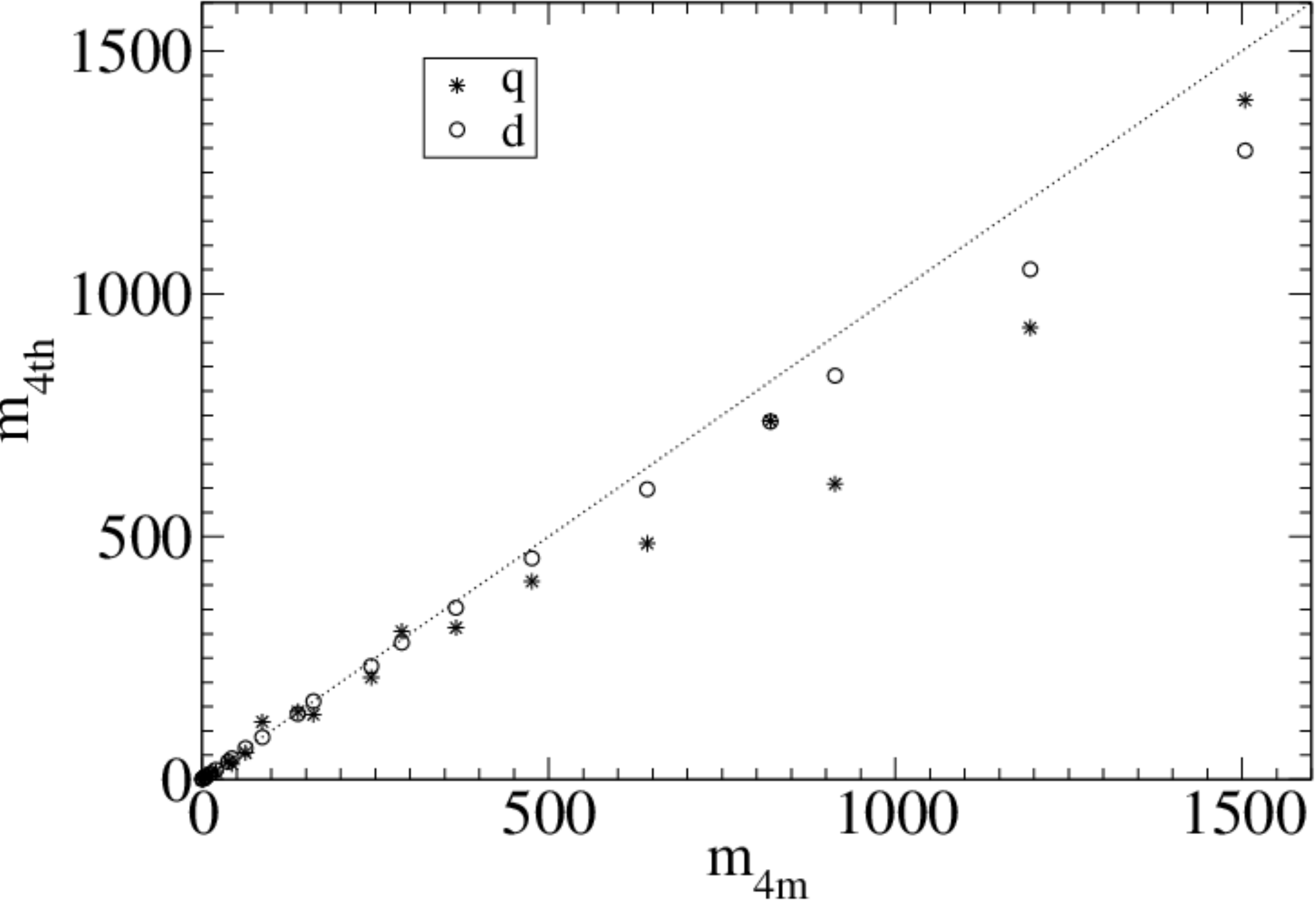}}

\put(-44,121.6){\includegraphics[width=163.9\unitlength,height=113\unitlength]{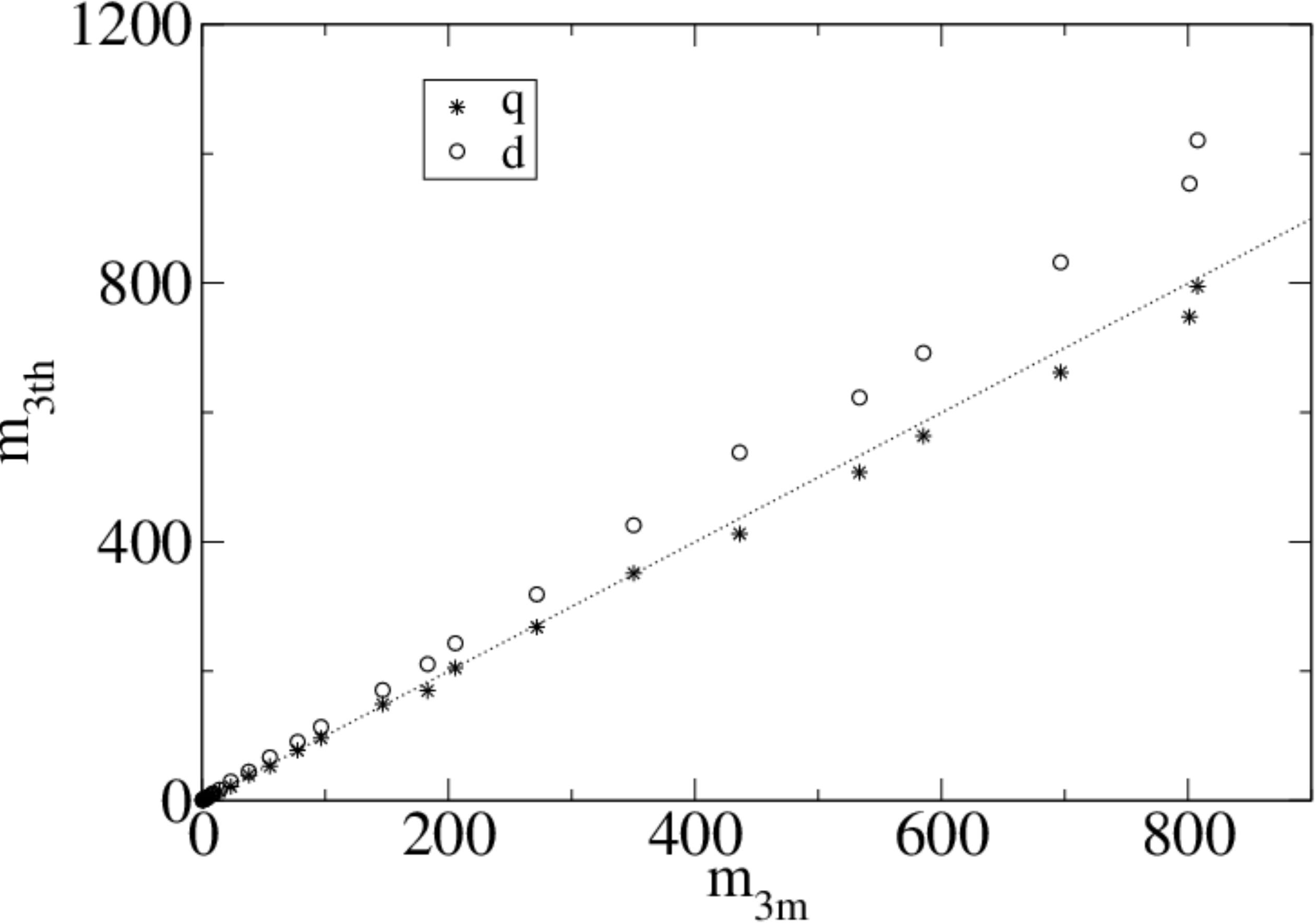}}
\put(126,122){\includegraphics[width=172.5\unitlength,height=111.5\unitlength]{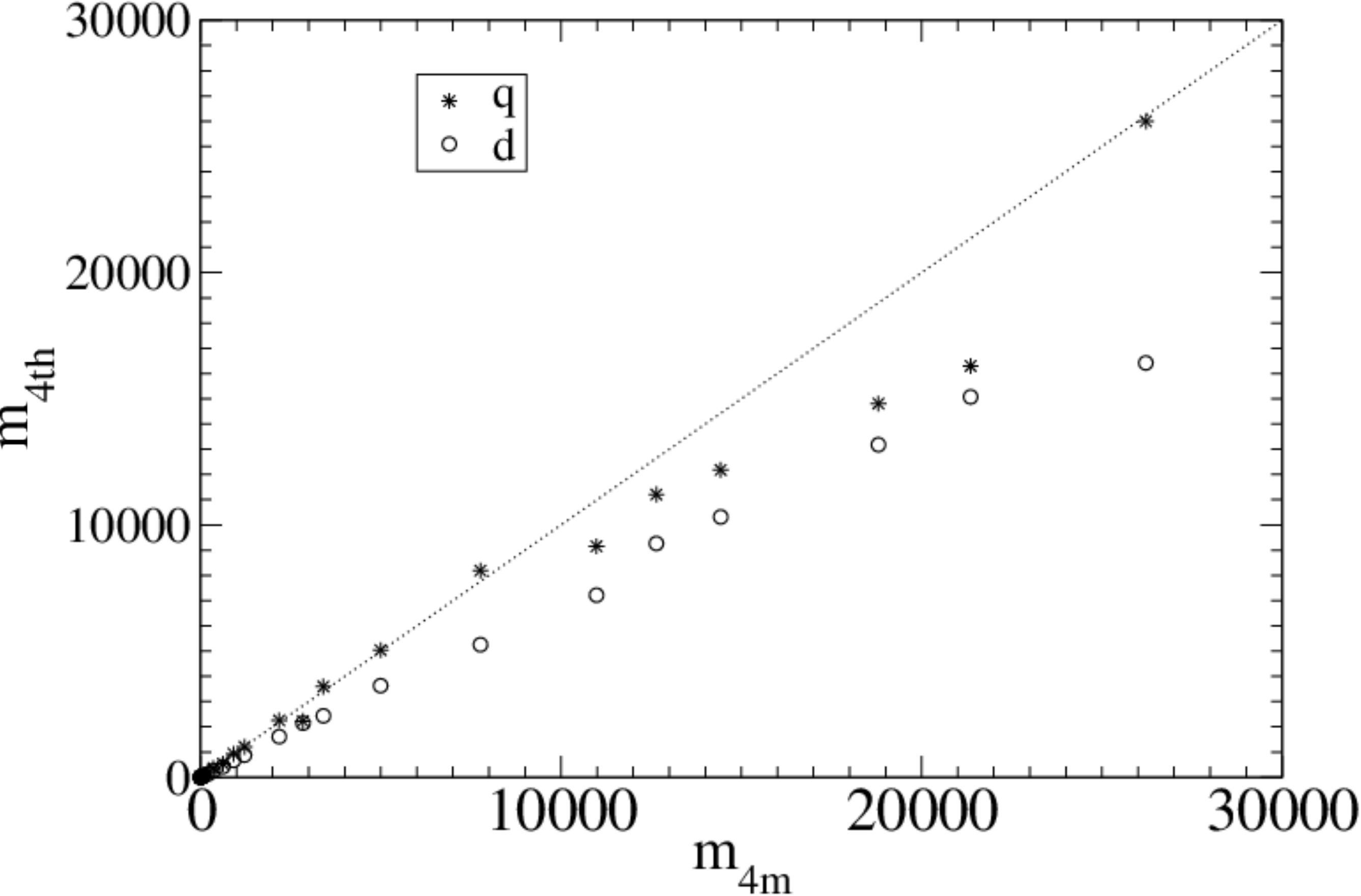}}

\put(-40,0){\includegraphics[width=166\unitlength,height=111\unitlength]{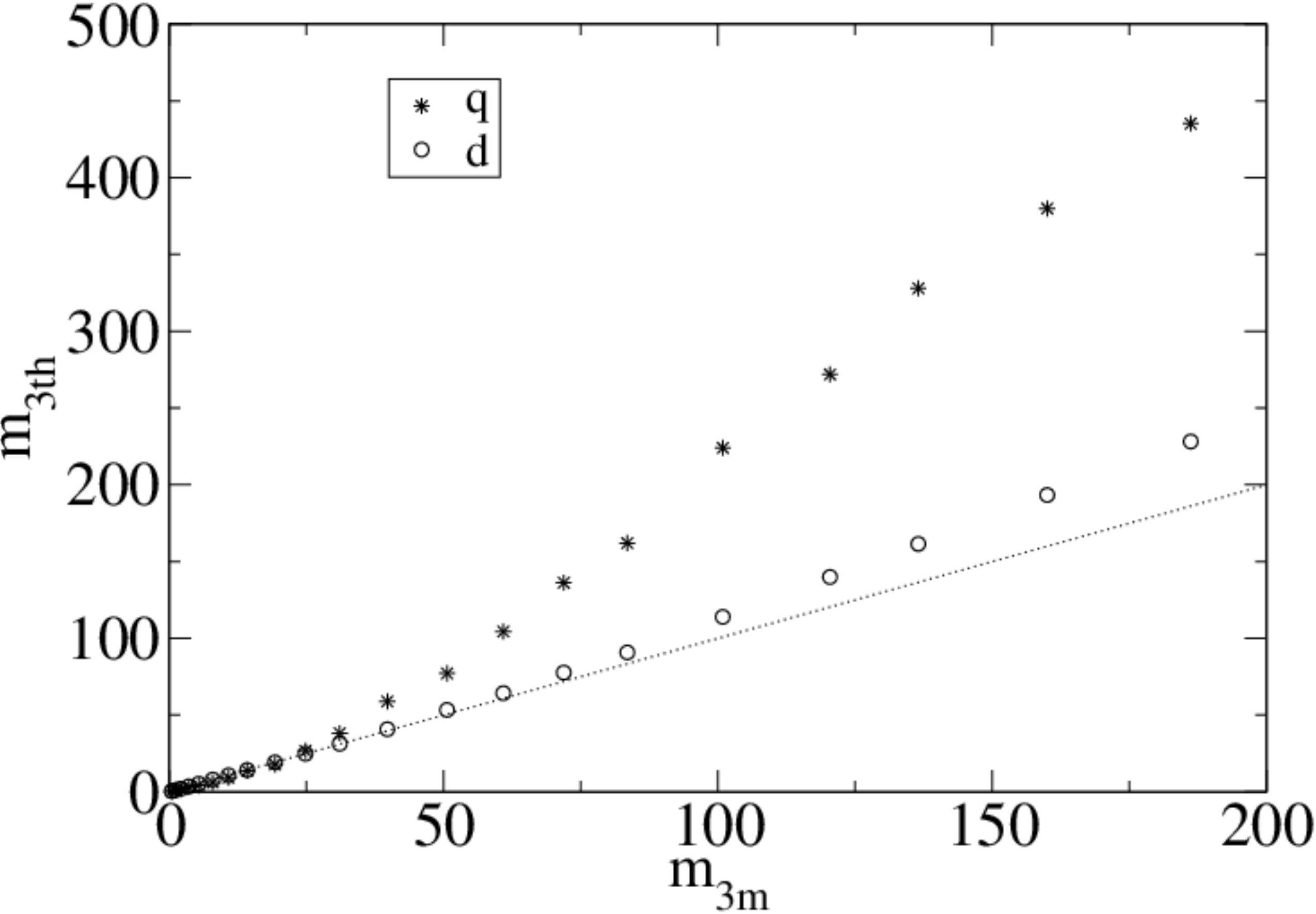}}
\put(125,0){\includegraphics[width=164\unitlength,height=110.7\unitlength]{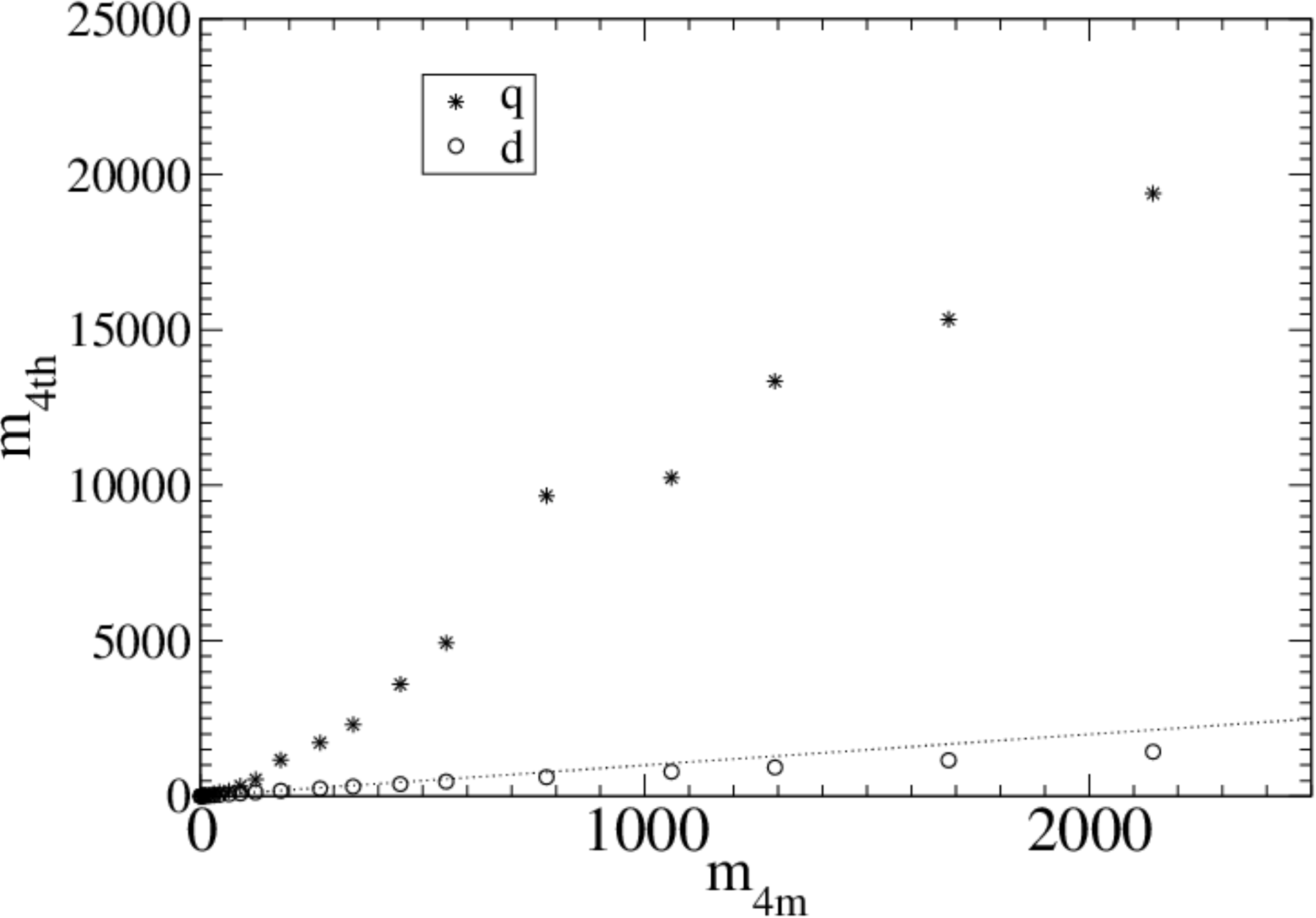}}
\end{picture}
\vspace*{-1mm}

\caption{
Predicted versus real $m_3$ (left) and $m_4$ (right) 
for random bi-partite graphs with $N=3000$ and $\alpha=0.5$ genetated via 
random wiring (top panels), q preferential (middle panels) and d preferential (bottom panel), 
calculated by using formulae (\ref{eq:m3_q}), (\ref{eq:m4_q}), (\ref{eq:m4_d}) and 
obsevables appearing in the formulae computed directly from the network. }

\label{fig:m3m4}
\end{figure}

\begin{figure}[t]
\setlength{\unitlength}{0.315mm}
\hspace*{20mm}
\begin{picture}(300,163)
\put(-80,20){\includegraphics[width=185\unitlength,height=135\unitlength]{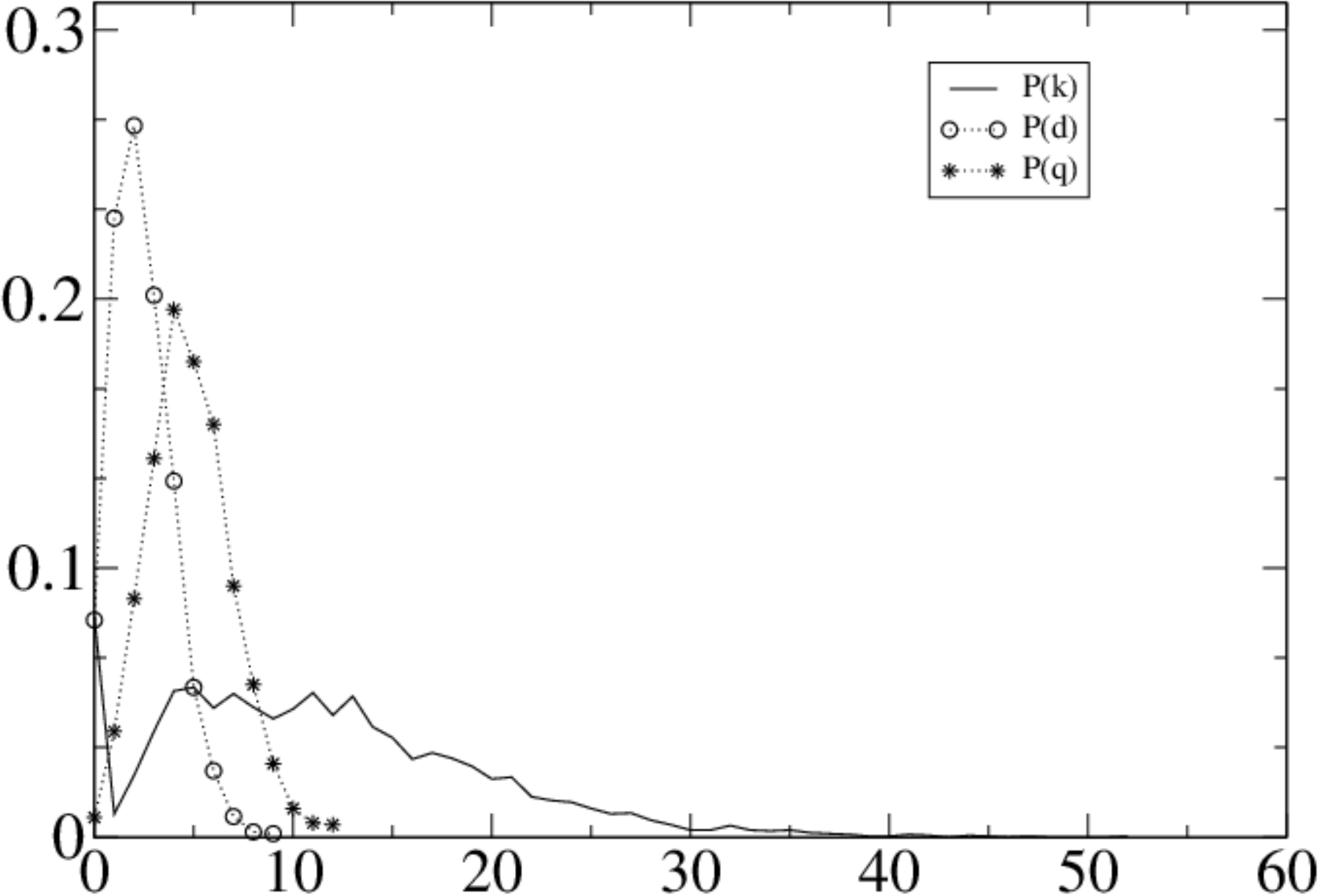}}
\put(0,5){\footnotesize $q,k,d$}
\put(109,20){\includegraphics[width=190\unitlength,height=135\unitlength]{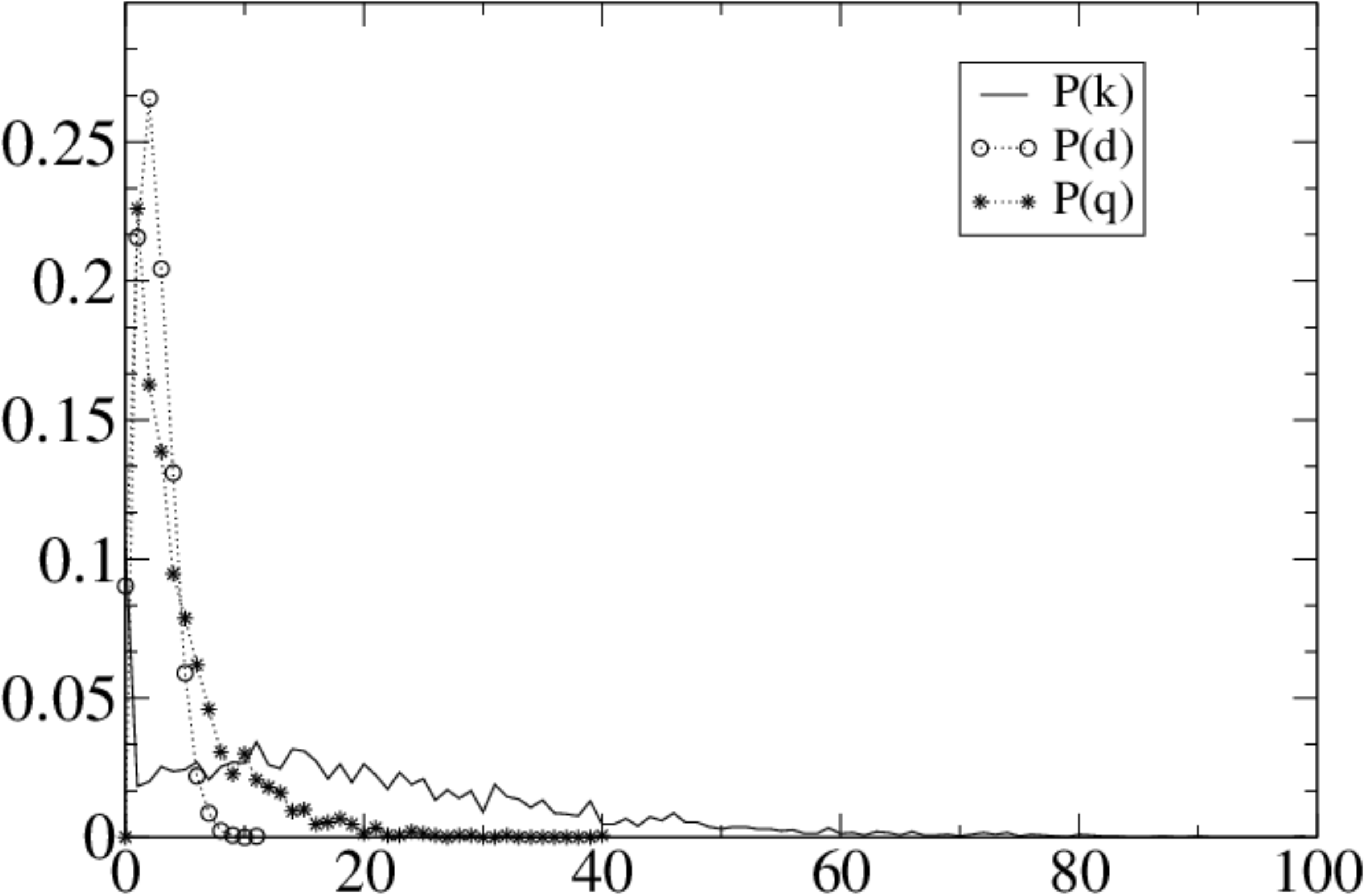}}
\put(190,5){\footnotesize $q,k,d$}
\put(305,20){\includegraphics[width=185\unitlength,height=139\unitlength]{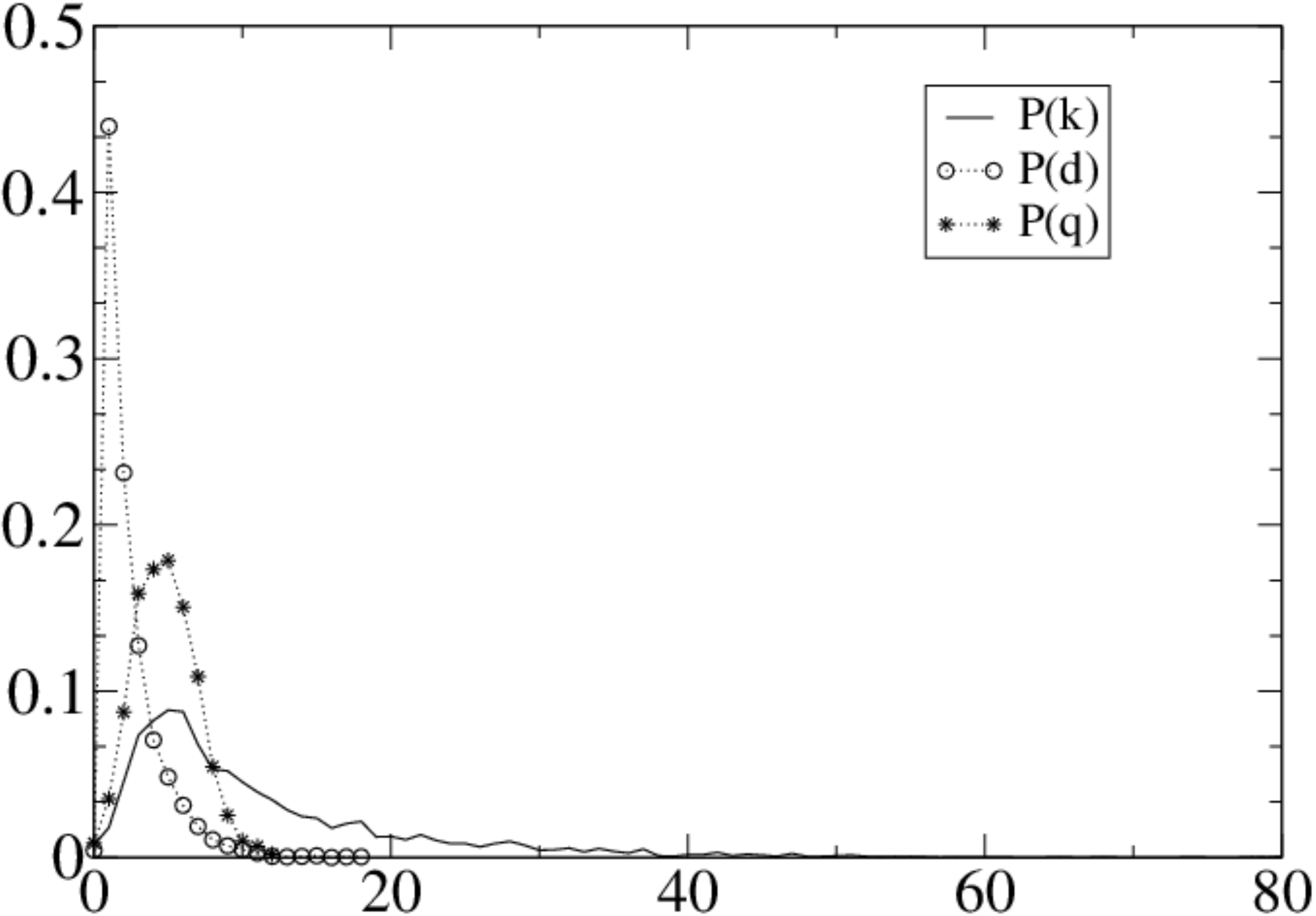}}
\put(380,5){\footnotesize $q,k,d$}

\end{picture}

\caption{
Distributions $P(q)$  of complex sizes, $P(d)$ or protein promiscuities, and $p(k)$  of the degrees in $\ba$  (distinguished by markers whom in the panel legends), 
for random bi-partite graphs with $N=3000$, $\alpha=0.5$ and $\qav=4.8$, which have been generated either via 
random wiring (left), via $q$-preferential attachment (middle), or via $d$-preferential attachment (right). 
}

\label{fig:pk}
\end{figure}

In Figure \ref{fig:m3m4} we show the values of 
$m_3$ and $m_4$ predicted by those formulae that involve only measurable graph observables, 
for the synthetically generated graphs used in Figure \ref{fig:m3}. The prediction of 
$m_3$ is now obtained from (\ref{eq:m3_q}) and (\ref{eq:m4_d}), for the $q$- and $d$- ensembles respectively, 
and $m_4$ is evaluated using (\ref{eq:m4_q}) and (\ref{eq:m4_d}).
In figure \ref{fig:pk} we plot the degree distribution $p(k)$ of graphs with identical values for the  number of nodes ($N=3000$) and the number of links $L=N\alpha \qav$,
generated  synthetically via the three chosen protocols, together with the distributions $P(q)$ of complex sizes and $P(d)$ of protein promiscuities. 
As explained in Section \ref{sec:mix}, tails in the degree distribution $p(k)\sim k^{-\mu}$ can arise either from a complex size distribution  $P(q)\sim q^{-\mu-1}$ and 
 a homogeneous promiscuity distribution, or 
from having an equally fat tail in the promiscuity distribution $P(d)\sim d^{-\mu}$ together with less heterogeneous complex sizes $P(q)\sim q^{-\alpha-1}$ with $\alpha>\mu$.

\section{Test against experimental protein interaction data}

\label{sec:real}
In this section we apply the results of  our analyses to real publicly available protein interaction datasets, 
 obtained via MS (mass spectrometry) and Y2H (yeast 2-hybrid) experiments. The detailed quantitative features of the various data sets and their 
references are listed in Table~\ref{tab:properties}.

\begin{table}[t]
\centering
\label{tab:properties}
\begin{tabular}{lccccr}
\\
Species & $N$ & $\bra k\ket$ & $k_{\rm max}$ & Method & Reference
\\
\hline
{\em C.elegans} & 2528 & 2.96 & 99 & Y2H & \cite{Simonis08}\\
{\em C.jejuni}    & 1324  & 17.5 & 207 & Y2H & \cite{Parrish07}\\
{\em E.coli    }    & 2457  &  7.05  & 641 & MS & \cite{Arifuzzaman06}\\
{\em H.pylori  }  & 724   &   3.87   & 55 & Y2H & \cite{Rain01}\\
{\em H.sapiens} I & 1499 & 3.37  & 125 & Y2H & \cite{Rual05}\\
{\em H.sapiens} II & 1655  & 3.71  & 95 & Y2H & \cite{Stelzl05}\\
{\em H.sapiens} III & 2268  & 5.67 & 314 & MS & \cite{Ewing07}\\
{\em M.loti      }     & 1803  & 3.43  & 401 & Y2H & \cite{Shimoda08}\\
{\em P.falciparum} & 1267  & 4.17 & 51 & Y2H & \cite{Lacount05}\\
{\em S.cerevisiae} I & 991   & 1.82  &  24 & Y2hH & \cite{Uetz00}\\
{\em S.cerevisiae} II & 787  & 1.91 & 55 & Y2H & \cite{Itocore01}\\
{\em S.cerevisiae} III & 3241  & 2.69 & 279 & Y2H & \cite{Itocore01}\\
{\em S.cerevisiae} IV & 1576  & 4.58 & 62 & MS & \cite{Ho02}\\
{\em S.cerevisiae} VI & 1358 & 4.73  & 53 & MS & \cite{Gavin02}\\
{\em S.cerevisiae} VIII & 2551 & 16.77 & 955 & MS & \cite{Gavin06}\\
{\em S.cerevisiae} IX &  2708 & 5.25 &  141 & MS & \cite{Krogan06}\\
{\em Synechocystis} & 1903  & 3.25 &  51 & Y2H & \cite{Sato07}\\
{\em T.pallidum} & 724 & 10.01  &  285 & Y2H & \cite{Titz08}\\
\hline
\end{tabular}\vspace*{4mm}

\caption{List of the publicly available experimental protein interaction data sets as used in the present study, together with their main quantitative characteristics (number of proteins $N$, average degree $\bra k\ket$, and largest degree $k_{\rm max}$) and references.}
\end{table}

\subsection{Mass spectrometry datasets}

\begin{figure}[t]
\setlength{\unitlength}{0.315mm}
\hspace*{20mm}

\begin{picture}(200,175)
\put(0,20){\includegraphics[width=200\unitlength,height=150\unitlength]{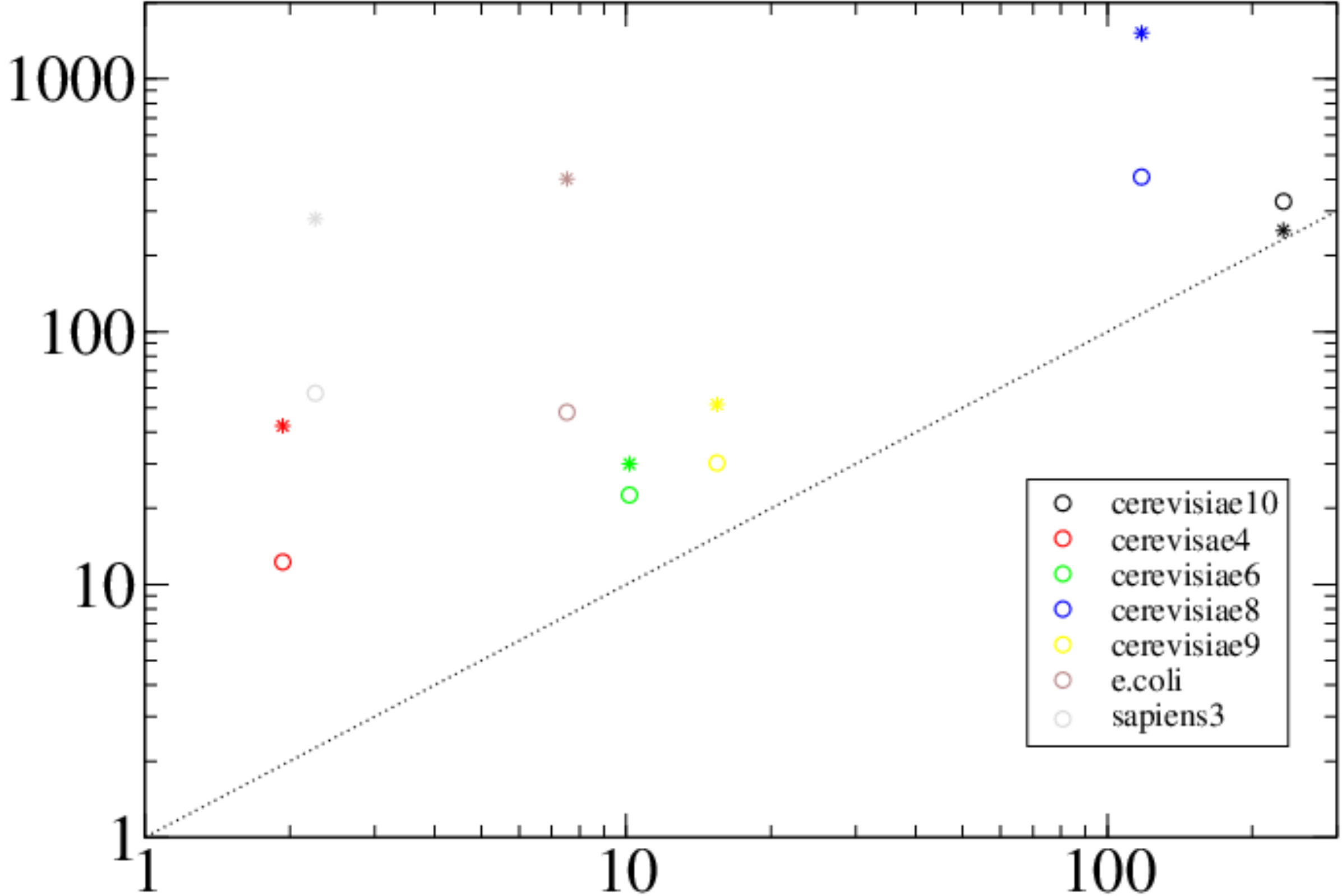}}
\put(100,7){\footnotesize $m_{3{\rm m}}$} \put(-25,120){\footnotesize $m_{3{\rm th}}$}

\put(250,-82){\includegraphics[width=310\unitlength,height=490\unitlength]{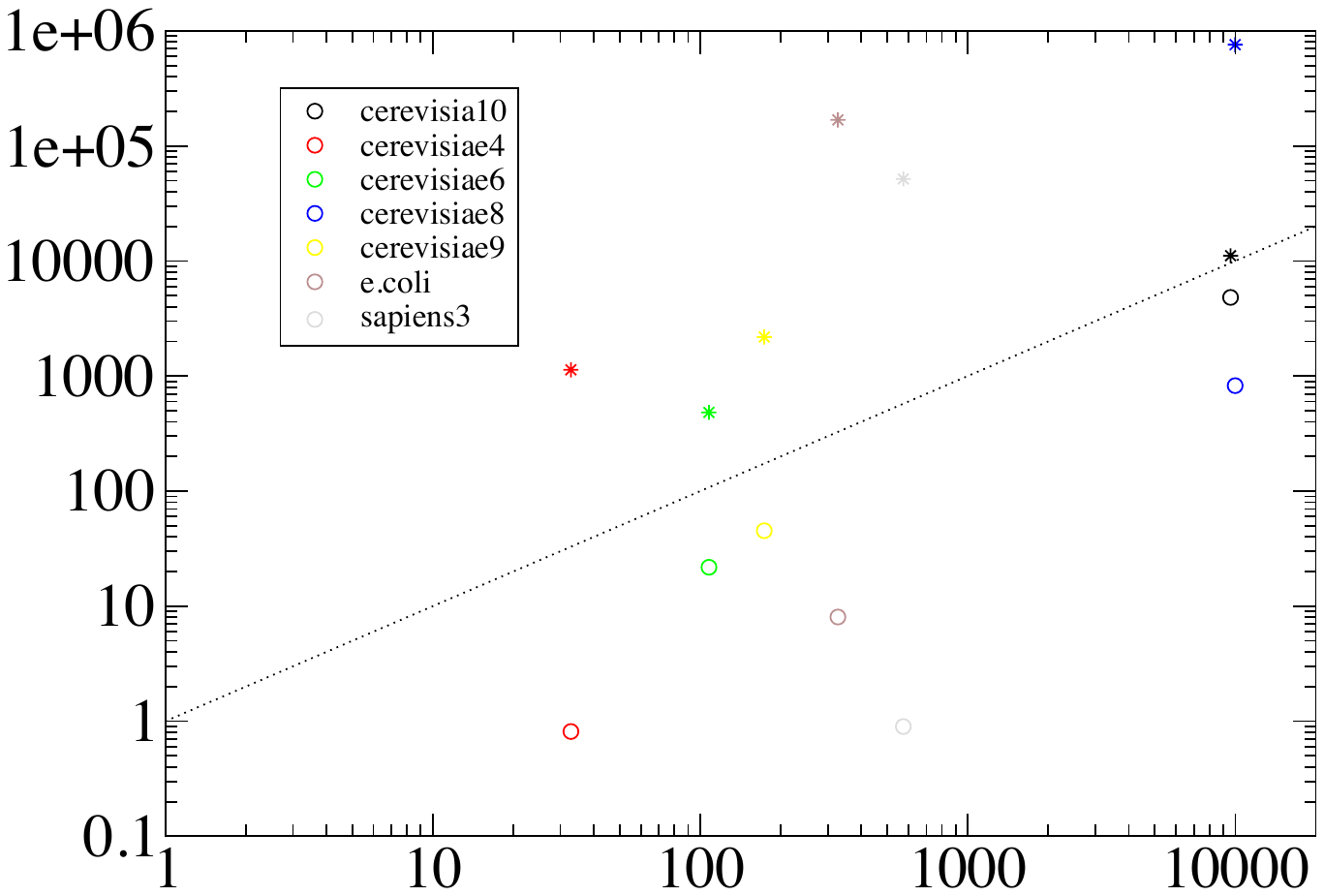}}
\put(370,7){\footnotesize $m_{4{\rm m}}$}\put(230,120){\footnotesize $m_{4{\rm th}}$}
\end{picture}\vspace*{-1mm}

\caption{Left: theoretical predictions $m_{3{\rm th}}$ for the densities  of length-3 loops in the PINs, as obtained from  the $q$-ensemble (stars)
and the $d$-ensemble (circles), plotted versus the values $m_{3{\rm m}}$ measured in the 
different MS datasets. Right: theoretical predictions $m_{4{\rm th}}$ for the densities  of length-4 loops in the same PINs, obtained from  the $q$-ensemble (stars)
and the $d$-ensemble (circles), plotted versus the measured values $m_{4{\rm m}}$. The diagonals are shown as guides to the eye. }
\label{fig:m3m4MS}

\end{figure}

Seven of the experimental PIN datasets in Table \ref{tab:properties} were obtained by MS experiments, and they involved three distinct biological species, namely 
{\it S. cerevisiae}, 
{\it H.sapiens} and {\it E.coli}. Each set takes the form of an 
 $N\times N$ matrix 
of binary entries $a_{ij}$, but with different values of $N$.

In Figure \ref{fig:m3m4MS}
we show the results of our analytical predictions for the densities of length-3 and length-4 loops, as given by the formulae for the bipartite $q$- and $d$-ensembles,
versus their measured values in the MS datasets. 
The $q$-ensemble leads to values of the number of short loops consistently higher than those predicted by the $d$-ensemble. This could have been expected, since  
the $q$-ensemble induces large cliques in the protein interaction networks $\bc$ and $\ba$, which boosts 
short loops.  
In contrast, the $d$-ensemble induces a homogeneous distribution for the complex sizes, and thereby suppresses the presence of large cliques in the protein interaction networks.

Remarkably, the values for lenght-$4$ loop densities of all the MS data sets are in between those 
of the $d$-ensemble (which thereby acts as a lower bound) and those of 
the $q$-ensemble (which acts as an upper bound). 
This suggests a compatibility of data from MS experiments with the 
expected separable form of the proteome network. However, the measured length-3 densities are consistently lower than the values compatible with a separable 
structure of the proteome.

\subsection{Yeast 2-hybrid datasets}

\begin{figure}[t]
\setlength{\unitlength}{0.325mm}
\hspace*{20mm}

\begin{picture}(200,185)
\put(0,20){\includegraphics[width=200\unitlength,height=170\unitlength]{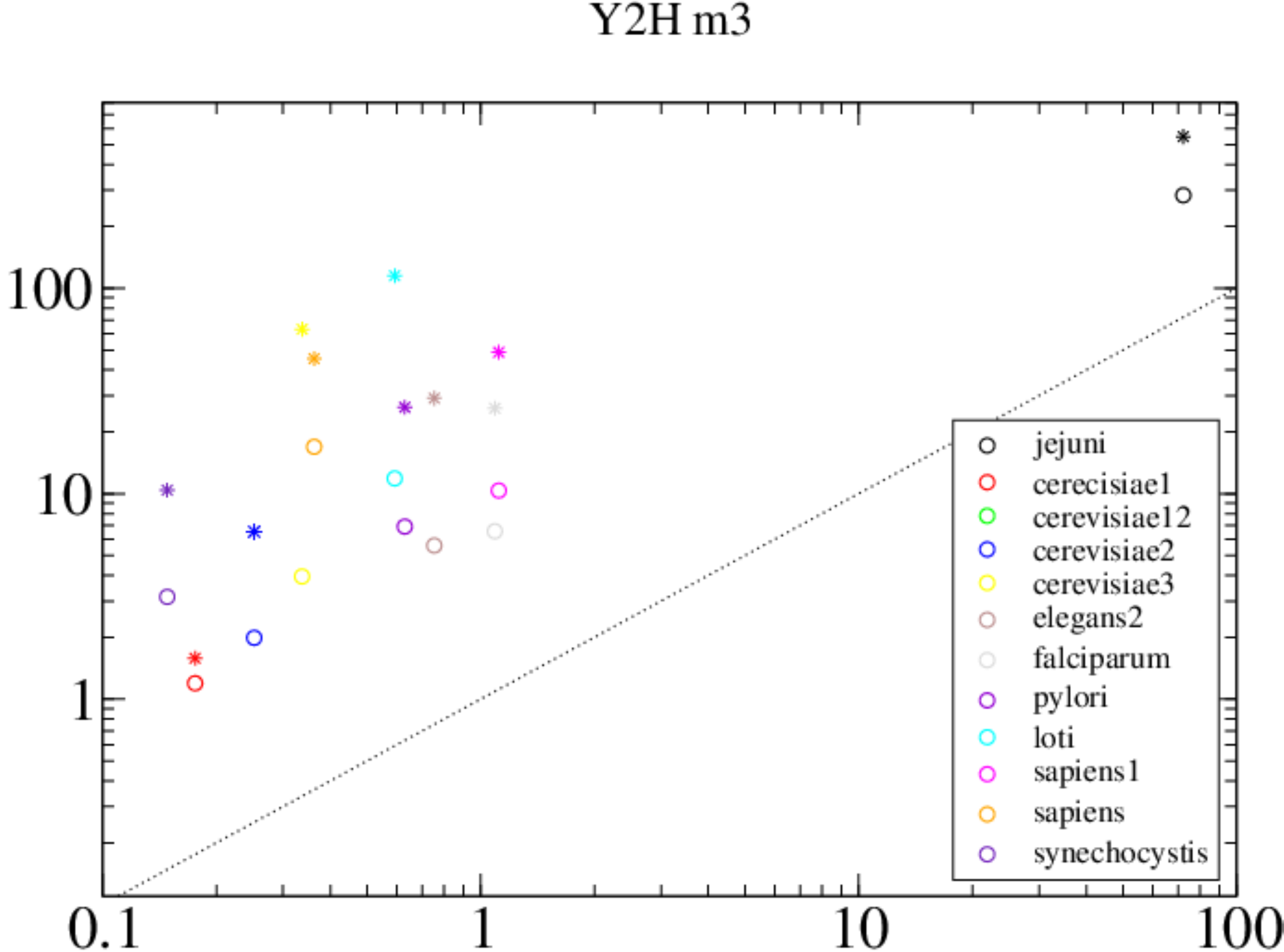}}
\put(100,7){\footnotesize $m_{3{\rm m}}$} \put(-25,120){\footnotesize $m_{3{\rm th}}$}

\put(250,22){\includegraphics[width=190\unitlength,height=169\unitlength]{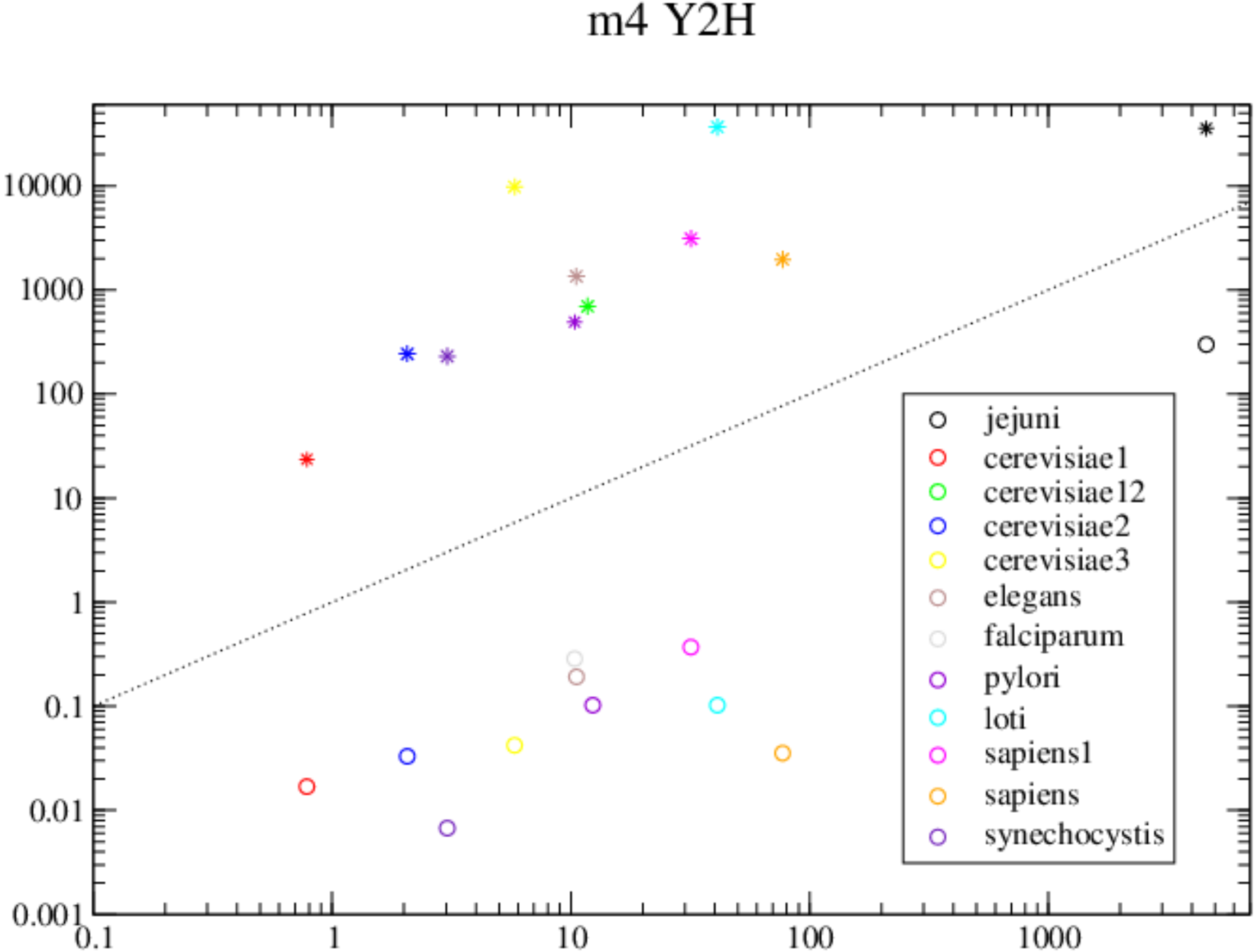}}
\put(345,7){\footnotesize $m_{4{\rm m}}$}\put(220,120){\footnotesize $m_{4{\rm th}}$}
\end{picture}\vspace*{-1mm}

\caption{
Left: theoretical predictions $m_{3{\rm th}}$ for the densities  of length-3 loops in the PINs, as obtained from  the $q$-ensemble (stars)
and the $d$-ensemble (circles), plotted versus the values $m_{3{\rm m}}$ measured in the 
different Y2H datasets. Right: theoretical predictions $m_{4{\rm th}}$ for the densities  of length-4 loops in the same PINs, obtained from  the $q$-ensemble (stars)
and the $d$-ensemble (circles), plotted versus the measured values $m_{4{\rm m}}$. The diagonals are shown as guides to the eye. }	

\label{fig:m3m4Y2H}
\end{figure}

We tested similarly the compatibility of Y2H data with a 
separable structure of the proteome, by checking whether the measured values 
for the network observables $m_3$ and $m_4$ fall within what appeared to be (in MS data) theoretical 
bounds set by the $q$- and $d$-ensembles. We now used the $12$ 
PIN datasets in Table \ref{tab:properties} that were obtained from 
Y2H experiments. 
Results are shown in Figure \ref{fig:m3m4Y2H}. We observe that Y2H datasets exhibit  generally fewer short loops 
than MS dataset. This may be due to the fact that Y2H 
experiments mostly detect direct binding domain contacts in protein interactions, leading to an undersampling 
of links (and thereby to an underestimation of connectivity and loops).
However, Y2H data sets still show the same level of compatibility with a separable 
structure of the proteome as the MS datasets did, with measured values of $m_4$ that are fully compatible, and values for $m_3$ that fall below those predicted by the   $d$-ensemble. 
This is quite remarkable, since MS and Y2H experiments are known to measure interactions in very 
different ways. 

%

\section{Conclusions}
\label{sec:conclusion}

In this paper we propose a bipartite network representation of protein interactions, where the two node types represent proteins and complexes, respectively.
A protein-protein interaction network can then be regarded as the result of 
a `marginalization'  of the bipartite network, whereby the complexes are integrated out (i.e. summed over). 
This leads to a weighted protein interaction network $\bc$ with a separable structure. Adjacency matrices of protein interaction networks $\ba$ are then simply the binary versions of the separable $\bc$, obtained by the entry  truncations $a_{ij}=\theta(c_{ij})$, with the convention $\theta(0)=0$.
One of the central results of this work is that for sufficiently large networks there is an equivalence between the two graph ensembles $p(\bc)$ and 
$p(\ba)$,  inasmuch as macroscopic statistical properties are concerned, such as densities of 
short loops and degree distributions. 
This allows us to regard the conventional protein interaction adjacency matrices
as if they were to have a separable structure, and induces 
precise relations between expectation  values of macroscopic graph observables
which, remarkably, only depend 
on measurable quantities and on the underlying mechanism with which proteins and complexes recruit each other. They 
 are independent of inaccessible microscopic details of proteins and their complexes.

We considered the two extreme complex recruitment scenarios, one  where 
recruitment is either driven solely by protein promiscuities, and one where  
it is driven by complex sizes.
Preferential attachment to large complexes (the $q$-ensemble) favours the presence of 
large cliques in PINs, which 
boosts the number of short loops. Hence we can reasonably expect that the 
predictions on short loop densities from the $q$-ensemble will over-estimate
the real number of loops. Conversely, preferential attachment based only on 
protein promiscuities (the $d$-ensemble) leads to homogeneous complex sizes, which suppresses 
large cliques in PINs, leading to an underestimation 
of short loop densities.
Remarkably, real protein interaction data from mass-spectronomy and yeast 2-hybrid experiments show a 
density of length-$4$ loops 
in between the predictions of the $d$-ensemble and 
those of  the $q$-ensemble, suggesting a degree of compatibility of these experimental data with a 
separable structure of the proteome. 
In contrast, both MS and Y2H dataset show densities or length-$3$ loops that are consistently smaller than all our theoretical predictions. 

We believe that, by providing a systematic and practical framework for understanding protein interaction experiments,  our approach may
represent a valuable step towards  establishing a more solid connection between protein interaction datasets 
and the underlying biology,  
Universal bounds on observables in PINs 
may become powerful tools for data quality testing.
Improved versions of the present models, with fit the experimental data better, may
open  a route  to infer quantities such as the ratio $\alpha$, and the distributions of protein promiscuities and complex sizes.  
Such quantities are not available in the current PIN data sets, and are difficult to access experimentally. The present work has revealed 
that the asymptotic forms of these distributions can be extracted  from the tails of the PIN degree distributions. 
Finally, our method my shed some light on the way protein and complexes recruit one another, in particular, whether this recruitment is driven by proteins or by complexes, and may enable us  
to discriminate between `party hub'  and `date hub' interactions. 

\section{Aknowledgements}

AA aknowledges Alessandro Pandini and Sun Chung for providing protein interaction datasets. 
Kate Roberts is aknowledged for interesting discussions during the early stages of this work.
ACCC is grateful for support from the UK's Biotechnology and Biological Sciences Research Council (BBSRC) .

\appendix

\section{Link probabilities in the weighted protein interaction network}
\label{app:pij}

In this appendix we derive the likelihood to have a link in the weighted protein interaction network $c_{ij}=\sum_\mu \xi_i^\mu \xi_j^\mu$, when the $\xi_i^\mu$ 
are drawn from the ensembles (\ref{eq:quenched_q},\ref{eq:quenched_d},\ref{eq:quenched_qd}).

\subsection{The $q$ ensemble}

In the $q$-ensemble we have
\begin{eqnarray}
\hspace*{-15mm}
p(c_{ij})&=& \Big\bra \delta_{c_{ij},\sum_{\mu\leq \alpha N}\xi_i^\mu\xi_j^\mu}\Big\ket_{\bxi}
= \int_{-\pi}^\pi\!\frac{\rmd\omega}{2\pi}\rme^{\rmi\omega c_{ij}}\prod_{\mu=1}^{\alpha N}\Big\bra \rme^{-\rmi\omega\xi^\mu_i\xi^\mu_j}\Big\ket_{\bxi}
\nonumber
\\
\hspace*{-15mm}
&=& \int_{-\pi}^\pi\!\frac{\rmd\omega}{2\pi}\rme^{\rmi\omega c_{ij}}\prod_{\mu=1}^{\alpha N}\Big\{\frac{q_\mu^2}{N^2} \rme^{-\rmi\omega}+(1-\frac{q_\mu^2}{N^2} )\Big\}
=\int_{-\pi}^\pi\!\frac{\rmd\omega}{2\pi}\rme^{\rmi\omega c_{ij}+\sum_{\mu=1}^{\alpha N}\frac{q_\mu^2}{N^2} [\rme^{-\rmi\omega}-1]-\frac{1}{2}\sum_{\mu=1}^{\alpha N}\frac{q_\mu^4}{N^4} [\rme^{-\rmi\omega}-1]^2}
\nonumber\\
\hspace*{-15mm}
&=&
\int_{-\pi}^\pi\!\frac{\rmd\omega}{2\pi}\rme^{\rmi\omega c_{ij}}\left[
1+\frac{\alpha \bra q^2\ket}{N}(\rme^{-\rmi\omega}-1)
-\frac{1}{2}\frac{\alpha \bra q^4\ket}{N^3}(\rme^{-\rmi\omega}-1)^2
+\frac{\alpha^2 \bra q^2\ket^2}{2N^2}(\rme^{-\rmi\omega}-1)^2
\right.
\nonumber\\
\hspace*{-15mm}
&&\hspace*{60mm}+\left.\frac{\alpha^3 \bra q^2\ket^3}{6N^3}(\rme^{-\rmi\omega}-1)^3
+\order{(N^{-4})}
\right]
\nonumber\\
\hspace*{-15mm}
&=&
\delta_{\cij,0}+\frac{\alpha \bra q^2\ket}{N}(\delta_{\cij,1}-\delta_{\cij,0})
+\left(\frac{\alpha^2 \bra q^2\ket^2}{2N^2}-\frac{1}{2}\frac{\alpha \bra q^4\ket}{N^3}\right)(\delta_{\cij,2}-2\delta_{\cij,1}+\delta_{\cij,0})
\nonumber\\
\hspace*{-15mm}
&&\hspace*{20mm}+\frac{\alpha^3 \bra q^2\ket^3}{6N^3}(\delta_{\cij,3}-3\delta_{\cij,2}+3\delta_{\cij,1}-\delta_{\cij,0})
+\order{(N^{-4})}
\end{eqnarray}
From this one reads off directly the values of $p(c_{ij}=0)$, $p(c_{ij}=1)$ and $p(c_{ij}\geq 2)$. 
The density of triangles is obtained writing (\ref{eq:m3_def}) as 
\begin{eqnarray}
m_3&=&(N\!-\!1)(N\!-\!2)\sum_{\mu\nu\rho=1}^{\alpha N} \bra\xi^\mu\xi^\nu\ket\bra\xi^\nu\xi^\rho\ket\bra
\xi^\rho\xi^\mu\ket
\label{eq:m3_def}
\end{eqnarray}
and using 
\begin{eqnarray}
\bra\xi^\mu\xi^\nu\ket&=& \bra\xi^\mu\ket\bra \xi^\nu\ket+\delta_{\mu\nu}\bra\xi^\mu\ket(1- \bra\xi^\mu\ket)= \frac{q_\mu q_\nu}{N^2} +\delta_{\mu\nu}\frac{q_\mu}{N}\big(1-\frac{q_\mu}{N}\big)
\label{eq:wick}
\end{eqnarray}
This gives
\begin{eqnarray}
m_3&=& \frac{1}{N}\Big[1\!+\!\order(\frac{1}{N})\Big]\sum_{\mu\nu\rho=1}^{\alpha N}q_\mu q_\nu q_\rho\Big[\frac{q_\nu}{N} \!+\!\delta_{\mu\nu}(1\!-\!\frac{q_\nu}{N})\Big]\Big[\frac{q_\rho}{N} \!+\!\delta_{\nu\rho}(1\!-\!\frac{q_\rho}{N})\Big]\Big[\frac{q_\mu}{N} \!+\!\delta_{\rho\mu}(1\!-\!\frac{q_\mu}{N})\Big]
\nonumber
\\
&=&\frac{1}{N}(1\!+\!\order(\frac{1}{N}))\sum_{\mu\nu\rho=1}^{\alpha N}q_\mu q_\nu q_\rho\Big\{
\frac{q_\mu q_\nu q_\rho}{N^3}+3\delta_{\mu\nu}\frac{q_\rho q_\mu}{N^2}(1\!-\!\frac{q_\mu}{N})+3\delta_{\mu\nu}\delta_{\nu\rho}
\frac{q_\mu}{N}(1\!-\!\frac{q_\mu}{N})^2
\nonumber
\\
&&\hspace*{95mm}
+\delta_{\mu\nu}\delta_{\nu\rho}\delta_{\rho\mu}(1\!-\!\frac{q_\mu}{N})^3\Big\}
\nonumber
\\[-1mm]
&=&\frac{1}{N}\sum_{\mu=1}^{\alpha N}(1\!-\!\frac{q_\mu}{N})^3 q_\mu^3+\order(\frac{1}{N})
=\alpha \bra q^3\ket  +\order(N^{-1})
\end{eqnarray}

\subsection{The $d$-ensemble}

In the $d$-ensemble we obtain
\bea
\hspace*{-15mm}
p(c_{ij})&=&\bra \delta_{c_{ij},\sum_\mu \xi_i^\mu\xi_j^\mu}
\ket=\int_{-\pi}^\pi \!\frac{\rmd\omega}{2\pi}\rme^{\rmi\omega c_{ij}+\frac{d_id_j}{\alpha N}(\rme^{-\rmi\omega}-1)-\half \frac{d_i^2 d_j^2}{\left(\alpha N\right)^3}(\rme^{-\rmi\omega}-1)^2}
\nonumber\\
\hspace*{-15mm}
&=&
\int_{-\pi}^\pi \frac{\rmd\omega}{2\pi}\rme^{\rmi\omega c_{ij}}
\Big[1+\frac{d_i d_j}{\alpha N}(\rme^{-\rmi\omega}-1)
+\half \left(\frac{d_i d_j}{\alpha N}\right)^2(\rme^{-\rmi\omega}-1)^2
-\half \frac{(d_i d_j)^2}{(\alpha N)^3}(\rme^{-\rmi\omega}-1)^2
\nonumber\\
\hspace*{-15mm}
&&
\hspace*{40mm}
+\frac{1}{6}\left(\frac{d_i d_j}{\alpha N}\right)^3(\rme^{-\rmi\omega}-1)^3
+\ldots
\Big]
\eea
which gives
\bea
p(c_{ij}=0)&=&1-\frac{d_i d_j}{\alpha N}+\half \left(\frac{d_i d_j}{\alpha N}\right)^2-\frac{1}{6}\left(\frac{d_i d_j}{\alpha N}\right)^3-\half \frac{d_i^2 d_j^2}{\left(\alpha N\right)^3}
\nonumber\\
p(c_{ij}=1)&=&\frac{d_i d_j}{\alpha N}-\left(\frac{d_i d_j}{\alpha N}\right)^2+\half \left(\frac{d_i d_j}{\alpha N}\right)^3+\frac{d_i^2 d_j^2}{\left(\alpha N\right)^3}
\nonumber\\
p(\cij\geq 2)&=&\order{(N^{-2})}
\eea

\subsection{The mixed ensemble}

For the mixed ensemble, the link likelihood is found to be 
\bea
p(c_{ij})&=&\bra \delta_{c_{ij},\sum_\mu \xi_i^\mu\xi_j^\mu}
\ket=\int_{-\pi}^\pi \!\frac{\rmd\omega}{2\pi}\rme^{\rmi\omega c_{ij}
+\sum_\mu\frac{d_id_j q_\mu^2}{\alpha^2 \qav^2 N^2}(\rme^{-\rmi\omega}-1)}
=
\int_{-\pi}^\pi\! \frac{\rmd\omega}{2\pi}\rme^{\rmi\omega c_{ij}+\frac{d_id_j \bra q^2\ket}{\alpha \qav^2 N}(\rme^{-\rmi\omega}-1)}
\nonumber\\
&=&
\rme^{-\frac{d_id_j \bra q^2\ket}{\alpha \qav^2 N}}\int_{-\pi}^\pi\! \frac{\rmd\omega}{2\pi}\rme^{\rmi\omega c_{ij}}
\Big[1+\frac{d_i d_j \bra q^2\ket}{\alpha N \qav^2}\rme^{-\rmi\omega}
+\half\Big(\frac{d_i d_j\bra q^2\ket}{\alpha N\qav^2}\Big)^2\rme^{-2\rmi\omega}+\ldots
\Big]
\eea
giving
\bea
p(c_{ij}=0)&=&1-\frac{d_i d_j \bra q^2\ket}{\alpha \qav^2 N}+\half 
\left(\frac{d_i d_j \bra q^2\ket}{\alpha \qav^2 N}\right)^2+\order{(N^{-3})}
\nonumber\\
p(c_{ij}=1)&=&\frac{d_i d_j \bra q^2\ket}{\alpha \qav^2 N}\left(1-
\frac{d_i d_j \bra q^2\ket}{\alpha \qav^2 N}\right)+\order{(N^{-3})}
\nonumber\\
p(\cij\geq 2)&=&\order{(N^{-2})}
\eea

\section{Calculation of the degree distribution $p(k)$}
\label{app:pk}
In this appendix we calculate the degree distribution of the weighted protein interaction network $c_{ij}=\sum_\mu \xi_i^\mu \xi_j^\mu$, in which the entries $\xi_i^\mu$ 
are drawn from the bipartite ensembles (\ref{eq:quenched_q},\ref{eq:quenched_d},\ref{eq:quenched_qd}), respectively.

\subsection{The $q$-ensemble}

In the $q$-ensemble, we can calculate $p(k)$ as follows:
\begin{eqnarray}
p(k)&=& \int_{-\pi}^\pi \!\frac{\rmd\omega}{2\pi}\rme^{\rmi\omega k}\Big\bra \frac{1}{N}\sum_i \rme^{-\rmi\omega\sum_j c_{ij}}\Big\ket_{\bxi}
= \int_{-\pi}^\pi \!\frac{\rmd\omega}{2\pi}\rme^{\rmi\omega k}\Big\bra \rme^{-\rmi\omega\sum_{j>1} c_{1j}}\Big\ket_{\bxi}\nonumber
\\
&=& \int_{-\pi}^\pi \!\frac{\rmd\omega}{2\pi}\rme^{\rmi\omega k}\Big\bra \rme^{-\rmi\omega\sum_\mu \xi_1^\mu\sum_{j>1} \xi_j^\mu}\Big\ket_{\bxi}
=  \int_{-\pi}^\pi \!\frac{\rmd\omega}{2\pi}\rme^{\rmi\omega k}\prod_\mu \Big\bra \rme^{-\rmi\omega \xi_1^\mu\sum_{j>1} \xi_j^\mu}\Big\ket_{\bxi}
\nonumber
\\
&=& \int_{-\pi}^\pi \!\frac{\rmd\omega}{2\pi}\rme^{\rmi\omega k}\prod_\mu \Big\{
1+\frac{q_\mu}{N}\Big[
\Big\bra \rme^{-\rmi\omega \xi^\mu}\Big\ket^{N-1}_{\xi^\mu}-1\Big]\Big\}
\nonumber
\nonumber
\\
&=& \int_{-\pi}^\pi \!\frac{\rmd\omega}{2\pi}\rme^{\rmi\omega k}\prod_\mu \Big\{
1+\frac{q_\mu}{N}\Big[
\Big( 1+\frac{q_\mu}{N}(\rme^{-\rmi\omega}-1)\Big)^{N-1}-1\Big]\Big\}
\nonumber
\\
&=& \int_{-\pi}^\pi \!\frac{\rmd\omega}{2\pi}\rme^{\rmi\omega k+\sum_\mu
\frac{q_\mu}{N}\Big[
\exp[q_\mu(\rme^{-\rmi\omega}-1)]-1\Big]+\order(N^{-1})}
\nonumber
\\
&=& \int_{-\pi}^\pi \!\frac{\rmd\omega}{2\pi}\rme^{\rmi\omega k+\alpha \Big\bra 
q\Big[
\exp[q(\rme^{-\rmi\omega}-1)]-1\Big]\Big\ket} +\order(N^{-1})
\nonumber
\\
&=& \rme^{-\alpha\bra q\ket}\int_{-\pi}^\pi \!\frac{\rmd\omega}{2\pi}\rme^{\rmi\omega k+\alpha \Big\bra 
q\rme^{-q}
\exp[q \rme^{-\rmi\omega}]\Big\ket} +\order(N^{-1})
\nonumber
\\
&=& \rme^{-\alpha\bra q\ket}\sum_{\ell\geq 0}\frac{\alpha^\ell}{\ell !}\int_{-\pi}^\pi \!\frac{\rmd\omega}{2\pi}\rme^{\rmi\omega k}
\Big\bra 
q\rme^{-q}
\exp[q \rme^{-\rmi\omega}]\Big\ket^\ell+\order(N^{-1})
\nonumber
\\
&=& e^{-\alpha\bra q\ket}\sum_{\ell\geq 0}\frac{\alpha^\ell}{\ell !}
\Big\bra \prod_{r\leq \ell}(q_{r}e^{-q_r})
\int_{-\pi}^\pi \!\frac{\rmd\omega}{2\pi}\rme^{\rmi\omega k}
\rme^{\rme^{-\rmi\omega}\sum_{r\leq \ell }q_r }\Big\ket_{q_1\ldots q_\ell}+\order(N^{-1})
\nonumber
\\
&=& \rme^{-\alpha\bra q\ket}\sum_{\ell\geq 0}\frac{\alpha^\ell}{\ell !}
\Big\bra \prod_{r\leq \ell}(q_{r}e^{-q_r})\sum_{s\geq 0}\frac{(\sum_{r\leq \ell }q_r )^s}{s!}
\int_{-\pi}^\pi \!\frac{\rmd\omega}{2\pi}\rme^{\rmi\omega k-i\omega s}\Big\ket_{q_1\ldots q_\ell}+\order(N^{-1})
\nonumber
\\
&=& \rme^{-\alpha\bra q\ket}\sum_{\ell\geq 0}\frac{\alpha^\ell}{\ell !}
\Big\bra \prod_{r\leq \ell}(q_{r}\rme^{-q_r})\frac{(\sum_{r\leq \ell }q_r )^k}{k!}
\Big\ket_{q_1\ldots q_\ell}+\order(N^{-1})
\end{eqnarray}
Hence, for large network sizes $N\to\infty$ we obtain
\begin{eqnarray}
\lim_{N\to\infty}p(k)&=& \rme^{-\alpha\bra q\ket}\sum_{\ell\geq 0}\frac{\alpha^\ell}{\ell !}
\Big\bra \left(\prod_{r\leq \ell}q_{r}\right)\rme^{-\sum_{r\leq \ell }q_r }\frac{(\sum_{r\leq \ell }q_r )^k}{k!}
\Big\ket_{q_1\ldots q_\ell}
\nonumber
\\
&=& \rme^{-\alpha\bra q\ket}\sum_{\ell\geq 0}\frac{\alpha^\ell}{\ell !}
\sum_{q_1\ldots q_\ell\geq 0} p(q_1)\ldots p(q_\ell)q_1\ldots q_\ell
\rme^{-\sum_{r\leq \ell }q_r }\frac{(\sum_{r\leq \ell }q_r )^k}{k!}
\end{eqnarray}
We can rewrite this 
in terms of the distribution $W(q)=qP(q)/\bra q\ket$, which 
denotes the likelihood to draw a link attached to a node of degree $q$ 
in the bi-partite graph,
\begin{eqnarray}
\lim_{N\to\infty}p(k)&=& \rme^{-\alpha\bra q\ket}\sum_{\ell\geq 0}\frac{(\alpha\bra q\ket)^\ell}{\ell !}
\sum_{q_1\ldots q_\ell\geq 0} W(q_1)\ldots W(q_\ell)
\rme^{-\sum_{r\leq \ell }q_r }\frac{(\sum_{r\leq \ell }q_r )^k}{k!}
\end{eqnarray}
and upon defining 
\bea
P(y)&=& \rme^{-\alpha\bra q\ket}\sum_{\ell\geq 0}\frac{(\alpha\bra q\ket)^\ell}{\ell !}
\sum_{q_1\ldots q_\ell\geq 0} W(q_1)\ldots W(q_\ell)~\delta[y-\sum_{r\leq \ell }q_r ]
\eea
we finally get to
\begin{eqnarray}
\lim_{N\to\infty}p(k)
&=& \int_0^\infty\!\rmd y~P(y)~\rme^{-y}y^k/k!
\end{eqnarray}
The interpretation is that if we draw $\ell$ from a Poisson distribution with $\bra \ell\ket=\alpha \bra q\ket$, and then draw $\ell$ variables $q_r$ 
from $W(q_r)$, we 
find $k$ as a Poissonian variable with $\bra k\ket=\sum_{r\leq \ell }q_r$.
Clearly $p(k)$ is normalised, and for its first moment we find:
\begin{eqnarray}
\bra k\ket&=& \int_0^\infty\!\rmd y~P(y)~y
= \rme^{-\alpha\bra q\ket}\sum_{\ell\geq 0}\frac{(\alpha\bra q\ket)^\ell}{\ell !}
\sum_{q_1\ldots q_\ell\geq 0} W(q_1)\ldots W(q_\ell)~\sum_{r\leq \ell }q_r 
\nonumber
\\
&=& \rme^{-\alpha\bra q\ket}\sum_{\ell> 0}\frac{(\alpha\bra q\ket)^\ell}{(\ell-1) !}
\sum_{q} W(q)q= \alpha \bra q^2\ket
\end{eqnarray}
For the second moment we obtain
\begin{eqnarray}
\bra k^2\ket&=& \bra k\ket+\int_0^\infty\!\rmd y~P(y)~y^2
\nonumber
\\
&=& \alpha \bra q^2\ket+
\rme^{-\alpha\bra q\ket}\sum_{\ell\geq 0}\frac{(\alpha\bra q\ket)^\ell}{\ell !}
\sum_{q_1\ldots q_\ell\geq 0} W(q_1)\ldots W(q_\ell)~\sum_{r,s\leq \ell }q_r q_s
\nonumber
\\
&=& \alpha \bra q^2\ket+
\rme^{-\alpha\bra q\ket}\Big(\sum_q W(q)q\Big)^2\sum_{\ell> 0}\frac{(\alpha\bra q\ket)^\ell}{\ell !}\ell^2
\nonumber
\\
&&
+
\rme^{-\alpha\bra q\ket}\Big[\sum_q W(q)q^2-\Big(\sum_q W(q)q\Big)^2\Big]\sum_{\ell>0}\frac{(\alpha\bra q\ket)^\ell}{(\ell\!-\!1) !}
\nonumber
\\
&=& \alpha \bra q^2\ket+
\rme^{-\alpha\bra q\ket}\sum_{\ell\geq 0}\frac{(\alpha\bra q\ket)^\ell}{\ell !}
\sum_{q_1\ldots q_\ell\geq 0} W(q_1)\ldots W(q_\ell)~\sum_{r,s\leq \ell }q_r q_s
\nonumber
\\
&=& 
\alpha \bra q^2\ket+\alpha
\Big[\bra q^3\ket-\frac{\bra q^2\ket^2}{\bra q\ket}\Big]+
\frac{\bra q^2\ket^2}{\bra q\ket^2}e^{-\alpha\bra q\ket}
\sum_{\ell> 0}\frac{(\alpha\bra q\ket)^\ell}{\ell !}\ell^2
\nonumber
\\
&=& 
\alpha \bra q^2\ket+\alpha
\Big[\bra q^3\ket-\frac{\bra q^2\ket^2}{\bra q\ket}\Big]+
\frac{\bra q^2\ket^2}{\bra q\ket^2}\Big[\alpha^2\bra q\ket^2+\alpha\bra q\ket\Big]
\nonumber
\\
&=& 
\alpha \bra q^2\ket+\alpha
\bra q^3\ket+
\alpha^2\bra q^2\ket^2
\end{eqnarray}
This is in agreement with results from a direct calculation:
\begin{eqnarray}
\bra k^2\ket&=& \frac{1}{N}\sum_{i\neq j\neq k} \bra c_{ij} c_{k\ell}\ket =\frac{1}{N}\sum_{i\neq j}  \bra c_{ij} c_{ji}\ket+\frac{1}{N}\sum_{[ijk]}  \bra c_{ij} c_{jk}\ket\nonumber
\\
&=&\frac{1}{N}\sum_{i\neq j}  \sum_{\mu \nu}\bra \xi_i^\mu \xi_j^\mu \xi_i^\nu \xi_j^\nu\ket+\frac{1}{N}\sum_{[ijk]}  \sum_{\mu \nu}\bra \xi_i^\mu \xi_j^\mu \xi_j^\nu \xi_k^\nu\ket\nonumber\nonumber\\
&=& \frac{1}{N}\sum_{i\neq j}  \sum_\mu\bra \xi_i^\mu \xi_j^\mu\ket+\frac{1}{N}\sum_{[ijk]}  \sum_{\mu \neq \nu}\bra \xi_i^\mu \xi_j^\mu\ket \bra \xi_j^\nu \xi_k^\nu\ket+\frac{1}{N}\sum_{[ijk]} \sum_\mu \bra \xi_i^\mu \xi_j^\mu \xi_k^\nu
\ket+\order(N^{-1})\nonumber\\
&=& \frac{1}{N}\sum_{i\neq j}  \sum_\mu \frac{q_\mu^2}{N^2}+\frac{1}{N}\sum_{[ijk]}  \sum_{\mu \neq \nu}\frac{q_\mu^2}{N^2}\frac{q_\nu^2}{N^2}+ \frac{1}{N}\sum_{[ijk]} \sum_\mu \frac{q_\mu^3}{N^3}+\order(N^{-1})\nonumber\\
&=& \alpha \bra q^2\ket+(\alpha \bra q^2\ket)^2+\alpha \bra q^3\ket +\order(N^{-1})=\bra k \ket +\bra k \ket^2 +\alpha \bra q^3 \ket+\order(N^{-1})
\end{eqnarray}

\subsection{The $d$-ensemble}

\renewcommand{\intw}{\int_{-\pi}^\pi\! \frac{\rmd\omega}{2\pi}}
We can calculate the asymptotic degree distribution in the $d$-ensemble as follows
\bea
p(k)&=&\lim_{N\to\infty}\frac{1}{N}\sum_i\bra  \delta_{k,\sum_j c_{ij}}\ket_{\bxi}=\frac{1}{N}\sum_i \intw
\rme^{\rmi\omega k}\bra \rme^{\rm-i\omega \sum_\mu \xi_i^\mu \sum_j\xi_j^\mu}\ket_{\bxi}
\nonumber\\
&=&\lim_{N\to\infty}\frac{1}{N}\sum_i \intw  \rme^{i\omega k}\prod_\mu\Big[
1+\frac{d_i}{\alpha N}\Big(\prod_j \bra \rme^{-\rmi\omega \xi_j^\mu}\ket -1
\Big)
\Big]
\nonumber\\
&=&\lim_{N\to\infty}\frac{1}{N}\sum_i \intw \rme^{\rmi\omega k}\prod_\mu\Big[
1+\frac{d_i}{\alpha N}\Big(
\rme^{\frac{\bra d \ket}{\alpha}(\rme^{-\rmi\omega}-1)}-1
\Big)
\Big]
\nonumber\\
&=&\lim_{N\to\infty}\frac{1}{N}\sum_i \intw \rme^{\rmi\omega k+d_i \Big(\rme^{\frac{\dav}{\alpha}(\rme^{-\rmi\omega}-1)}-1\Big)}
\nonumber\\
&=&
\sum_d P(d)\intw e^{i\omega k+d\left(\rme^{\frac{\dav}{\alpha}(\rme^{-\rmi\omega}-1)}-1\right)}
\nonumber\\
&=&
\sum_d P(d)\rme^{-d}\sum_\ell \frac{d^\ell}{\ell !}\rme^{-\ell \frac{\dav}{\alpha}}\intw \rme^{\rmi\omega k+\ell \frac{\dav}{\alpha}\rme^{-\rmi\omega}}
\nonumber\\
&=&\sum_d P(d)\sum_\ell \rme^{-d}\frac{d^\ell}{\ell!}\rme^{-\ell\frac{\dav}{\alpha}}\frac{\left(\frac{\ell \dav}{\alpha}\right)^k}{k!}
\eea

\subsection{The mixed ensemble}

In the mixed ensemble we have the asymptotic degree distribution
\bea
p(k)&=&\lim_{N\to\infty}\frac{1}{N}\sum_i \bra \delta_{k,\sum_j c_{ij}}\ket_{\bxi}=\frac{1}{N}\sum_i \intw \rme^{\rmi\omega k}\bra \rme^{-\rmi\omega \sum_\mu \xi_i^\mu \sum_j\xi_j^\mu}\ket_{\bxi}
\nonumber\\
&=&\lim_{N\to\infty}\frac{1}{N}\sum_i \intw \rme^{\rmi\omega k}\prod_\mu\Big[
1+\frac{d_i q_\mu}{\alpha \qav N}
\Big(\prod_j \bra \rme^{-\rmi\omega \xi_j^\mu}\ket -1
\Big)
\Big]
\nonumber\\
&=&\lim_{N\to\infty}\frac{1}{N}\sum_i \intw \rme^{\rmi\omega k}\prod_\mu\Big[
1+\frac{d_i q_\mu}{\alpha \qav N}\Big(
\rme^{q_\mu(\rme^{-\rmi\omega}-1)}-1
\Big)
\Big]
\nonumber\\
&=&\lim_{N\to\infty}\frac{1}{N}\sum_i \intw \rme^{\rmi\omega k+
\sum_\mu \frac{d_i q_\mu}{\alpha \qav N} 
\Big(\rme^{q_\mu(\rme^{-\rmi\omega}-1)}-1\Big)}
\nonumber\\
&=&
\sum_d P(d)\intw \rme^{\rmi\omega k+\frac{d}{\qav}
\bra q(\rme^{q(\rme^{-\rmi\omega}-1)}-1)\ket_q}
\nonumber\\
&=&
\sum_d P(d) \rme^{-d}
\int_{-\pi}^\pi \!\frac{\rmd\omega}{2\pi}\rme^{\rmi\omega k+\frac{d}{\qav} \bra 
q\rme^{-q}
\exp[q \rme^{-\rmi\omega}]\ket_q} 
\nonumber
\\
&=& 
\sum_d P(d) \rme^{-d}\sum_{\ell\geq 0}\frac{(d/\qav)^\ell}{\ell !}\int_{-\pi}^\pi \!\frac{\rmd\omega}{2\pi}
\rme^{\rmi\omega k}
\bra 
q\rme^{-q}
\exp[q \rme^{-\rmi\omega}]\ket_q^\ell
\nonumber
\\
&=& 
\sum_d P(d) \rme^{-d}\sum_{\ell\geq 0}\frac{d^\ell}{\ell !}
\Big\bra \prod_{r\leq \ell}\left(\frac{q_{r}e^{-q_r}}{\qav}\right)
\frac{(\sum_{r\leq \ell }q_r )^k}{k!}
\Big\ket_{q_1\ldots q_\ell}
\end{eqnarray}
We can rewrite this expression in terms of the associated distribution $W(q)=qP(q)/\bra q\ket$ as:
\begin{eqnarray}
p(k)&=&\sum_d P(d)\rme^{-d} \sum_{\ell\geq 0}\frac{d^\ell}{\ell !}
\Big\bra \prod_{r\leq \ell}\Big(\frac{q_{r}\rme^{-q_r }}{\qav}\Big)\frac{(\sum_{r\leq \ell }q_r )^k}{k!}
\Big\ket_{q_1\ldots q_\ell}
\nonumber
\\
&=& \sum_d P(d)\rme^{-d}
\sum_{\ell\geq 0}\frac{d^\ell}{\ell !}
\sum_{q_1\ldots q_\ell\geq 0} W(q_1)\ldots W(q_\ell)
\rme^{-\sum_{r\leq \ell }q_r }\frac{(\sum_{r\leq \ell }q_r )^k}{k!}
\end{eqnarray}
or, equivalently, as 
\begin{eqnarray}
p(k)
&=& \int_0^\infty\!\rmd y~P(y)~\rme^{-y}y^k/k!
\eea
where
\bea
P(y)&=& 
 \sum_d P(d)\rme^{-d}
\sum_{\ell\geq 0}\frac{d^\ell}{\ell !}
\sum_{q_1\ldots q_\ell\geq 0} W(q_1)\ldots W(q_\ell)
~\delta[y-\sum_{r\leq \ell }q_r ]
\end{eqnarray}
The first two moments of $p(k)$ are 
\begin{eqnarray}
\bra k\ket&=& \int_0^\infty\!\rmd y~P(y)~y
= \sum_d P(d) \rme^{-d}\sum_{\ell\geq 0}\frac{d^\ell}{\ell !}
\sum_{q_1\ldots q_\ell\geq 0} W(q_1)\ldots W(q_\ell)~\sum_{r\leq \ell }q_r 
\nonumber
\\
&=&  \sum_d P(d) \rme^{-d}\sum_{\ell> 0}\frac{d^\ell}{(\ell\!-\!1) !}
\sum_{q} W(q)q= \dav \frac{\bra q^2\ket}{\qav}=
\alpha \bra q^2\ket
\\
\bra k^2\ket&=& \bra k\ket+\int_0^\infty\!\rmd y~P(y)~y^2
\nonumber
\\
&=& \alpha \bra q^2\ket+
 \sum_d P(d) e^{-d}\sum_{\ell\geq 0}\frac{d^\ell}{\ell !}
\sum_{q_1\ldots q_\ell\geq 0} W(q_1)\ldots W(q_\ell)~\sum_{r,s\leq \ell }q_r q_s
\nonumber
\\
&=& \alpha \bra q^2\ket+ \sum_d P(d)
\rme^{-d}\sum_{\ell> 0}\frac{d^\ell}{\ell !}\Big[\ell \sum_q W(q) q^2
+\ell(\ell-1) \Big(\sum_q W(q)q
\Big)^2
\Big]
\nonumber
\\
&=&
\alpha \bra q^2\ket +\frac{\bra q^3\ket}{\qav} \dav+
\bra d^2\ket \frac{\bra q^2\ket^2}{\qav^2} 
= \alpha \bra q^2\ket+ \alpha \bra q^3\ket+\frac{\bra d^2\ket}{\dav^2}
\kav^2
\end{eqnarray}

\section{The link between observables in the $\ba$ and $\bc$ networks}

\label{app:a_c}
In this appendix we inspect  the relation between expectation values of various observables in the ensembles $p(\ba)$ and  $p(\bc)$.

\subsection{The $q$-ensemble}

Denoting averages in the $\ba$ ensemble as $\bra \ldots\ket_a$, we have, for the $q$-ensemble
of bipartite graphs:
\begin{eqnarray}
\bra k\ket_a&=& \frac{1}{N}\sum_{ij}\bra a_{ij}\ket_a=\frac{1}{N}\sum_{ij}\bra \theta[c_{ij}-\frac{1}{2}]\ket
\nonumber
\\
&=& \frac{1}{N}\sum_{ij}[1-\bra \delta_{c_{ij},0}\ket]= \alpha\bra q^2\ket+\order(N^{-1})
= \bra k\ket+\order(N^{-1})
\\
\bra k^2\ket_a&=& \frac{1}{N}\sum_{i\neq j\neq k}\bra a_{ij}a_{jk}\ket
= \frac{1}{N}\sum_{ij}\bra a_{ij}\ket+\frac{1}{N}\sum_{i\neq j\neq k(\neq i)}
\bra a_{ij}a_{jk}\ket
\nonumber\\ 
&=&\frac{1}{N}\sum_{ij}\bra (1-\delta_{c_{ij},0})\ket+\frac{1}{N}\sum_{i\neq j\neq k(\neq i)}
\bra (1-\delta_{c_{ij},0})(1-\delta_{c_{jk},0})\ket
\nonumber\\ 
&=&\frac{1}{N}\sum_{ij}\frac{\alpha \bra q^2\ket}{N}
+\frac{1}{N}\sum_{i\neq j\neq k(\neq i)}
(1-2\bra \delta_{c_{ij},0}\ket +\bra \delta_{c_{ij},0}
\delta_{c_{jk},0}\ket)
\nonumber\\
&=&(N-1)(N-2)-2(N-1)(N-2)\Big(1-\frac{\alpha \bra q^2\ket}{N}+
\frac{\alpha^2\bra q^2\ket^2}{2N^2}
\Big)\nonumber\\
&&+(N-1)(N-2)\Big(1-2\frac{\alpha\bra q^2\ket}{N}
+\frac{\alpha \bra q^3\ket}{N^2}+
2\frac{\alpha^2\bra q^2\ket^2}{N^2}
\Big)+\alpha \bra q^2\ket
\nonumber\\
&=&\alpha \bra q^2\ket+\alpha \bra q^3\ket +\alpha^2 \bra q^2\ket^2\equiv \bra k^2\ket
\eea
\renewcommand{\intww}{\int_{-\pi}^\pi\!\frac{\rmd\omega\rmd\omega^\prime}{4\pi^2}}
where we used 
\bea
&&\frac{1}{N(N-1)(N-2)}\sum_{i\neq j\neq k(\neq i)}\bra \delta_{c_{ij},0}
\delta_{c_{jk},0}\ket
\nonumber\\
&=&\frac{1}{N(N-1)(N-2)}\sum_{i\neq j\neq k(\neq i)}
\intww \prod_\mu \bra \rme^{\rmi\xi_j^\mu (\xi_i^\mu \omega +\xi_k^\mu \omega')}
\ket
\nonumber\\
&=&\intww \prod_\mu \Big\{
1+\frac{q_\mu^2}{N^2}\Big[
(\rme^{\rmi\omega}+\rme^{\rmi\omega'}-2)+\frac{q_\mu}{N}(\rme^{\rmi(\omega+\omega')}-\rme^{\rmi\omega}
-\rme^{\rmi\omega'}+1)
\Big]
\Big\}
\nonumber\\
&=&
\intww \rme^{
\frac{\alpha \bra q^2\ket}{N}
(\rme^{\rmi\omega}+\rme^{\rmi\omega'}-2)+
\frac{\alpha \bra q^3\ket}{N^2}(\rme^{\rmi(\omega+\omega')}-
\rme^{\rmi\omega}-\rme^{\rmi\omega'}+1)-
\frac{\alpha \bra q^4\ket}{2N^3}(\rme^{\rmi\omega}+\rme^{\rmi\omega'}-2)^2}
\nonumber\\
&=&
1-2\frac{\alpha \bra q^2\ket}{N}+\frac{\alpha \bra q^3\ket}{N^2}+2
\frac{\alpha^2 \bra q^2\ket^2}{N^2}
-2\frac{\alpha \bra q^4\ket}{N^3}
-2\frac{\alpha^2 \bra q^2\ket \bra q^3\ket}{N^3}
-\frac{4}{3}\frac{\alpha^3 \bra q^2\ket^3}{N^3}
\eea
For loops of length $3$ we proceed in the same way, obtaining
\bea
m_3^a&=&\frac{1}{N}\sum_{i\neq j\neq k(\neq i)}\bra a_{ij}a_{jk}a_{ki}\ket
=\frac{1}{N}\sum_{i\neq j\neq k(\neq i)}\bra (1\!-\!\delta_{c_{ij},0})(
1\!-\!\delta_{c_{jk},0})(1\!-\!\delta_{c_{ki},0})\ket
\nonumber\\
&=&\frac{1}{N}\sum_{i\neq j\neq k(\neq i)}(1-3\bra \delta_{c_{ij},0} \ket
+3\bra \delta_{c_{ij},0}\delta_{c_{jk},0}\ket -
\bra \delta_{c_{ij},0}\delta_{c_{jk},0}\delta_{c_{ki},0})\ket
\nonumber\\
&=&(N-1)(N-2)-3(N-1)(N-2)\Big(
1-\frac{\alpha\bra q^2\ket}{N}
+2\frac{\alpha^2\bra q^2\ket^2}{2N^2}
\Big)
\nonumber\\
&&+3(N-1)(N-2)\Big(
1-2\frac{\alpha\bra q^2\ket}{N}
+\frac{\alpha \bra q^3\ket}{N^2}+
2\frac{\alpha^2\bra q^2\ket^2}{N^2}
\Big)
\nonumber\\
&&
-(N-1)(N-2)\Big(
1-3\frac{\alpha \bra q^2\ket}{N}+2\frac{\alpha \bra q^3\ket}{N^2}+
\frac{9}{2}\frac{\alpha^2 \bra q^2\ket^2}{N^2}
\Big)
=\alpha \bra q^3\ket \equiv m_3^c
\eea
\renewcommand{\intwww}{\int_{-\pi}^\pi\!\frac{\rmd\omega\rmd\omega^\prime\rmd\omega^\pprime}{8\pi^3}}
where we used 
\bea
\hspace*{-15mm}
&&\hspace*{-10mm} \frac{1}{N(N-1)(N-2)}\sum_{i\neq j\neq k(\neq i)}\bra \delta_{c_{ij},0}
\delta_{c_{jk},0}\delta_{c_{ki},0}\ket
\nonumber\\
\hspace*{-15mm}
&=&\frac{1}{N(N-1)(N-2)}\sum_{i\neq j\neq k(\neq i)}
\intwww \!\prod_\mu \bra \rme^{ \rmi\xi_i^\mu (\xi_j^\mu \omega +\xi_k^\mu \omega^\dprime)+i\xi_j^\mu \xi_k^\mu \omega^\prime}
\ket
\nonumber\\
\hspace*{-15mm}
&=&\intwww \prod_\mu \Big\{
1+\frac{q_\mu^2}{N^2}\Big[
( \rme^{ \rmi\omega}\!+ \!\rme^{ \rmi\omega^\prime}\!+ \!\rme^{ \rmi\omega^\dprime}\!-\!3)
\!+\!\frac{q_\mu}{N}( \rme^{ \rmi(\omega+\omega^\prime+\omega^\dprime)}\!- \!\rme^{ \rmi\omega}
\!- \!\rme^{ \rmi\omega^\prime}\!- \!\rme^{ \rmi\omega^\dprime}+2)
\Big]
\Big\}
\nonumber\\
\hspace*{-15mm}
&=&
\intwww  \rme^{\sum_\mu 
\frac{q_\mu^2}{N^2}( \rme^{ \rmi\omega}+ \rme^{ \rmi\omega^\prime}+ \rme^{ \rmi\omega^\dprime}-3)
+\sum_\mu 
\frac{q_\mu^3}{N^3}( \rme^{ \rmi(\omega+\omega^\prime+\omega^\dprime)}- \rme^{ \rmi\omega}
- \rme^{ \rmi\omega^\prime}- \rme^{ \rmi\omega^\dprime}+2)
-\sum_\mu 
\frac{q_\mu^4}{2N^4}( \rme^{ \rmi\omega}+ \rme^{ \rmi\omega^\prime}+ \rme^{ \rmi\omega^\dprime}-3)^2
}
\nonumber\\
\hspace*{-15mm}
&=&
\intwww  \rme^{\frac{\alpha \bra q^2\ket}{N}( \rme^{ \rmi\omega}+ \rme^{ \rmi\omega^\prime}+ \rme^{ \rmi\omega^\dprime}-3)
+\frac{\alpha \bra q^3\ket}{N^2}( \rme^{ \rmi(\omega+\omega^\prime+\omega^\dprime)}- \rme^{ \rmi\omega}
- \rme^{ \rmi\omega^\prime}- \rme^{ \rmi\omega^\dprime}+2)
-\frac{\alpha \bra q^4\ket}{2N^3}( \rme^{ \rmi\omega}+ \rme^{ \rmi\omega^\prime}+ \rme^{ \rmi\omega^\dprime}-3)^2
}
\nonumber\\
\hspace*{-15mm}
&=&
1-3\frac{\alpha \bra q^2\ket}{N}+2\frac{\alpha \bra q^3\ket}{N^2}+
\frac{9}{2}\frac{\alpha^2 \bra q^2\ket^2}{N^2}
-\frac{9}{2}\frac{\alpha \bra q^4\ket}{N^3}
-6\frac{\alpha^2 \bra q^2\ket \bra q^3\ket}{N^3}
-\frac{9}{2}\frac{\alpha^3 \bra q^2\ket^3}{N^3}
\eea
Finally for loops of length $4$, we have
\bea
m_4^a&=&\frac{1}{N}\sum_{[i,j,k,\ell]}\bra a_{ij}a_{jk}a_{k\ell}a_{\ell i}\ket
\nonumber\\
&=&\frac{1}{N}\sum_{[i,j,k,\ell]}\bra (1-\delta_{c_{ij},0})(
1-\delta_{c_{jk},0})(1-\delta_{c_{k\ell},0})(1-\delta_{c_{\ell i},0})\ket
\nonumber\\
&=&\frac{1}{N}\sum_{[i,j,k,\ell]}(1-4\bra \delta_{c_{ij},0} \ket
+4\bra \delta_{c_{ij},0}\delta_{c_{jk},0}\ket 
+2\bra \delta_{c_{ij},0}\ket \bra\delta_{c_{jk},0}\ket
-4
\bra \delta_{c_{ij},0}\delta_{c_{jk},0}\delta_{c_{k\ell},0})\ket
\nonumber\\
&&+
\bra \delta_{c_{ij},0}\delta_{c_{jk},0}\delta_{c_{k\ell},0}
\delta_{c_{\ell i},0})\ket
\nonumber\\
&=&(N-1)(N-2)(N-3)\left\{1-4
\Big(
1-\frac{\alpha\bra q^2\ket}{N}
+\frac{\alpha^2\bra q^2\ket^2}{2N^2}-\frac{\alpha^3\bra q^2\ket^3}{6N^3}
-\frac{\alpha\bra q^4\ket}{2N^3}
\Big)\right.
\nonumber\\
&&+4\Big(
1-2\frac{\alpha\bra q^2\ket}{N}
+\frac{\alpha \bra q^3\ket}{N^2}+
2\frac{\alpha^2\bra q^2\ket^2}{N^2}
-\frac{4}{3}\frac{\alpha^3\bra q^2\ket^3}{N^3}
-2\frac{\alpha\bra q^4\ket}{N^3}
-2\frac{\alpha^2\bra q^2\ket \bra q^3\ket}{N^3}
\Big)
\nonumber\\
&&+2\Big(
1-\frac{\alpha\bra q^2\ket}{N}
+\frac{\alpha^2\bra q^2\ket^2}{2N^2}-\frac{\alpha^3\bra q^2\ket^3}{6N^3}
-\frac{\alpha\bra q^4\ket}{2N^3}
\Big)^2
\nonumber\\
&&
-4\Big(
1-3\frac{\alpha \bra q^2\ket}{N}+2\frac{\alpha \bra q^3\ket}{N^2}+
\frac{9}{2}\frac{\alpha^2 \bra q^2\ket^2}{N^2}
-\frac{9}{2}\frac{\alpha^3\bra q^2\ket^3}{N^3}
-\frac{9}{2}\frac{\alpha\bra q^4\ket}{N^3}
-6\frac{\alpha^2\bra q^2\ket \bra q^3\ket}{N^3}
\Big)
\nonumber\\
&&\hspace*{-6mm}
+\left.\Big(
1-4\frac{\alpha \bra q^2\ket}{N}+4\frac{\alpha \bra q^3\ket}{N^2}
-9\frac{\alpha}{N^3}\bra q^4\ket+8\frac{\alpha^2 \bra q^2\ket^2}{N^2}
-16\frac{\alpha^2 \bra q^2\ket\bra q^3\ket }{N^3}-
\frac{32}{3}\frac{\alpha^3 \bra q^2\ket^3}{N^3}
\Big)\right\}
\hspace*{-10mm}
\nonumber\\
&=&\alpha\bra q^4\ket
\equiv m_4^c
\eea
where we used 
\bea
\hspace*{-15mm}
&&\hspace*{-10mm}
\frac{1}{N(N-1)(N-2)(N-3)}\sum_{i\neq j\neq k(\neq i)}\bra \delta_{c_{ij},0}
\delta_{c_{jk},0}\delta_{c_{k\ell},0}\ket
\nonumber\\
\hspace*{-15mm}
&=&\frac{1}{N(N-1)(N-2)(N-3)}\sum_{i\neq j\neq k(\neq i)}
\intwww \prod_\mu \bra  \rme^{ \rmi\xi_j^\mu (\xi_i^\mu \omega +\xi_k^\mu \omega^\prime)+i \xi_\ell^\mu \xi_k^\mu \omega^\dprime}
\ket
\nonumber\\
\hspace*{-15mm}
&=&\intwww \prod_\mu 
\Big\{1+\frac{q_\mu^2}{N^2}\Big[
( \rme^{ \rmi\omega}+ \rme^{ \rmi\omega^\prime}+ \rme^{ \rmi\omega^\dprime}-3)
+\frac{q_\mu}{N}( \rme^{ \rmi\omega^\prime}-1)( \rme^{ \rmi\omega}+ \rme^{ \rmi\omega^\dprime}-2)
\nonumber\\[-2mm]
&&
\hspace*{50mm}
+\frac{q_\mu^2}{N^2} \rme^{ \rmi\omega^\prime}( \rme^{ \rmi\omega}-1)( \rme^{ \rmi\omega^\dprime}-1)
\Big]
\Big\}
\nonumber\\[1mm]
\hspace*{-15mm}
&=&
\intwww  \rme^{\sum_\mu 
\frac{q_\mu^2}{N^2}( \rme^{ \rmi\omega}+ \rme^{ \rmi\omega^\prime}+ \rme^{ \rmi\omega^\dprime}-3)
+\sum_\mu 
\frac{q_\mu^3}{N^3}( \rme^{ \rmi\omega^\prime}-1)( \rme^{ \rmi\omega}+ \rme^{ \rmi\omega^\dprime}-2))
-\sum_\mu 
\frac{q_\mu^4}{2N^4}( \rme^{ \rmi\omega}+ \rme^{ \rmi\omega^\prime}+ \rme^{ \rmi\omega^\dprime}-3)^2
}\hspace*{-5mm}
\nonumber\\[-2mm]
\hspace*{-15mm}
&&\hspace*{50mm}
\times e^{\frac{q_\mu^4}{N^4}e^{i\omega^\prime}(e^{i\omega}-1)(e^{i\omega^\dprime}-1)}
\nonumber\\[1mm]
\hspace*{-15mm}
&=&
\intwww  \rme^{\frac{\alpha \bra q^2\ket}{N}( \rme^{ \rmi\omega}+ \rme^{ \rmi\omega^\prime}+ \rme^{ \rmi\omega^\dprime}-3)
+\frac{\alpha \bra q^3\ket}{N^2}( \rme^{ \rmi\omega^\prime}-1)( \rme^{ \rmi\omega}+ \rme^{ \rmi\omega^\dprime}-2))
-\frac{\alpha \bra q^4\ket}{2N^3}( \rme^{ \rmi\omega}+ \rme^{ \rmi\omega^\prime}+ \rme^{ \rmi\omega^\dprime}-3)^2
}\nonumber\\[-2mm]
\hspace*{-15mm}
&&\hspace*{50mm}
\times  \rme^{\alpha \frac{\bra q^4\ket}{N^3} \rme^{ \rmi\omega^\prime}( \rme^{ \rmi\omega}-1)( \rme^{ \rmi\omega^\dprime}-1)}
\nonumber\\[1mm]
\hspace*{-15mm}
&=&
1-3\frac{\alpha \bra q^2\ket}{N}+2\frac{\alpha \bra q^3\ket}{N^2}+
\frac{9}{2}\frac{\alpha^2 \bra q^2\ket^2}{N^2}
-\frac{9}{2}\frac{\alpha \bra q^4\ket}{N^3}
-6\frac{\alpha^2 \bra q^2\ket \bra q^3\ket}{N^3}
-\frac{9}{2}\frac{\alpha^3 \bra q^2\ket^3}{N^3}
\eea
\renewcommand{\intwwww}{\int_{-\pi}^\pi\!\frac{\rmd\omega\rmd\omega^\prime\rmd\omega^\pprime\rmd\omega^{\pprime\prime}}{16\pi^4}}
and
\bea
\hspace*{-15mm}
&&
\hspace*{-10mm}
\frac{1}{N(N-1)(N-2)(N-3)}\sum_{[ijk\ell]}\bra \delta_{c_{ij},0}
\delta_{c_{jk},0}\delta_{c_{k\ell},0}\delta_{c_{\ell i},0}\ket
\nonumber\\
\hspace*{-15mm}
&=&\frac{1}{N(N-1)(N-2)(N-3)}\sum_{[ijk\ell]}
\intwwww \prod_\mu \bra  \rme^{ \rmi\xi_i^\mu (\xi_j^\mu \omega +\xi_\ell^\mu \omega^\tprime)+\xi_k^\mu (\xi_j^\mu \omega^\prime+\xi_\ell^\mu \omega^\dprime)}
\ket
\nonumber\\
\hspace*{-15mm}
&=&\intwwww \prod_\mu \left\{\left(1-\frac{q_\mu}{N}\right)\left\{
\frac{q_\mu}{N}\left[\frac{q_\mu^2}{N^2}\rme^{\rmi(\omega+\omega')}
+\frac{q_\mu}{N}\left(1\!-\!\frac{q_\mu}{N}\right)(\rme^{\rmi\omega^\prime}
\!+\!\rme^{\rmi\omega^\dprime})+\left(1\!-\!\frac{q_\mu}{N}\right)^2\right]
\right.\right.
\nonumber\\
\hspace*{-15mm}
&&\hspace*{50mm}
+\left.\left(1-\frac{q_\mu}{N}\right)
\right\}
+\frac{q_\mu}{N}\left\{\frac{q_\mu^2}{N^2}
\rme^{\rmi(\omega+\omega^\tprime)}\left(1-\frac{q_\mu}{N}+\frac{q_\mu}{N}
\rme^{\rmi(\omega^\prime+\omega^\dprime)}\right)
\right.
\nonumber\\
\hspace*{-15mm}
&&
+\left.\frac{q_\mu}{N}\left(1-\frac{q_\mu}{N}\right)
\left[\rme^{\rmi\omega}\left(1-\frac{q_\mu}{N}+\frac{q_\mu}{N}\rme^{\rmi\omega^\prime}
\right)
+\rme^{\rmi\omega^\tprime}\left(1-\frac{q_\mu}{N}+\frac{q_\mu}{N}\rme^{\rmi\omega^\dprime}
\right)
\right]+\left(1-\frac{q_\mu}{N}\right)^2
\right\}
\nonumber\\
\hspace*{-15mm}
&=&
\prod_\mu \left\{
\frac{q_\mu}{N}\left(1-\frac{q_\mu}{N}\right)^2
+\left(1-\frac{q_\mu}{N}\right)\left[
\frac{q_\mu}{N}\left(1-\frac{q_\mu}{N}\right)^2+\left(1-\frac{q_\mu}{N}\right)
\right]
\right\} 
\nonumber\\
\hspace*{-15mm}
&=&
\prod_\mu \left\{
1-4\frac{q_\mu^2}{N^2}+4\frac{q_\mu^3}{N^3}-\frac{q_\mu^4}{N^4}
\right\}
=\rme^{-4\frac{\alpha \bra q^2\ket}{N}
+4\frac{\alpha \bra q^3\ket}{N^2}
-9\frac{\alpha \bra q^4\ket}{N^3}
+\order(N^{-4})}
\nonumber\\
\hspace*{-15mm}
&=&
1-4\frac{\alpha \bra q^2\ket}{N}+4\frac{\alpha \bra q^3\ket}{N^2}+
8\frac{\alpha^2 \bra q^2\ket^2}{N^2}
-9\frac{\alpha \bra q^4\ket}{N^3}
-16\frac{\alpha^2 \bra q^2\ket \bra q^3\ket}{N^3}
-\frac{32}{3}\frac{\alpha^3 \bra q^2\ket^3}{N^3}+\order(N^{-4})
\eea
Again, the square brackets underneath the summations indicate that 
all indices are different, to exclude backtracking in the 
counting of loops of length $4$. 

\subsection{The $d$-ensemble}

For the $d$-ensemble, denoting averages relating to $\ba$ as $\bra \ldots\ket_a$, we have:
\begin{eqnarray}
\bra k\ket_a&=& \frac{1}{N}\sum_{ij}\bra a_{ij}\ket_a=\frac{1}{N}\sum_{ij}[1-\bra \delta_{c_{ij},0}\ket]= 
\nonumber\\
&=&
\frac{1}{N}\sum_{ij}\left[\frac{d_i d_j}{\alpha N}-\half \left(\frac{d_i d_j}{\alpha N}\right)^2+\frac{1}{6}\left(\frac{d_i d_j}{\alpha N}\right)^3+\half \frac{d_i^2 d_j^2}{\left(\alpha N\right)^3}\right]
\nonumber\\
&=&\frac{\bra d\ket^2}{\alpha}+\order(N^{-1})
= \bra k\ket+\order(N^{-1})
\\
\bra k^2\ket_a&=& \frac{1}{N}\sum_{i\neq j\neq k}\bra a_{ij}a_{jk}\ket
= \frac{1}{N}\sum_{ij}\bra a_{ij}\ket+\frac{1}{N}\sum_{i\neq j\neq k(\neq i)}
\bra a_{ij}a_{jk}\ket
\nonumber\\ 
&=&\frac{\bra d\ket^2}{\alpha}
+\frac{1}{N}\sum_{i\neq j\neq k(\neq i)}
\bra (1-\delta_{c_{ij},0})(1-\delta_{c_{jk},0})\ket
\nonumber\\ 
&=&
\frac{\bra d\ket^2}{\alpha}
+\frac{1}{N}\sum_{i\neq j\neq k(\neq i)}
(1-2\bra \delta_{c_{ij},0}\ket +\bra \delta_{c_{ij},0}
\delta_{c_{jk},0}\ket)
\nonumber\\
&=&\frac{\bra d\ket^2}{\alpha}
+(N-1)(N-2)-\frac{2}{N}\sum_{[ijk]}\left(
1-\frac{d_i d_j}{\alpha N}+\half \left(\frac{d_i d_j}{\alpha N}\right)^2
\right)+\frac{1}{N}\sum_{[ijk]}\bra \delta_{c_{ij},0}
\delta_{c_{jk},0}\ket
\nonumber\\
&=&\frac{\bra d\ket^2}{\alpha}
+2\frac{\bra d \ket^2}{\alpha}N-2\frac{\bra d \ket^2}{\alpha}-\frac{\bra d^2\ket^2}{\alpha^2}-2N \frac{\bra d \ket^2}{\alpha}+2\frac{\bra d \ket^2}{\alpha}
+\frac{\bra d\ket^3}{\alpha^2}+
\frac{\bra d^2\ket^2}{\alpha^2}
-\frac{\bra d^2\ket \bra d \ket^2}{\alpha^2}
\nonumber\\
&&=\frac{\bra d\ket^2}{\alpha}
+\frac{\bra d\ket^3}{\alpha^2}
+\frac{\bra d^2\ket \bra d \ket^2}{\alpha^2}\equiv \bra k^2\ket
\nonumber\\
\eea
where we used 
\bea
\hspace*{-15mm}
&&\hspace*{-10mm}
\frac{1}{N}\sum_{i\neq j\neq k(\neq i)}\bra \delta_{c_{ij},0}
\delta_{c_{jk},0}\ket
=\frac{1}{N}\sum_{i\neq j\neq k(\neq i)}
\intww \prod_\mu \bra \rme^{\rmi\xi_j^\mu (\xi_i^\mu \omega +\xi_k^\mu \omega')}
\ket
\nonumber\\
\hspace*{-15mm}
&&\hspace*{-1cm}=\frac{1}{N}\sum_{[ijk]}\intww \prod_\mu \left\{
1+\frac{d_j}{\alpha N}\left[\frac{d_i}{\alpha N}
(\rme^{\rmi\omega}-1)+\frac{d_k}{\alpha N}(\rme^{\rmi\omega'}-1)+\frac{d_i d_k}{(\alpha N)^2}(\rme^{\rmi(\omega+\omega')}-\rme^{\rmi\omega}
-\rme^{\rmi\omega'}+1)
\right]
\right\}
\nonumber\\
\hspace*{-15mm}
&&\hspace*{-0cm}=\frac{1}{N}\sum_{[ijk]}
\intww 
\left\{
1+\frac{d_j}{\alpha N}\left[d_i
(\rme^{\rmi\omega}-1)+d_k(\rme^{\rmi\omega'}-1)+\frac{d_i d_k}{\alpha N}(\rme^{\rmi(\omega+\omega')}-\rme^{\rmi\omega}
-\rme^{\rmi\omega'}+1)
\right]\right.
\nonumber\\
\hspace*{-15mm}
&&\hspace*{50mm}+\left.\half \left(\frac{d_j}{\alpha N}[d_i
(\rme^{\rmi\omega}-1)+d_k(\rme^{\rmi\omega'}-1)]
\right)^2
-\frac{d_i d_j^2 d_k}{(\alpha N)^3}(d_i+d_k)
\right\}
\nonumber\\
\hspace*{-15mm}
&&\hspace*{-0cm}=\frac{1}{N}\sum_{[ijk]}
\left\{
1+\frac{d_j}{\alpha N}\left[-d_i-d_k+\frac{d_i d_k}{\alpha N}
\right]+\half \left(\frac{d_j}{\alpha N}\right)^2(d_i^2+d_k^2+2 d_i d_k)
-\frac{d_i^2 d_j^2 d_k}{(\alpha N)^3}-\frac{d_i d_j^2 d_k^2}{(\alpha N)^3}
\right\}
\nonumber\\
\hspace*{-15mm}
&&\hspace*{-0cm}=
(N-1)(N-2)-2N \frac{\bra d \ket^2}{\alpha}+2\frac{\bra d \ket^2}{\alpha}
+\frac{\bra d\ket^3}{\alpha^2}+
\frac{\bra d^2\ket^2}{\alpha^2}
+\frac{\bra d^2\ket \bra d \ket^2}{\alpha^2}
\eea
For loops of length $3$ we have:
\bea
m_3^a&=&\frac{1}{N}\sum_{i\neq j\neq k(\neq i)}\bra a_{ij}a_{jk}a_{ki}\ket
=\frac{1}{N}\sum_{i\neq j\neq k(\neq i)}\bra (1-\delta_{c_{ij},0})(
1-\delta_{c_{jk},0})(1-\delta_{c_{ki},0})\ket
\nonumber\\
&=&\frac{1}{N}\sum_{[ijk]}(1-3\bra \delta_{c_{ij},0} \ket
+3\bra \delta_{c_{ij},0}\delta_{c_{jk},0}\ket -
\bra \delta_{c_{ij},0}\delta_{c_{jk},0}\delta_{c_{ki},0})\ket
\nonumber\\
&=&(N-1)(N-2)-3\frac{1}{N}\sum_{[ijk]}\left(
1-\frac{d_i d_j}{\alpha N}+\half \left(\frac{d_i d_j}{\alpha N}\right)^2
\right)
\nonumber\\
&&
+3\left[
(N-1)(N-2)-2N \frac{\bra d \ket^2}{\alpha}+2\frac{\bra d \ket^2}{\alpha}
+\frac{\bra d\ket^3}{\alpha^2}+
\frac{\bra d^2\ket^2}{\alpha^2}
-\frac{\bra d^2\ket \bra d \ket^2}{\alpha^2}
\right]
\nonumber\\
&&-(N-1)(N-2)+3N\frac{\bra d \ket^2}{\alpha}-3\frac{\bra d \ket^2}{\alpha}
-2\frac{\bra d \ket^3}{\alpha^2}-\frac{3}{2}\frac{\bra d^2\ket^2}{\alpha^2}
-3\frac{\bra d^2\ket \bra d\ket^2}{\alpha^2}
\nonumber\\
&=&3\frac{\bra d \ket^2}{\alpha}N-3\frac{\bra d \ket^2}{\alpha}
-\frac{3}{2}\frac{\bra d^2\ket^2}{\alpha^2}
\nonumber\\
&&
+3\left[
-2N \frac{\bra d \ket^2}{\alpha}+2\frac{\bra d \ket^2}{\alpha}
+\frac{\bra d\ket^3}{\alpha^2}+
\frac{\bra d^2\ket^2}{\alpha^2}
+\frac{\bra d^2\ket \bra d \ket^2}{\alpha^2}
\right]
\nonumber\\
&&+3N\frac{\bra d \ket^2}{\alpha}-3\frac{\bra d \ket^2}{\alpha}
-2\frac{\bra d \ket^3}{\alpha^2}-\frac{3}{2}\frac{\bra d^2\ket^2}{\alpha^2}
-3\frac{\bra d^2\ket \bra d\ket^2}{\alpha^2}
=\frac{\bra d\ket^3}{\alpha^2} \equiv m_3^c
\eea
where we used 
\bea
\hspace*{-5mm}&&\hspace*{-15mm}
\frac{1}{N}\sum_{[ijk]}\bra \delta_{c_{ij},0}
\delta_{c_{jk},0}\delta_{c_{ki},0}\ket
=\frac{1}{N}\sum_{[ijk]}
\intwww \prod_\mu \bra \rme^{\rmi\xi_i^\mu (\xi_j^\mu \omega +\xi_k^\mu \omega^\dprime)+i\xi_j^\mu \xi_k^\mu \omega^\prime}
\ket
\nonumber\\
\hspace*{-5mm}
&=&\frac{1}{N}\sum_{[ijk]}
\intwww \prod_\mu \left\{
\frac{d_i}{\alpha N}\bra \rme^{\rmi(\xi_j^\mu \omega +\xi_k^\mu \omega^\dprime+\xi_j^\mu \xi_k^\mu \omega^\prime}\ket
+\left(1-\frac{d_i}{\alpha N}\right)\bra \rme^{\rmi\xi_j^\mu \xi_k^\mu \omega^\prime}\ket
\right\}
\nonumber\\
\hspace*{-15mm}
&=&\frac{1}{N}\sum_{[ijk]}\intwww \left(
1+\frac{d_j d_k}{(\alpha N)^2}(\rme^{\rmi\omega^\prime}-1)+\frac{d_i}{\alpha N}
\left\{-1-\frac{d_j d_k}{(\alpha N)^2}(\rme^{\rmi\omega^\prime}-1)
\right.\right.
\nonumber\\
\hspace*{-15mm}
&&\left.\left.
+
\frac{d_j}{\alpha N}\rme^{\rmi\omega}
\left[
1+\frac{d_k}{\alpha N}(\rme^{\rmi(\omega^\dprime+\omega^\prime)}-1)
\right]+\left(
1-\frac{d_j}{\alpha N}
\right)
\left[1+\frac{d_k}{\alpha N}(\rme^{\rmi\omega^\dprime}-1)
\right]
\right\}\right)^{\alpha N}
\nonumber\\
\hspace*{-15mm}
&=&\frac{1}{N}\sum_{[ijk]}
\left(
1-\frac{d_j d_k}{(\alpha N)^2}
+\frac{d_i}{\alpha N}\left\{
-\frac{d_j}{\alpha N}
-\frac{d_k}{\alpha N}+2\frac{d_j d_k}{(\alpha N)^2}
\right\}\right)^{\alpha N}
\nonumber\\
\hspace*{-15mm}
&=&\frac{1}{N}\sum_{[ijk]}
\left[
1-\frac{d_j d_k}{\alpha N}
+\frac{d_i}{\alpha N}\left\{
-d_j
-d_k+2\frac{d_j d_k}{\alpha N}
\right\}+\half\left(\frac{d_i}{\alpha N}
\right)^2(d_j^2+d_k^2+2d_j d_k)
\right.
\nonumber\\
\hspace*{-15mm}
&&\left.
+\half 
\frac{d_j^2 d_k^2}{(\alpha N)^2}
+\frac{d_id_jd_k}{(\alpha N)^2}(d_j+d_k)
\right]
\nonumber\\
\hspace*{-15mm}
&=&
(N-1)(N-2)-3N\frac{\bra d \ket^2}{\alpha}+3\frac{\bra d \ket^2}{\alpha}
+2\frac{\bra d \ket^3}{\alpha^2}+\frac{3}{2}\frac{\bra d^2\ket^2}{\alpha^2}
+3\frac{\bra d^2\ket \bra d\ket^2}{\alpha^2}
\eea
Finally, for loops of length $4$ we have
\bea
\hspace*{-15mm}
m_4^a&=&\frac{1}{N}\sum_{[i,j,k,\ell]}\bra a_{ij}a_{jk}a_{k\ell}a_{\ell i}\ket
=\frac{1}{N}\sum_{[i,j,k,\ell]}\bra (1-\delta_{c_{ij},0})(
1-\delta_{c_{jk},0})(1-\delta_{c_{k\ell},0})(1-\delta_{c_{\ell i},0})\ket
\nonumber\\
\hspace*{-15mm}
&=&\frac{1}{N}\sum_{[i,j,k,\ell]}(1-4\bra \delta_{c_{ij},0} \ket
+4\bra \delta_{c_{ij},0}\delta_{c_{jk},0}\ket 
+2\bra \delta_{c_{ij},0}\ket \bra\delta_{c_{k\ell},0}\ket
-4
\bra \delta_{c_{ij},0}\delta_{c_{jk},0}\delta_{c_{k\ell},0})\ket
\nonumber\\
\hspace*{-15mm}
&&\hspace*{50mm} +
\bra \delta_{c_{ij},0}\delta_{c_{jk},0}\delta_{c_{k\ell},0}
\delta_{c_{\ell i},0})\ket
\nonumber\\
\hspace*{-15mm}
&=& (N-1)(N-2)(N-3) - 
4 \left[ (N-1)(N-2)(N-3) -N^2 \frac{\bra d\ket^2}{\alpha} 
\right.
\nonumber\\
\hspace*{-15mm}
&&
\hspace*{50mm}
\left. +\frac{1}{2}N\frac{\bra d^2 \ket^2}{{\alpha}^2} -
-\frac{1}{2}\frac{\bra d^2 \ket^2}{{\alpha}^3} - \frac{1}{6}\frac{\bra d^3\ket^2}{\alpha^3} \right]
\nonumber\\
\hspace*{-15mm}
&&+
4 \left[ (N-1)(N-2)(N-3) -2N^2\frac{\bra d\ket^2}{\alpha} +N\frac{\bra d \ket^3}{\alpha^2} - \frac{\bra d^2 \ket^2}{\alpha^3}
\right.
\nonumber\\
\hspace*{-15mm}
&&
\hspace*{0mm}
\left.
-\frac{\bra d^2 \ket \bra d\ket^2}{\alpha^3} +N\frac{\bra d^2 \ket^2}{\alpha^2} +N\frac{\bra d^2 \ket\bra d\ket^2}{\alpha^2} -2\frac{\bra d^2 \ket^2 \bra d\ket}{\alpha^3} -\frac{1}{3}\frac{\bra d^3 \ket^2}{\alpha^3}
-\frac{\bra d^3\ket\bra d^2\ket\bra d\ket}{\alpha^3} \right]
\nonumber\\
\hspace*{-15mm}
&&+
2 \left[(N-1)(N-2)(N-3)-2N^2\frac{\bra d\ket^2}{\alpha} +N\frac{\bra d^2 \ket^2}{\alpha^2} +N\frac{\bra d \ket^4}{\alpha^2}
\right.
\nonumber\\
\hspace*{-15mm}
&&
\left.\hspace*{50mm}
- \frac{\bra d^2 \ket^2}{\alpha^3} -\frac{1}{3}\frac{\bra d^3 \ket^2}{\alpha^3} -\frac{\bra d^2 \ket^2 \bra d\ket^2}{\alpha^3} \right]
\nonumber\\
\hspace*{-15mm}
&&-
4\left[ (N-1)(N-2)(N-3) -3N^2\frac{\bra d\ket^2}{\alpha} +2N\frac{\bra d \ket^3}{\alpha^2} - \frac{3}{2}\frac{\bra d^2 \ket^2}{\alpha^3}
\right.
\nonumber\\
\hspace*{-15mm}
&&
\left.
-2\frac{\bra d^2 \ket \bra d\ket^2}{\alpha^3} +\frac{3}{2}N\frac{\bra d^2 \ket^2}{\alpha^2} -\frac{\bra d\ket^4}{\alpha^3} +2N\frac{\bra d^2 \ket \bra d\ket^2}{\alpha^2} +N\frac{\bra d\ket^4}{\alpha^2}
\right.
\nonumber\\
\hspace*{-15mm}
&&
\left.
-4\frac{\bra d^2 \ket^2 \bra d \ket}{\alpha^3} - 2\frac{\bra d^2 \ket\bra d\ket^3}{\alpha^3} -\frac{1}{2}\frac{\bra d^3 \ket^2}{\alpha^3} -2\frac{\bra d^3\ket\bra d^2\ket\bra d\ket}{\alpha^3} -\frac{\bra d^2 \ket^2 \bra d\ket^2}{\alpha^3} \right]
\nonumber\\
\hspace*{-15mm}
&&+
(N-1)(N-2)(N-3) -4N^2\frac{\bra d\ket^2}{\alpha} +4N\frac{\bra d \ket^3}{\alpha^2} -2\frac{\bra d^2 \ket^2}{\alpha^3}
\nonumber\\
\hspace*{-15mm}
&&-
4\frac{\bra d^2 \ket \bra d\ket^2}{\alpha^3} +2N\frac{\bra d^2 \ket^2}{\alpha^2} -3\frac{\bra d\ket^4}{\alpha^3} +4N\frac{\bra d^2 \ket \bra d\ket^2}{\alpha^2} +2N\frac{\bra d\ket^4}{\alpha^2}
\nonumber\\
\hspace*{-15mm}
&&-
8\frac{\bra d^2 \ket^2 \bra d \ket}{\alpha^3} - 8\frac{\bra d^2 \ket\bra d\ket^3}{\alpha^3} -\frac{2}{3}\frac{\bra d^3 \ket^2}{\alpha^3} -4\frac{\bra d^3\ket\bra d^2\ket\bra d\ket}{\alpha^3} -2\frac{\bra d^2 \ket^2 \bra d\ket^2}{\alpha^3}
\nonumber\\
\hspace*{-15mm}
&&=
\frac{\bra d\ket^4}{\alpha^3} \equiv m_4^c
\eea
where we used
\bea
\hspace*{-15mm}
&&\hspace*{-10mm}\frac{1}{N}\sum_{[ijk\ell]}\bra \delta_{c_{ij},0}
\delta_{c_{jk},0}\delta_{c_{k\ell},0} \ket
=\frac{1}{N}\sum_{[ijk\ell]}
\intwww \prod_\mu \bra \rme^{\rmi\xi_j^\mu (\xi_i^\mu \omega +\xi_k^\mu \omega^\prime)+\xi_k^\mu \xi_\ell^\mu \omega^\dprime}
\ket
\nonumber\\
\hspace*{-15mm}
&=&\frac{1}{N}\sum_{[ijk\ell]}
\intwww \prod_\mu \left\{
\frac{d_j}{\alpha N}\bra \rme^{\rmi(\xi_i^\mu \omega +\xi_k^\mu \omega^\prime)+i\xi_k^\mu \xi_\ell^\mu \omega^\dprime}\ket
+\left(1-\frac{d_j}{\alpha N}\right)\bra e^{i\xi_k^\mu \xi_\ell^\mu \omega^\dprime}\ket
\right\}
\nonumber\\
\hspace*{-15mm}
&=&\frac{1}{N}\sum_{[ijk\ell]}\intwww \left(
1+\frac{d_i d_j}{(\alpha N)^2}(\rme^{\rmi\omega}-1)+\frac{d_j d_k}{(\alpha N)^2}(e^{i\omega^\prime}-1)+\frac{d_k d_\ell}{(\alpha N)^2}(\rme^{\rmi\omega^\dprime}-1)
\right.
\nonumber\\
\hspace*{-15mm}
&&
\left. +\frac{d_i d_j d_k}{(\alpha N)^3}(\rme^{\rmi(\omega+\omega^\prime)}-\rme^{\rmi\omega}-\rme^{\rmi\omega^\prime}+1)+\frac{d_j d_k d_\ell}{(\alpha N)^3} (\rme^{\rmi(\omega^\prime+\omega^\dprime)}-\rme^{\rmi\omega^\prime}-\rme^{\rmi\omega^\dprime}+1) 
\right.
\nonumber\\
\hspace*{-15mm}
&&
\left. +\frac{d_i d_j d_k d_\ell}{(\alpha N)^4} (\rme^{\rmi(\omega+\omega^\prime+\omega^\dprime)} - \rme^{\rmi(\omega + \omega^\prime)} -  \rme^{\rmi(\omega + \omega^\dprime)} + \rme^{\rmi\omega^\prime}) \right) ^{\alpha N}
\nonumber\\
\hspace*{-15mm}
&=&
\frac{1}{N}\sum_{[ijk\ell]} \left\{ 1- \frac{d_i d_j}{\alpha N}-\frac{d_j d_k}{\alpha N}-\frac{d_k d_\ell}{\alpha N} + \frac{d_i d_j d_k}{(\alpha N)^2} +  \frac{d_j d_k d_\ell}{(\alpha N)^2} 
\right.
\nonumber\\
\hspace*{-15mm}
&&
\left.  -\frac{1}{2}\left[ \frac{d_i^2 d_j^2}{(\alpha N)^3} +\frac{d_j^2 d_k^2}{(\alpha N)^3} +\frac{d_k^2 d_\ell^2}{(\alpha N)^3} + 2\frac{d_i d_j^2 d_k}{(\alpha N)^3}+ 2\frac{d_i d_j d_k d_\ell}{(\alpha N)^3}+ 2\frac{d_j d_k^2 d_\ell}{(\alpha N)^3} \right] 
\right.
\nonumber\\
\hspace*{-15mm}
&&
\left. +\frac{1}{2} \left[ \frac{d_i^2 d_j^2}{(\alpha N)^2} +\frac{d_j^2 d_k^2}{(\alpha N)^2} +\frac{d_k^2 d_\ell^2}{(\alpha N)^2} + 2\frac{d_i d_j^2 d_k}{(\alpha N)^2}+2\frac{d_i d_j d_k d_\ell}{(\alpha N)^2}
+2\frac{d_j d_k^2 d_\ell}{(\alpha N)^2} -2\frac{d_i^2 d_j^2 d_k}{(\alpha N)^3} 
\right.\right.
\nonumber\\
\hspace*{-15mm}
&&
\left.\left. -2\frac{d_i^2 d_j d_k d_\ell}{(\alpha N)^3} -2\frac{d_i d_j^2 d_k^2}{(\alpha N)^3} -2\frac{d_i d_j d_k^2 d_\ell}{(\alpha N)^3} 
 -2\frac{d_j^2 d_k^2 d_\ell}{(\alpha N)^3} -2\frac{d_j d_k^2 d_\ell^2}{(\alpha N)^3} \right] -\frac{1}{6} \left[ \frac{d_i^3 d_j^3}{(\alpha N)^3} +\frac{d_j^3 d_k^3}{(\alpha N)^3} 
\right.\right.
\nonumber\\
\hspace*{-15mm}
&&
\left.\left. +\frac{d_k^3 d_\ell^3}{(\alpha N)^3}+3\frac{d_i^2 d_j^3 d_k}{(\alpha N)^3} +3\frac{d_i d_j^3 d_k^2}{(\alpha N)^3} +3\frac{d_i^2 d_j^2 d_k d_\ell}{(\alpha N)^3} +3\frac{d_i d_j d_k^2 d_\ell^2}{(\alpha N)^3} +3\frac{d_j^2 d_k^3 d_\ell}{(\alpha N)^3} + 3\frac{d_j d_k^3 d_\ell ^2}{(\alpha N)^3} \right] \right\}
\nonumber\\
\hspace*{-15mm}
&=&
(N-1)(N-2)(N-3) -3N^2\frac{\bra d\ket^2}{\alpha} +2N\frac{\bra d \ket^3}{\alpha^2} -\frac{3}{2}\frac{\bra d^2 \ket^2}{\alpha^3} - 2 \frac{\bra d^2\ket \bra d \ket^2}{\alpha^3} - \frac{\bra d \ket^4}{\alpha^3}
\nonumber\\
\hspace*{-15mm}
&&
\frac{3}{2}N \frac{\bra d^2 \ket ^2}{\alpha^2} +2N \frac{\bra d^2 \ket \bra d \ket^2}{\alpha^2} + N\frac{\bra d \ket^4}{\alpha^2} - 4\frac{\bra d^2 \ket^2 \bra d \ket}{\alpha^3} -2 \frac{\bra d^2 \ket \bra d \ket^3}{\alpha^3}
\nonumber\\
\hspace*{-15mm}
&&
-\frac{1}{2} \frac{\bra d^3 \ket^2}{\alpha^3} -2\frac{\bra d^2 \ket \bra d^2 \ket \bra d \ket}{\alpha^3} - \frac{\bra d^2 \ket^2 \bra d \ket^2}{\alpha^3}
\eea
and
\bea
\hspace*{-15mm}
&&\hspace*{-10mm}\frac{1}{N}\sum_{[ijk\ell]}\bra \delta_{c_{ij},0}
\delta_{c_{jk},0}\delta_{c_{k\ell},0} \delta_{c_{\ell i},0}\ket
\nonumber\\
\hspace*{-15mm}
&=&\frac{1}{N}\sum_{[ijk\ell]}
\intwwww \prod_\mu \bra \rme^{\rmi\xi_i^\mu (\xi_j^\mu \omega +\xi_\ell^\mu \omega^\tprime)+\xi_k^\mu (\xi_j^\mu \omega^\prime + \xi_\ell^\mu \omega^\dprime)}
\ket
\nonumber\\
\hspace*{-15mm}
&=&\frac{1}{N}\sum_{[ijk\ell]}
\intwwww \prod_\mu \left\{
\frac{d_i}{\alpha N}\bra \rme^{\rmi(\xi_j^\mu \omega +\xi_\ell^\mu \omega^\tprime)+i\xi_k^\mu(\xi_j^\mu \omega^\prime+\xi_\ell^\mu \omega^\dprime)}\ket
+\left(1-\frac{d_i}{\alpha N}\right)\bra \rme^{\rmi\xi_k^\mu (\xi_j^\mu \omega^\prime+\xi_\ell^\mu \omega^\dprime)}\ket
\right\}
\nonumber\\
\hspace*{-15mm}
&=&\frac{1}{N}\sum_{[ijk\ell]}\intwwww \left(
1+\frac{d_i d_j}{(\alpha N)^2}(\rme^{\rmi\omega}-1)+\frac{d_j d_k}{(\alpha N)^2}(\rme^{\rmi\omega^\prime}-1)+\frac{d_k d_\ell}{(\alpha N)^2}(\rme^{\rmi\omega^\dprime}-1)
\right.
\nonumber\\
\hspace*{-15mm}
&&
\left. +\frac{d_i d_\ell}{(\alpha N)^2}(\rme^{\rmi\omega^\tprime}-1) +\frac{d_i d_j d_k}{(\alpha N)^3}(\rme^{\rmi(\omega+\omega^\prime)}-\rme^{\rmi\omega}-\rme^{\rmi\omega^\prime}+1)+\frac{d_i d_k d_\ell}{(\alpha N)^3} (\rme^{\rmi(\omega^\dprime+\omega^\tprime)}-\rme^{\rmi\omega^\dprime}-\rme^{\rmi\omega^\tprime}+1) 
\right.
\nonumber\\
\hspace*{-15mm}
&&
\left. +\frac{d_i d_j d_\ell}{(\alpha N)^3}(\rme^{\rmi(\omega+\omega^\tprime)}-\rme^{\rmi\omega}-\rme^{\rmi\omega^\tprime}+1)+\frac{d_j d_k d_\ell}{(\alpha N)^3} (\rme^{\rmi(\omega^\prime+\omega^\dprime)}-\rme^{\rmi\omega^\prime}-\rme^{\rmi\omega^\dprime}+1) 
\right.
\nonumber\\
\hspace*{-15mm}
&&
\left. +\frac{d_i d_j d_k d_\ell}{(\alpha N)^4} (\rme^{\rmi(\omega+\omega^\prime+\omega^\dprime+\omega^\tprime)} - \rme^{\rmi(\omega + \omega^\prime)} - \rme^{\rmi(\omega^\dprime + \omega^\tprime)} - \rme^{\rmi(\omega + \omega^\tprime)}- \rme^{\rmi(\omega^\prime + \omega^\dprime)}
\right.
\nonumber\\
\hspace*{-15mm}
&&
\left. +\rme^{\rmi\omega}+\rme^{\rmi\omega^\prime}+\rme^{\rmi\omega^\dprime}+\rme^{\rmi\omega^\tprime} -1) \right) ^{\alpha N}
\nonumber\\
\hspace*{-15mm}
&=&
\frac{1}{N}\sum_{[ijk\ell]} \left\{ 1- \frac{d_i d_j}{\alpha N}-\frac{d_j d_k}{\alpha N}-\frac{d_k d_\ell}{\alpha N}-\frac{d_i d_\ell}{\alpha N} + \frac{d_i d_j d_k}{(\alpha N)^2} + \frac{d_i d_k d_\ell}{(\alpha N)^2} + 
\frac{d_i d_j d_\ell}{(\alpha N)^2} + \frac{d_j d_k d_\ell}{(\alpha N)^2} 
\right.
\nonumber\\
\hspace*{-15mm}
&&
\left. -\frac{d_i d_j d_k d_\ell}{(\alpha N)^3} -\frac{1}{2}\left[ \frac{d_i^2 d_j^2}{(\alpha N)^3} +\frac{d_j^2 d_k^2}{(\alpha N)^3} +\frac{d_k^2 d_\ell^2}{(\alpha N)^3}+ \frac{d_i^2 d_\ell^2}{(\alpha N)^3} + 2\frac{d_i d_j^2 d_k}{(\alpha N)^3}+ 2\frac{d_i d_j d_k d_\ell}{(\alpha N)^3}+ 
2\frac{d_i^2 d_j d_\ell}{(\alpha N)^3}
\right.\right.
\nonumber\\
\hspace*{-15mm}
&&
\left. \left. +2\frac{d_j d_k^2 d_\ell}{(\alpha N)^3} +2\frac{d_i d_j d_k d_\ell}{(\alpha N)^3}+2\frac{d_i d_k d_\ell^2}{(\alpha N)^3} \right] +\frac{1}{2} \left[ \frac{d_i^2 d_j^2}{(\alpha N)^2} +\frac{d_j^2 d_k^2}{(\alpha N)^2} +\frac{d_k^2 d_\ell^2}{(\alpha N)^2} +\frac{d_i^2 d_\ell^2}{(\alpha N)^2} + 2\frac{d_i d_j^2 d_k}{(\alpha N)^2} 
\right.\right.
\nonumber\\
\hspace*{-15mm}
&&
\left. \left. +2\frac{d_i d_j d_k d_\ell}{(\alpha N)^2}+2\frac{d_i^2 d_j d_\ell}{(\alpha N)^2} +2\frac{d_j d_k^2 d_\ell}{(\alpha N)^2} +2\frac{d_i d_j d_k d_\ell}{(\alpha N)^2} +2\frac{d_i d_k d_\ell^2}{(\alpha N)^2} -2\frac{d_i^2 d_j^2 d_k}{(\alpha N)^3} -2\frac{d_i^2 d_j^2 d_\ell}{(\alpha N)^3} 
\right.\right. 
\nonumber\\
\hspace*{-15mm}
&&
\left.\left.  -2\frac{d_i^2 d_j d_k d_\ell}{(\alpha N)^3}-2\frac{d_i d_j^2 d_k d_\ell}{(\alpha N)^3} 
-2\frac{d_i d_j^2 d_k^2}{(\alpha N)^3} 
-2\frac{d_i d_j d_k^2 d_\ell}{(\alpha N)^3} 
-2\frac{d_i d_j^2 d_k d_\ell}{(\alpha N)^3} -2\frac{d_j^2 d_k^2 d_\ell}{(\alpha N)^3} -2\frac{d_i d_j d_k^2 d_\ell}{(\alpha N)^3} 
\right.\right.
\nonumber\\
\hspace*{-15mm}
&&
\left.  -2\frac{d_i d_k^2 d_\ell^2}{(\alpha N)^3}-2\frac{d_i d_j d_k d_\ell^2}{(\alpha N)^3}-2\frac{d_j d_k^2 d_\ell^2}{(\alpha N)^3} -2\frac{d_i^2 d_j d_k d_\ell}{(\alpha N)^3} -2\frac{d_i^2 d_k d_\ell^2}{(\alpha N)^3} -2\frac{d_i^2 d_j d_\ell^2}{(\alpha N)^3} -2\frac{d_i d_j d_k d_\ell^2}{(\alpha N)^3} \right]
\nonumber\\
\hspace*{-15mm}
&&
-\frac{1}{6} \left[ \frac{d_i^3 d_j^3}{(\alpha N)^3}  +\frac{d_j^3 d_k^3}{(\alpha N)^3} 
+\frac{d_k^3 d_\ell^3}{(\alpha N)^3}+\frac{d_i^3 d_\ell^3}{(\alpha N)^3} +3\frac{d_i^2 d_j^3 d_k}{(\alpha N)^3} +3\frac{d_i d_j^3 d_k^2}{(\alpha N)^3} +3\frac{d_i^2 d_j^2 d_k d_\ell}{(\alpha N)^3} 
\right.
\nonumber\\
\hspace*{-15mm}
&&
+3\frac{d_i d_j d_k^2 d_\ell^2}{(\alpha N)^3}
+3\frac{d_i^3 d_j^2 d_\ell}{(\alpha N)^3} 
+3\frac{d_i^3 d_j d_\ell^2}{(\alpha N)^3} 
+3\frac{d_j^2 d_k^3 d_\ell}{(\alpha N)^3}+3\frac{d_j d_k^3 d_\ell^2}{(\alpha N)^3} +3\frac{d_i d_j^2 d_k^2 d_\ell}{(\alpha N)^3} +3\frac{d_i^2 d_j d_k d_\ell^2}{(\alpha N)^3}
\nonumber\\
\hspace*{-15mm}
&&
\left.\left.
 +3\frac{d_i d_k^2 d_\ell^3}{(\alpha N)^3}
+3\frac{d_i^2 d_k d_\ell^3}{(\alpha N)^3} \right] \right\}
\nonumber\\
\hspace*{-15mm}
&=&
(N-1)(N-2)(N-3) -4N^2\frac{\bra d\ket^2}{\alpha} +4N\frac{\bra d \ket^3}{\alpha^2} -2\frac{\bra d^2 \ket^2}{\alpha^3} - \frac{\bra d \ket^4}{\alpha^3}
\nonumber\\
\hspace*{-15mm}
&&-
4\frac{\bra d^2 \ket \bra d\ket^2}{\alpha^3} +2N\frac{\bra d^2 \ket^2}{\alpha^2} -2\frac{\bra d\ket^4}{\alpha^3} +4N\frac{\bra d^2 \ket \bra d\ket^2}{\alpha^2}  +2N\frac{\bra d\ket^4}{\alpha^2}
\nonumber\\
\hspace*{-15mm}
&&
-8\frac{\bra d^2 \ket^2 \bra d \ket}{\alpha^3} - 8\frac{\bra d^2 \ket\bra d\ket^3}{\alpha^3} -\frac{2}{3}\frac{\bra d^3 \ket^2}{\alpha^3} -4\frac{\bra d^3\ket\bra d^2\ket\bra d\ket}{\alpha^3} -2\frac{\bra d^2 \ket^2 \bra d\ket^2}{\alpha^3}
\eea
\end{document}